\documentclass{article}
\usepackage{epubtk}
\usepackage{epsf}
\usepackage{longtable}
\usepackage{lscape}
\usepackage{amssymb}

\newcommand{\vsp}{\rule{0.0 em}{1.2 em}}
\newcommand{\z}{\phantom{0}}
\newcommand{\zz}{\phantom{00}}
\newcommand{\zzz}{\phantom{000}}
\newcommand{\zzzz}{\phantom{0000}}

\newcommand{\zzzzzzz}{\phantom{0000000}}
\newcommand{\dz}{\phantom{.0}}
\newcommand{\dzz}{\phantom{.00}}
\newcommand{\dzzz}{\phantom{.000}}

\begin{document}

\title{Binary and Millisecond Pulsars}

\author{\epubtkAuthorData{Duncan R.\ Lorimer}
        {University of Manchester\\
         Jodrell Bank Observatory \\
         Macclesfield\\
         Cheshire, SK11 9DL, U.K.}
        {Duncan.Lorimer@manchester.ac.uk}
        {http://www.jb.man.ac.uk/~drl}
}

\date{}
\maketitle


\begin{abstract}
  We review the main properties, demographics and applications of
  binary and millisecond radio pulsars. Our knowledge of these
  exciting objects has greatly increased in recent years, mainly due
  to successful surveys which have brought the known pulsar population
  to over 1700. There are now 80 binary and millisecond pulsars
  associated with the disk of our Galaxy, and a further 103 pulsars in
  24 of the Galactic globular clusters. Recent highlights have been
  the discovery of the first ever double pulsar system and a recent
  flurry of discoveries in globular clusters, in particular Terzan~5.
\end{abstract}

\epubtkKeywords{pulsars}

\newpage


\section{Introduction and Overview}
\label{sec:preamble}

Pulsars -- rapidly rotating highly magnetised neutron stars -- have
resulted in many applications in physics and astronomy. Striking
examples include the confirmation of the existence of gravitational
radiation~\cite{nobpr1993} as predicted by general
relativity~\cite{tw82, tw89}, the first detection of an extra-solar
planetary system~\cite{wf92, psrplanets} and the discovery of the first
double-pulsar binary system~\cite{bdp03, lbk04}. The diverse zoo of
radio pulsars currently known is summarized in Figure~\ref{fig:venn}.

\epubtkImage{venn.png}{
  \begin{figure}[htbp]
    \def\epsfsize#1#2{0.4#1}
    \centerline{\epsfbox{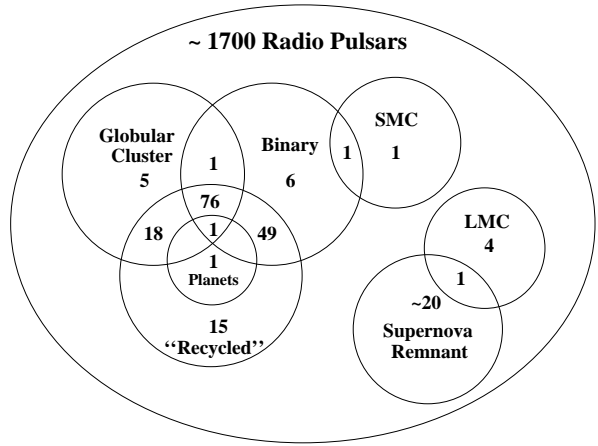}}
    \caption{\it Venn diagram showing the numbers and locations of the
      various types of radio pulsars known as of January 2005. The
      large and small Magellanic clouds are denoted by LMC and SMC.}
    \label{fig:venn}
  \end{figure}}

Pulsar research has proceeded at a rapid pace since the first two
versions of this article~\cite{lor98e, lor01}. Surveys mostly with the
Parkes radio telescope~\cite{mbeampsr}, but also at Green
Bank~\cite{gbt}, Arecibo~\cite{arecibo} and the Giant Metre Wave Radio
Telescope~\cite{gmrt} have more than doubled the number of pulsars
known back in 1997. The most exciting new results and discoveries from
these searches are discussed in this updated review.

We begin in Section~\ref{sec:intro} with an overview of the pulsar
phenomenon, the key observed population properties, the origin and
evolution of pulsars and the main search strategies. In
Section~\ref{sec:gal}, we review present understanding in pulsar
demography, discussing selection effects and their correction
techniques. This leads to empirical estimates of the total number of
normal and millisecond pulsars (see Section~\ref{sec:nmsppop}) and relativistic
binaries (see Section~\ref{sec:relpop}) in the Galaxy and has implications for
the detection of gravitational radiation from coalescing neutron 
star binaries
which these systems are the progenitors of. Our
review of pulsar timing in Section~\ref{sec:pultim} covers the basic techniques
(see Section~\ref{sec:tmodel}), timing stability (see Section~\ref{sec:tstab}), binary
pulsars (see Section~\ref{sec:tbin}), and 
using pulsars as sensitive detectors of long-period gravitational
waves (see Section~\ref{sec:gwdet}). We conclude with a brief
outlook to the future in Section~\ref{sec:future}. Up-to-date tables of
parameters of binary and millisecond pulsars are included in
Appendix~\ref{appendix}.

\newpage


\section{Pulsar Phenomenology}
\label{sec:intro}

Many of the basic observational facts about radio pulsars were
established shortly after their discovery~\cite{hbp68}
in 1967. Although there are still many open questions, the
basic model has long been established beyond all reasonable doubt,
i.e.\ pulsars are rapidly rotating, highly magnetised neutron stars
formed during the supernova explosions of massive
($\gtrsim 5 \mbox{--} 10\,M_\odot$) stars. In the following, we discuss the
observational properties that are most relevant to this review.


\subsection{The lighthouse model}
\label{sec:light}

Figure~\ref{fig:rotns} shows an animation depicting
the rotating neutron star or ``lighthouse'' model. 
As the neutron star spins, charged particles are accelerated out along
magnetic field lines in the magnetosphere (depicted by the light blue
cones). The accelerating particles emit electromagnetic
radiation, most readily detected at radio frequencies as a sequence of
observed pulses produced as the magnetic axis (and hence the radiation
beam) crosses the observer's line of sight each rotation. The
repetition period of the pulses is therefore simply the rotation
period of the neutron star. The moving ``tracker ball'' on the
pulse profile in the animation shows the relationship
between observed intensity and rotational phase of the neutron star.

\epubtkMovie{rotns.gif}{rotns.png}{
  \begin{figure}[htbp]
    \def\epsfsize#1#2{0.7#1}
    \centerline{\epsfbox{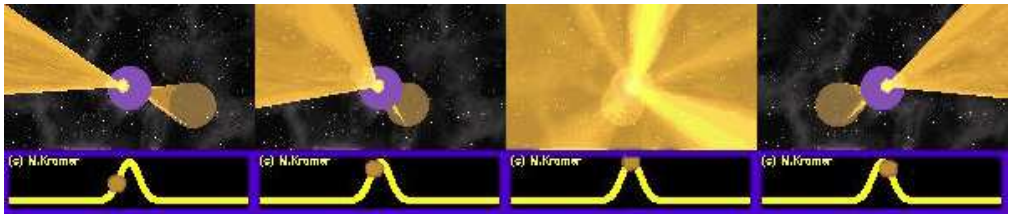}}
    \caption{\it GIF movie showing the rotating neutron star (or
      ``lighthouse'') model for pulsar emission. Animation designed by
      Michael Kramer.}
    \label{fig:rotns}
  \end{figure}}

Neutron stars are essentially large celestial flywheels with moments
of inertia $\sim 10^{38} \mathrm{\ kg\ m}^2$. The rotating neutron star
model~\cite{pac68, gol68} predicts a gradual slowdown and hence an
increase in the pulse
period as the outgoing radiation carries away rotational kinetic
energy. This model became universally accepted when a period increase
of $36.5 \mathrm{\ ns}$ per day was measured for the pulsar in the Crab
nebula~\cite{rc69b}, which implied that a rotating neutron star with a
large magnetic field must be the dominant energy supply for the
nebula~\cite{gol69}.


\subsection{Pulse periods and slowdown rates}
\label{sec:nms}

The present public-domain catalogue~\cite{mhth05}, available
on-line~\cite{psrcat}, contains up-to-date parameters for over 1500
pulsars. Most of these are ``normal'' with pulse periods $P \sim 0.5 \mathrm{\ s}$
which increase secularly at rates $\dot{P} \sim 10^{-15} \mathrm{\ s/s}$.
A growing fraction are ``millisecond pulsars'',
with $1.5 \mathrm{\ ms} \lesssim P \lesssim \lesssim 30 \mathrm{\ ms}$
and $\dot{P} \lesssim 10^{-19} \mathrm{\ s/s}$. For many years, the
most rapidly rotating neutron star known was the original millisecond
pulsar B1937+21~\cite{bkh82}, with $P=1.5578 \mathrm{\ ms}$. However,
one of the recent discoveries in the globular cluster Terzan 5~appears
to be rotating even more rapidly, with
$P = 1.4 \mathrm{\ ms}$~\cite{ran05}. Confirmation of this exciting
result should be published in due course. While the hunt for 
``sub-millisecond pulsars'' continues, and most neutron star
equations of state allow shorter periods, it has been
suggested~\cite{bil98, cmm03} that the lack of pulsars with
$P<1.5 \mathrm{\ ms}$ is caused by gravitational wave emission from R-mode
instabilities~\cite{ajks00}. 

As can be seen from the ``$P \mbox{--} \dot{P}$ diagram'' in
Figure~\ref{fig:ppdot}, normal and millisecond pulsars are distinct
populations. The differences in $P$ and $\dot{P}$ imply fundamentally
different magnetic field strengths and ages. Treating the pulsar as a
rotating magnetic dipole, one may show~\cite{lk05} that the surface
magnetic field strength $B \propto (P \dot{P})^{1/2}$ and the 
characteristic age $\tau_\mathrm{c} = P/(2\dot{P})$.

Lines of constant $B$ and $\tau_\mathrm{c}$ are drawn on
Figure~\ref{fig:ppdot}, from which we infer typical
values of $10^{12} \mathrm{\ G}$ and $10^{7} \mathrm{\ yr}$ for the normal
pulsars and $10^{8} \mathrm{\ G}$ and $10^{9} \mathrm{\ yr}$ for the
millisecond pulsars. For the rate of loss of kinetic energy, sometimes
called the spin-down luminosity, we have $\dot{E} \propto \dot{P}/P^3$. The lines of
constant $\dot{E}$ shown on Figure~\ref{fig:ppdot} show that the most
energetic objects are the very young normal pulsars and the most
rapidly spinning millisecond pulsars.

\epubtkImage{ppdot.png}{
  \begin{figure}[htbp]
    \def\epsfsize#1#2{0.6#1}
    \centerline{\epsfbox{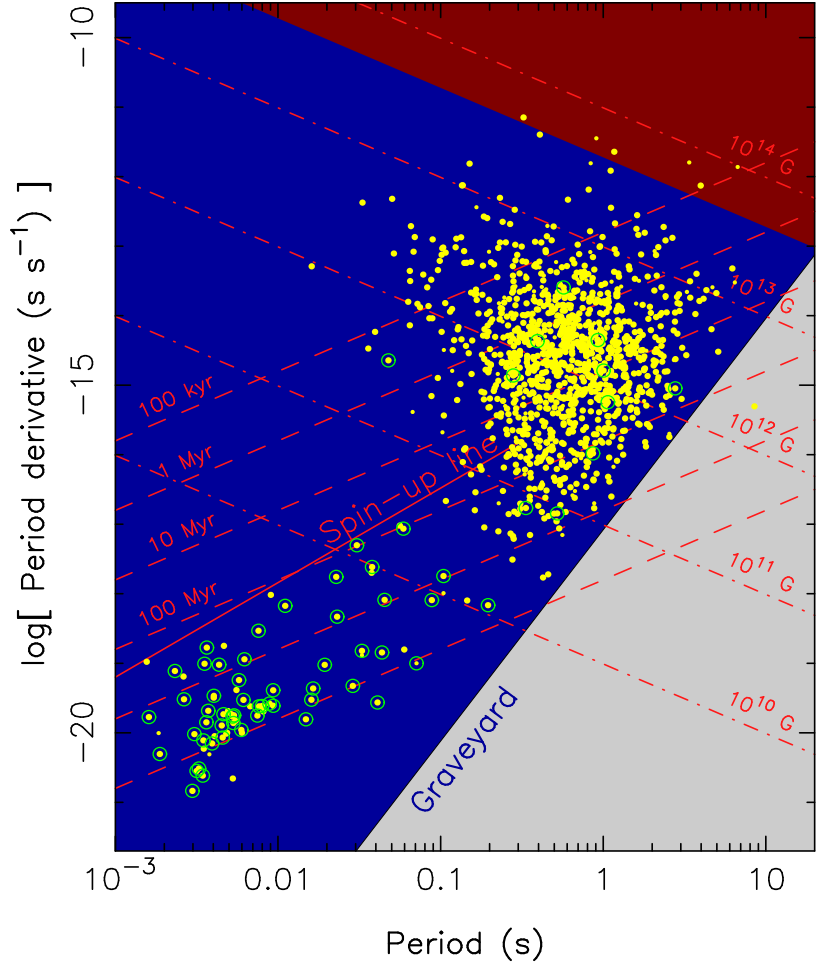}}
    \caption{\it The $P \mbox{--} \dot{P}$ diagram showing the current
      sample of radio pulsars. Binary pulsars are highlighted by open
      circles. Theoretical models~\cite{cr93} do not predict radio
      emission outside the dark blue region. Figure provided by
      Michael Kramer.}
    \label{fig:ppdot}
  \end{figure}}


\subsection{Pulse profiles}
\label{sec:profs}

Pulsars are weak radio sources. Measured intensities, usually quoted
in the literature for a radio frequency of $400 \mathrm{\ MHz}$, vary
between $0.1 \mathrm{\ mJy}$
and $5 \mathrm{\ Jy}$ ($1 \mathrm{\ Jy} \equiv 10^{-26} \mathrm{\ W\
  m}^{-2} \mathrm{\ Hz}^{-1}$). As a result, even with a large radio telescope, the
coherent addition of many hundreds or even thousands of pulses is
usually required in order to produce a detectable ``integrated
profile''. Remarkably, although the individual pulses vary
dramatically, the integrated profile at any particular observing
frequency is very stable and can be thought of as a fingerprint of the
neutron star's emission beam. Profile stability is of key importance
in pulsar timing measurements discussed in Section~\ref{sec:pultim}.

The selection of integrated profiles in Figure~\ref{fig:profs} shows a
rich diversity in morphology including two
examples of ``interpulses'' -- a secondary pulse separated by about
180 degrees from the main pulse. The most natural interpretation for
this phenomenon is that the two pulses originate from opposite
magnetic poles of the neutron star (see however~\cite{ml77}). Since
this is an unlikely viewing angle we would expect interpulses to be a
rare phenomenon. Indeed, the fraction of known
pulsars in which interpulses are observed in their pulse profiles is
only a few percent~\cite{kgm04}.

\epubtkImage{profs.png}{
  \begin{figure}[htbp]
    \def\epsfsize#1#2{0.55#1}
    \centerline{\epsfbox{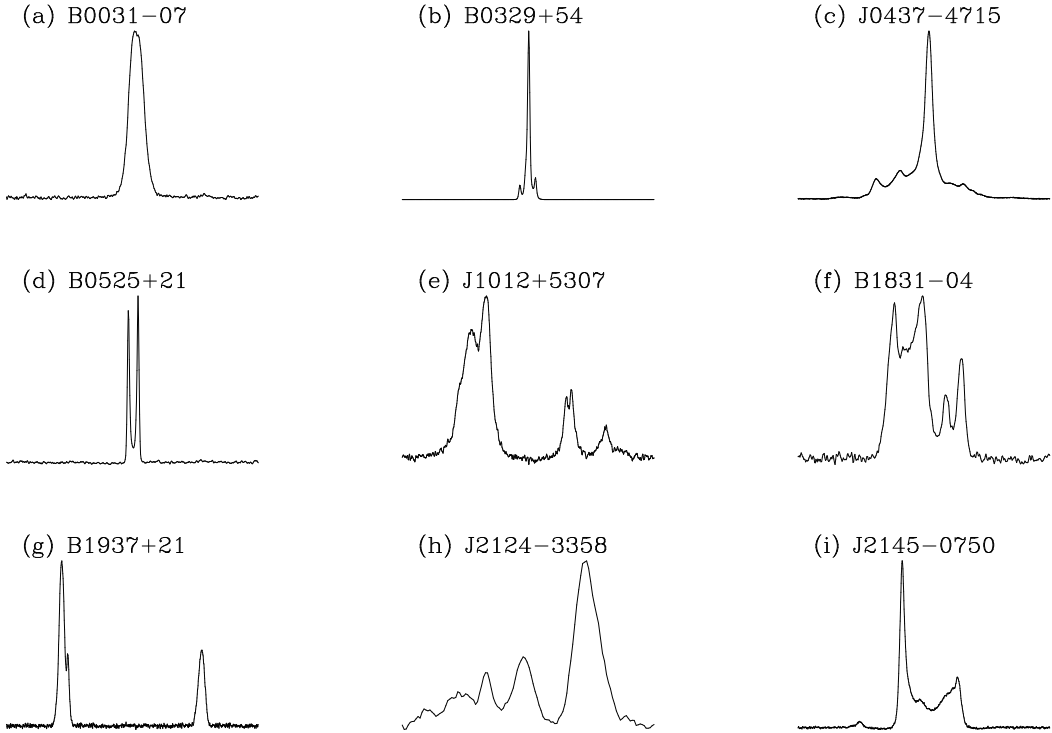}}
    \caption{\it A variety of integrated pulse profiles taken from the
      available literature. References: Panels~a, b, d, f~\cite{gl98},
      Panel~c~\cite{bjb97}, Panels~e, g, i~\cite{kxl98},
      Panel~h~\cite{bbm97}. Each profile represents 360 degrees of
      rotational phase. These profiles are freely available from an
      on-line database~\cite{epndb}.}
    \label{fig:profs}
  \end{figure}}

Two contrasting phenomenological models to explain the observed pulse
shapes are shown in Figure~\ref{fig:shapes}. The ``core and cone''
model~\cite{ran83} depicts the beam as a core surrounded by a series
of nested cones. Alternatively, the ``patchy beam''
model~\cite{lm88, hm01} has the beam populated by a series of
randomly-distributed emitting regions. Further work in this area is
necessary to improve our understanding of the shape and evolution of
pulsar beams and fraction of sky they cover. This is of key importance
to the results of population studies reviewed in Section~\ref{sec:corsamp}.

\epubtkImage{shapes.png}{
  \begin{figure}[htbp]
    \def\epsfsize#1#2{0.6#1}
    \centerline{\epsfbox{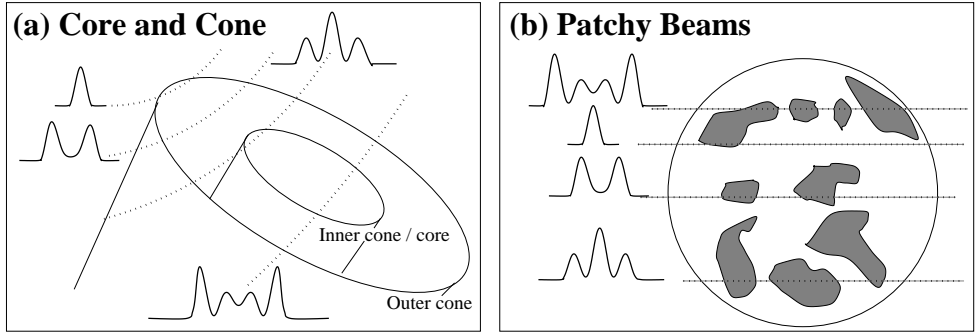}}
    \caption{\it Phenomenological models for pulse shape morphology
      produced by different line-of-sight cuts of the beam. Figure
      designed by Michael Kramer.}
    \label{fig:shapes}
  \end{figure}}


\subsection{Interstellar dispersion and the pulsar distance scale}
\label{sec:dist}

From the sky distribution shown in Figure~\ref{fig:aitoff} it is immediately 
apparent that pulsars are strongly concentrated along the Galactic plane. 
This indicates that pulsars populate the disk of our Galaxy. 
Quantitative estimates of the distance
to each pulsar can be made from the measurement of \emph{pulse dispersion} -- 
the delay in pulse arrival times across a finite bandwidth.
Dispersion occurs because the group velocity of the pulsed
radiation through the ionised component of the 
interstellar medium is frequency
dependent: Pulses emitted at lower radio frequencies travel slower
through the interstellar medium, arriving later than those emitted
at higher frequencies. The delay $\Delta t$ in arrival times between a
high frequency $\nu_\mathrm{hi}$ and a low frequency $\nu_\mathrm{lo}$
pulse is given~\cite{lk05} by
\begin{equation}
  \Delta t = 4.15 \mathrm{\ ms} \times
  \left[ \left( \frac{\nu_\mathrm{lo}}{\mathrm{GHz}} \right)^{-2} - 
  \left( \frac{\nu_\mathrm{hi}}{\mathrm{GHz}} \right)^{-2} \right] \times
  \left( \frac{\mathrm{DM}}{\mathrm{cm}^{-3}\mathrm{\ pc}} \right),
  \label{equ:defdt}
\end{equation}
where the dispersion measure is
\begin{equation}
  \mathrm{DM} = \int_0^d \!\!\! n_\mathrm{e} \, dl
  \label{equ:defdm}
\end{equation}
is the integrated free electron column density $n_\mathrm{e}$ out to the pulsar
at a distance $d$. This equation may be solved for $d$ given a 
measurement of DM and a model of the free electron distribution.

\epubtkImage{aitoff.png}{
  \begin{figure}[htbp]
    \def\epsfsize#1#2{0.7#1}
    \centerline{\epsfbox{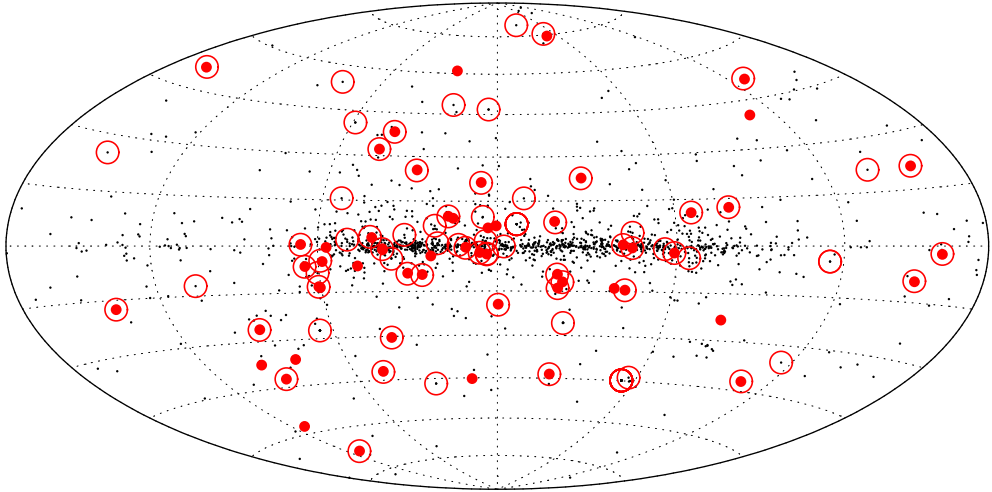}}
    \caption{\it The sky distribution of pulsars in Galactic
      coordinates. The plane of the Galaxy is the central horizontal
      line. The Galactic centre is the midpoint of this
      line. Millisecond pulsars are indicated in red. Binary pulsars
      are highlighted by the open circles. The more isotropic
      distribution of the millisecond and binary pulsars reflects the
      differences in detecting them out to large distances cf.\ the
      normal population (see Section~\ref{sec:gal}).}
    \label{fig:aitoff}
  \end{figure}}

In practice, the electron density model is calibrated from the 100 or
so pulsars with independent distance estimates and measurements of
scattering for lines of sight towards various Galactic and
extragalactic sources~\cite{tc93, wei96}. The most recent model of this
kind~\cite{cl02a, cl02b} provides distance estimates with an average 
uncertainty of $\sim 30$\%.


\subsection{Pulsars in binary systems}
\label{sec:bincomps}

As can be inferred from Figure~\ref{fig:venn}, only 4\% of all
known pulsars in the Galactic disk are members of binary systems.
Timing measurements (see Section~\ref{sec:pultim}) place useful constraints on
the masses of the companions which, supplemented by observations at
other wavelengths, tell us a great deal about their nature. The
present sample of orbiting companions are either white dwarfs, main
sequence stars or other neutron stars. Two notable hybrid systems
are the ``planet pulsars'' B1257+12 and B1620$-$26. PSR B1257+12 is a
$6.2 \mbox{-}\mathrm{ms}$ pulsar accompanied by at least three terrestrial-mass
bodies~\cite{wf92, psrplanets, wdk00} while B1620$-$26, an $11 \mbox{-}\mathrm{ms}$ pulsar
in the globular cluster M4, is part of a triple system with 
a $1 \mbox{--} 2\,M_\mathrm{Jupiter}$ planet~\cite{tat93, bfs93, tacl99, srh03}
orbiting a neutron star--white dwarf stellar binary system.
The current limits from pulsar timing 
favour a roughly $45 \mbox{-}\mathrm{yr}$ low-eccentricity ($e \sim 0.16$) orbit with semi-major
axis $\sim 25 \mathrm{\ AU}$. Despite several tentative claims over the years,
no other convincing cases for planetary companions to pulsars exist.
Orbiting companions are much more common around millisecond
pulsars ($\sim 80\%$ of the observed sample) than around the normal
pulsars ($\lesssim 1\%$). Binary systems below the
line have low-mass companions ($\lesssim 0.7\,M_\odot$
-- predominantly white dwarfs) and essentially circular orbits:
$10^{-5} \lesssim e \lesssim 0.01$. Binary pulsars with high-mass companions
($\gtrsim 1\,M_\odot$ -- massive white dwarfs, other neutron stars 
or main sequence stars) tend to have more eccentric orbits, 
$0.15 \gtrsim e \gtrsim 0.9$.


\subsection{Evolution of normal and millisecond pulsars}
\label{sec:evolution}

A simplified version of the presently favoured
model~\cite{bk74, fv75, sb76, acrs82} to explain the formation of the
various types of systems observed is shown in Figure~\ref{fig:bevol}.
Starting with a binary star system, a neutron star is formed during
the supernova explosion of the initially more massive star. From the
virial theorem, in the absence of any other factors,
the binary will be disrupted if more than half the
total pre-supernova mass is ejected from the system during the
(assumed symmetric) 
explosion~\cite{hil83, bv91}. In practice, the fraction of surviving
binaries is also affected by the magnitude and direction of any impulsive
``kick'' velocity the neutron star receives at
birth from a slightly asymmetric explosion~\cite{hil83, bai89}. 
Binaries that disrupt produce a
high-velocity isolated neutron star and an OB runaway
star~\cite{bla61}. The high probability of disruption explains
qualitatively why so few normal pulsars have companions.
There are currently four known normal radio pulsars with massive main
sequence companions in eccentric orbits which are examples of
binary systems which survived the supernova 
explosion~\cite{jml92, kjb94, sml01, sml03, lyn05}. Over the
next $10^{7 \mbox{--} 8} \mathrm{\ yr}$ after the explosion, the neutron star may be
observable as a normal radio pulsar spinning down to a period $\gtrsim$
several seconds. After this time, the energy output of the star
diminishes to a point where it no longer produces significant amounts
of radio emission.

\epubtkImage{bevol.png}{
  \begin{figure}[htbp]
    \def\epsfsize#1#2{0.21#1}
    \centerline{\epsfbox{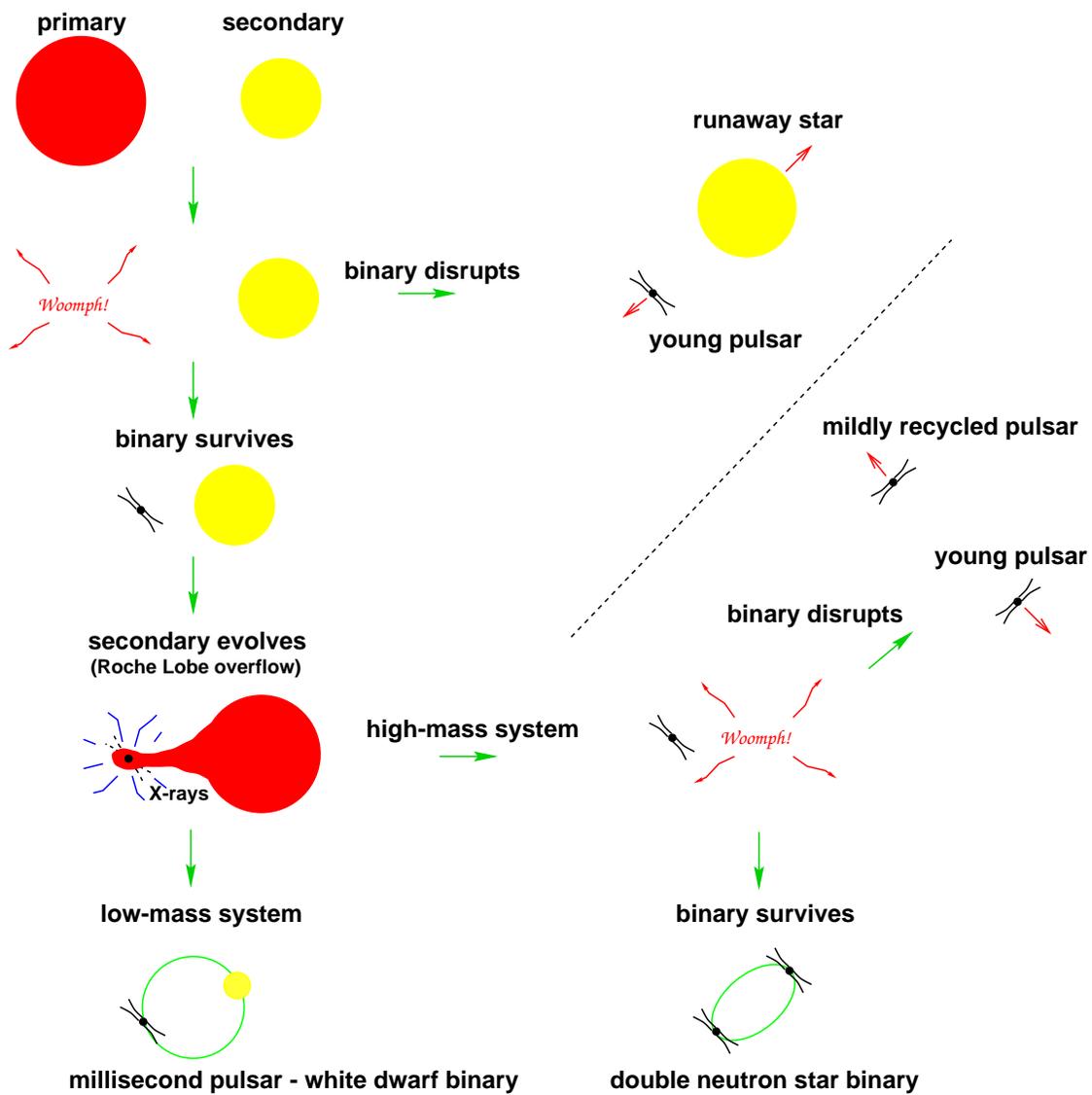}}
    \caption{\it Cartoon showing various evolutionary scenarios
      involving binary pulsars.}
    \label{fig:bevol}
  \end{figure}}

For those few binaries that remain bound, and in which the companion
is sufficiently massive to evolve into a giant and
overflow its Roche lobe, the old spun-down neutron star can gain a new
lease of life as a pulsar by accreting matter and angular momentum at
the expense of the orbital angular momentum of the binary
system~\cite{acrs82}. The term ``recycled pulsar'' is often used to
describe such objects. During this accretion phase, the X-rays
produced by the liberation of gravitational energy of the infalling
matter onto the neutron star mean that such a system is expected to be
visible as an X-ray binary. Two classes of X-ray binaries relevant to
binary and millisecond pulsars exist: neutron stars with high-mass or
low-mass companions. Detailed reviews of the X-ray binary
population, including systems likely to contain black holes,
can be found elsewhere~\cite{bv91}.

\epubtkImage{pbe.png}{
  \begin{figure}[htbp]
    \def\epsfsize#1#2{0.65#1}
    \centerline{\epsfbox{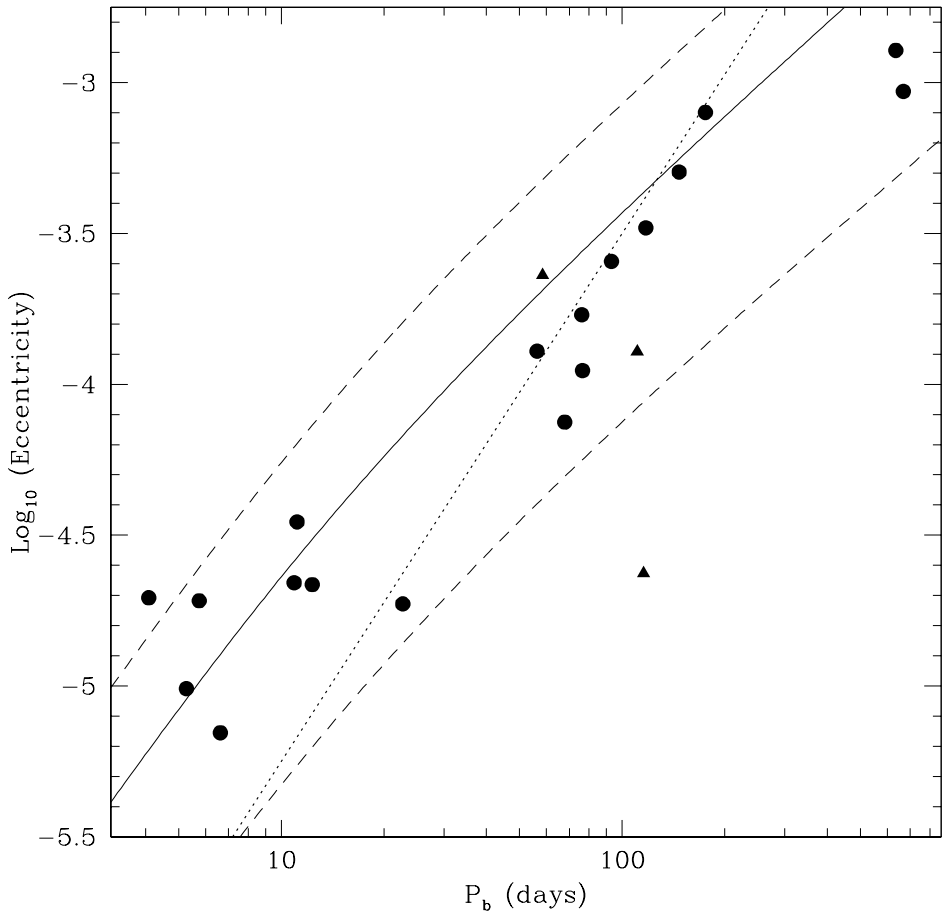}}
    \caption{\it Eccentricity versus orbital period for a sample of 21
      low-mass binary pulsars which are not in globular clusters, with
      the triangles denoting three recently discovered
      systems~\cite{sfl05}. The solid line shows the median of the
      predicted relationship between orbital period and
      eccentricity~\cite{phi92}. Dashed lines show 95\% the confidence
      limit about this relationship. The dotted line shows
      $P_\mathrm{b} \propto e^2$. Figure provided by Ingrid
      Stairs~\cite{sfl05} using an adaptation of the orbital
      period-eccentricity relationship tabulated by Fernando Camilo.}
    \label{fig:pbe}
  \end{figure}}

In a high-mass X-ray binary, the companion is massive enough that it
might also explode as a supernova, producing a second neutron star. If
the binary system is lucky enough to survive the explosion, the result
is a double neutron star binary. At least five and probably eight such
systems are now known, the original example being PSR
B1913+16~\cite{ht75a} -- a $59 \mbox{-}\mathrm{ms}$ radio pulsar which orbits its
companion every $7.75 \mathrm{\ hr}$~\cite{tw82, tw89}. In this formation scenario,
PSR B1913+16 is an example of the older, first-born, neutron star that
has subsequently accreted matter from its companion.

For many years, no clear example was known where the second-born
neutron star was observed as a pulsar. The discovery of the double pulsar
J0737$-$3039~\cite{bdp03, lbk04}, where a $22.7 \mbox{-}\mathrm{ms}$ recycled pulsar ``A''
orbits a $2.77 \mbox{-}\mathrm{s}$ normal pulsar ``B'' every $2.4 \mathrm{\ hr}$, has now provided a
dramatic confirmation of this evolutionary model in which we identify
A and B as the first and second-born neutron stars respectively.
Just how many more observable double pulsar systems exist in our Galaxy
is not clear. Although the population of double neutron star systems
in general is reasonably well understood (see Section~\ref{sec:nsns}),
given that the lifetime of the second born pulsar is less than
one tenth that of the recycled pulsar, and that its radio beam
is likely to be much smaller (see Section~\ref{sec:beaming}),
the prospects of ever finding
more than a few 0737-like systems are rather low. 

\epubtkImage{porbmwd.png}{
  \begin{figure}[htbp]
    \def\epsfsize#1#2{0.38#1}
    \centerline{\epsfbox{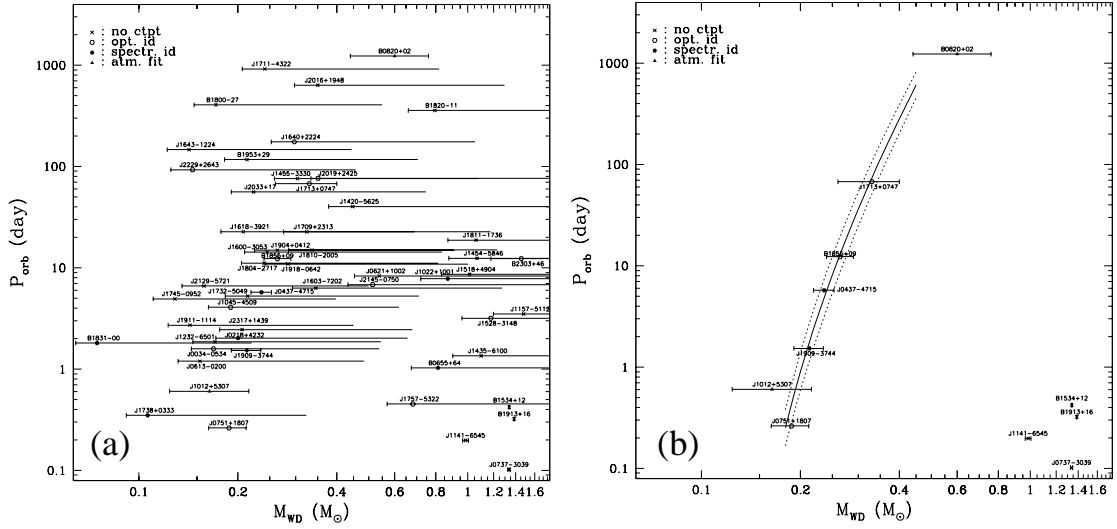}}
    \caption{\it Companion mass versus orbital period for binary pulsars
      showing the whole sample where, in the absence of mass
      determinations, statistical arguments based on a random
      distribution of orbital inclination angles (see
      Section~\ref{sec:tbin}) have been used to constrain the masses as
      shown (Panel~a), and only those with well determined companion
      masses (Panel~b). The dashed lines show the uncertainties in the
      predicted relation~\cite{ts99}. This relationship indicates that
      as these systems finished a period of stable mass transfer due to
      Roche-lobe overflow, the size and hence period of the orbit was
      determined by the mass of the evolved secondary star. Figure
      provided by Marten van Kerkwijk~\cite{vbjj05}.}
    \label{fig:porbmwd}
  \end{figure}}

The companion in a low-mass X-ray binary evolves and transfers
matter onto the neutron star on a much longer time-scale, spinning it
up to periods as short as a few ms~\cite{acrs82}. Tidal forces
during the accretion process serve to circularize the orbit.
At the end of the
spin-up phase, the secondary sheds its outer layers to become a white
dwarf in orbit around a rapidly spinning millisecond pulsar. This
model has gained strong support in recent years from the discoveries
of quasi-periodic kHz oscillations in a number of low-mass X-ray
binaries~\cite{wz97}, as well as Doppler-shifted $2.49 \mbox{-}\mathrm{ms}$ X-ray
pulsations from the transient X-ray burster SAX
J1808.4$-$3658~\cite{wv98, cm98}. Six other ``X-ray millisecond
pulsars'' are now known with spin rates and orbital periods ranging
between $185 \mbox{--} 600 \mathrm{\ Hz}$ and
$40 \mathrm{\ min} \mbox{--} 4.3 \mathrm{\ hr}$ respectively~\cite{wij05, mkv05a, mkv05b}.
Despite intensive searches~\cite{bbp03}, no radio pulsations have so
far been detected in these binaries. This could be a result of
free-free absorption of any radio waves by the thick accretion disk,
or perhaps quenching the accelerating potential in the neutron
star magnetosphere by infalling matter. More sensitive radio observations
are ultimately required to place stringent limits on the physical
processes responsible for the lack of emission.

Numerous examples of these systems in their post X-ray phase are now
seen as the millisecond pulsar--white dwarf binary systems.
Presently, 20 of these systems have compelling optical identifications
of the white dwarf companion, and upper limits or tentative detections
have been found in about 30 others~\cite{vbjj05}. Comparisons between
the cooling ages of the white dwarfs and the millisecond pulsars
confirm the age of these systems and suggest that the accretion rate
during the spin-up phase was well below the Eddington limit~\cite{hp98}.

Further support for the above evolutionary scenarios comes from two
correlations in the observed sample of low-mass binary pulsars.
Firstly, as seen in Figure~\ref{fig:pbe}, there is a strong correlation
between orbital period and eccentricity. The data are very good
agreement with a theoretical relationship which predicts a relic
orbital eccentricity due to convective eddy currents in the accretion
process~\cite{phi92}.
Secondly, as shown in Panel~b of Figure~\ref{fig:porbmwd}, where companion masses
have been measured accurately, through radio
timing~(see Section~\ref{sec:tbin}) and/or through optical
observations~\cite{vbjj05}, they are in good agreement with a relation
between companion mass and orbital period predicted by binary
evolution theory~\cite{ts99}. A word of caution is required in using
these models to make predictions, however. When confronted with a
larger ensemble of binary pulsars using statistical arguments to
constrain the companion masses (see Panel~a of Figure~\ref{fig:porbmwd}), current
models have problems in explaining the full range of orbital periods on this
diagram~\cite{sfl05}.

The range of white dwarf masses observed is becoming broader. Since
this article originally appeared~\cite{lor98e}, the number of
``intermediate-mass binary pulsars''~\cite{cam96c} has grown
significantly~\cite{clm01}. These systems are distinct to the
millisecond pulsar--white dwarf binaries in several ways:
\begin{enumerate}
\item The spin period of the radio pulsar is generally longer
  ($9 \mbox{--} 200 \mathrm{\ ms}$).
\item The mass of the white dwarf is larger (typically
  $\gtrsim 0.5\,M_\odot$).
\item The orbit, while still essentially circular, is often
  significantly more eccentric ($e \gtrsim 10^{-3}$).
\item They do not necessarily follow the mass--period or
  eccentricity--period relationships.
\end{enumerate}
It is not presently clear whether these systems
originated from low- or high-mass X-ray binaries. It was suggested by
van den Heuvel~\cite{vdh94} that they have more in common with
high-mass systems. More recently, it has been proposed~\cite{li02}
that a thermal-viscous instability in the accretion disk of a low-mass
X-ray binary could truncate the accretion phase and produce a more
slowly spinning neutron star.


\subsection{Isolated recycled pulsars}

The scenarios outlined qualitatively above represent a reasonable
understanding of binary evolution. There are, however, a number of
pulsars with spin properties that suggest a phase of recycling took
place but have no orbiting companions. While the existence of such
systems in globular clusters are more readily explained by the high
probability of stellar interactions compared to the disk~\cite{sp95},
it is somewhat surprising to find them in the Galactic disk. Out of a
total of 66 millisecond pulsars in the Galactic disk, 15 are isolated
(see Table~\ref{tab:imsps}). Although it has been proposed that these
millisecond pulsars have ablated their companion via their strong
relativistic winds~\cite{krst88} as may be happening in the
PSR B1957+20 system~\cite{fst88}, it is not clear whether the
energetics or time-scales for this process are
feasible~\cite{le91}. There is some observational evidence that
suggests that solitary millisecond pulsars are less luminous than
binary millisecond pulsars~\cite{bjb97, kxl98}. If confirmed by future
discoveries, this would need to be explained by any viable
evolutionary model.

There are two further ``anomalous'' isolated pulsars with periods in
the range $55 \mbox{--} 60 \mathrm{\ ms}$~\cite{cnt93, lma04}. When placed on the
$P \mbox{--} \dot{P}$ diagram, these objects populate the region occupied by the
double neutron star binaries. The most natural explanation for their
existence, therefore, is that they are ``failed double neutron star
binaries'' which disrupted during the supernova explosion of the
secondary~\cite{cnt93}. 
A simple calculation~\cite{lma04}, however, suggests that
for every double neutron star we should see of order ten such isolated
objects. Although some other examples of such pulsars are
known~\cite{kl00, eb01b, bur04}, exactly why so few are observed is
currently not clear.


\subsection{Pulsar velocities}
\label{sec:pvel}

Pulsars have long been known to have space velocities at least an order 
of magnitude larger than those of their main sequence progenitors, which 
have typical values between 10 and $50 \mathrm{\ km\ s}^{-1}$. The first direct 
evidence for large velocities came from optical observations of the Crab 
pulsar~\cite{tri68}, showing that the neutron star has a velocity 
in excess of $100 \mathrm{\ km\ s}^{-1}$. Proper motions for 233 pulsars have 
subsequently been measured largely by radio timing and interferometric
techniques~\cite{las82, bmk90b, fgl92, hla93, hllk05}. These data imply a broad
velocity spectrum ranging from 0 to over $1000 \mathrm{\ km\ s}^{-1}$~\cite{ll94}.

\epubtkMovie{init.gif}{init.png}{
  \begin{figure}[htbp]
    \def\epsfsize#1#2{0.8#1}
    \centerline{\epsfbox{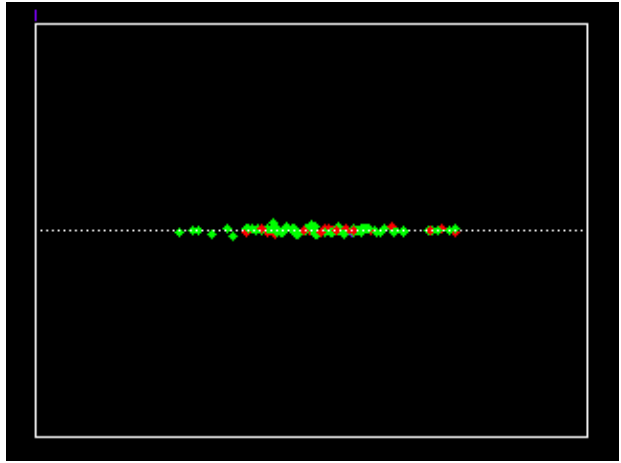}}
    \caption{\it GIF movie showing a simulation following the motion of
      100 pulsars in a model gravitational potential of our Galaxy for
      $ 200 \mathrm{\ Myr} $. The view is edge-on, i.e.\ the
      horizontal axis represents the Galactic plane
      ($ 30 \mathrm{\ kpc} $ across) while the vertical axis
      represents $\pm 10 \mathrm{\ kpc} $ from the plane. This snapshot
      shows the initial configuration of young neutron stars.}
    \label{fig:migrate}
  \end{figure}}

Such large velocities are perhaps not surprising, given the violent
conditions under which neutron stars are
formed. Shklovskii~\cite{shk70} demonstrated that, if the explosion is
only slightly asymmetric, an impulsive ``kick'' velocity of up to
$1000 \mathrm{\ km\ s}^{-1}$ can be imparted to the neutron star. In
addition, if the neutron star progenitor was a member of a binary
system prior to the explosion, the pre-supernova orbital velocity will
also contribute to the resulting speed of the newly-formed pulsar. As
Figure~\ref{fig:migrate} illustrates, high-velocity pulsars born close
to the Galactic plane quickly migrate to higher Galactic latitudes.
Given such a broad velocity spectrum, as many as half of all pulsars
will eventually escape the Galactic gravitational potential~\cite{ll94, cc98}.

The distribution of pulsar velocities remains an issue of contention.
For example, Monte Carlo simulations by Arzoumanian et
al.~\cite{acc02} strongly favour a bimodal distribution with low and
high velocity peaks. On the other hand, a recent study~\cite{hllk05}
found the mean birth velocity of normal pulsars to be consistent with
a Maxwellian distribution with a mean of $\sim 400 \mathrm{\ km\
s}^{-1}$. Regardless of the form of the distribution, however, we can
say that the mean velocity of young pulsars is significantly larger
than the $80 \mbox{--} 140 \mathrm{\ km\ s}^{-1}$ range for millisecond and binary
pulsars~\cite{lor95, cc97, lml98}. The main reasons for the lower
velocities are the fact that the kick must have been small to avoid
disruption, and the surviving neutron star has to pull the companion
along with it, thus slowing the system down.


\subsection{Current and future binary and millisecond pulsar search strategies}
\label{sec:wheretolook}


\subsubsection{All-sky searches}
\label{sec:allsky}

The oldest radio pulsars form a relaxed population of stars
oscillating in the Galactic gravitational potential~\cite{hp97}. The
scale height for such a population is at least $500 \mathrm{\ pc}$, about 10 times
that of the massive stars which populate the Galactic plane. Since the
typical ages of millisecond pulsars are several Gyr or more, we
expect, from our vantage point in the Galaxy, to be in the middle of
an essentially isotropic population of nearby sources. All-sky
searches for millisecond pulsars at high Galactic latitudes have been
very effective in probing this population. Much of the initial
interest and excitement in this area began with the discovery of two
recycled pulsars at high latitudes with the $305 \mbox{-}\mathrm{m}$ Arecibo telescope:
the double neutron star binary B1534+12~\cite{wol91a} and the
``planets pulsar'' B1257+12~\cite{wf92}. Surveys carried out at
Arecibo, Parkes, Jodrell Bank and Green Bank (using the 140 ft telescope)
by others in the 1990s
(summarised in three review papers~\cite{cam95, cam97, cam98}) have
found many other millisecond and recycled pulsars in this way.


\subsubsection{Searches close to the plane of our Galaxy}
\label{sec:plane}

Young pulsars are most likely to be found near to their places of
birth, before they have had time for their velocity to move them away, and
hence they lie close to the Galactic plane (see Figure~\ref{fig:aitoff}). This was
the target region of the main Parkes multibeam survey and has
already resulted in the discovery of over 700 new
pulsars~\cite{mlc01, mhl02, kbm03, hfs04, fsk04}, almost half the number
currently known! Such a large haul inevitably results in a number of
interesting individual objects such as PSR J1141$-$6545, a young
pulsar in a relativistic $4.8 \mbox{-}\mathrm{hr}$ orbit around a white
dwarf~\cite{klm00, obv02, bokh03, hbo05}, PSR J1740$-$3052, a young
pulsar orbiting an $\sim 11\,M_{\odot}$ star (probably a main sequence 
B-star~\cite{sml01, sml03}), PSR 1638$-$4715, a young pulsar in a $\sim
5 \mathrm{\ yr}$-eccentric orbit ($e \sim 0.9$) around around a
$10 \mbox{--} 20\,M_\odot$ companion~\cite{lyn05}, several intermediate-mass
binary pulsars~\cite{clm01}, and two double neutron star 
binaries~\cite{lcm00, fkl05}. Although the
main survey has now been completed, extensions of the survey region,
and re-analyses of existing data~\cite{fsk04} 
will ensure further discoveries in
the near future.


\subsubsection{Searches at intermediate Galactic latitudes}
\label{sec:intlat}

To probe more deeply into the population of millisecond and recycled
pulsars than possible at high Galactic latitudes, the Parkes multibeam
system was also used to survey intermediate
latitudes~\cite{edw00, eb01}. Among the 69 new pulsars found in the
survey, 8 are relatively distant recycled objects. Two of the new recycled
pulsars from this survey~\cite{eb01} are mildly relativistic neutron
star-white dwarf binaries. An analysis of the full results from this
survey should significantly improve our knowledge on the Galaxy-wide
population and birth-rate of millisecond pulsars. 
Arecibo surveys at intermediate latitudes also continue to find new
pulsars, such as the long-period binaries J2016+1948 and 
J0407+1607~\cite{naf03, lxf05}, and
the likely double neutron star system J1829+2456~\cite{clm04}.


\subsubsection{Targeted searches of globular clusters}
\label{sec:globs}

Globular clusters have long been known to be breeding grounds for
millisecond and binary pulsars~\cite{cr05}. The main reason for this
is the high stellar density and consequently high rate of stellar interaction
in globular clusters relative to most of
the rest of the Galaxy. As a result, low-mass X-ray binaries are
almost 10 times more abundant in clusters than in the Galactic
disk. In addition, exchange interactions between binary and multiple
systems in the cluster can result in the formation of exotic binary
systems~\cite{sp95}. To date, searches have revealed 103 pulsars in
24 globular clusters (see Table~\ref{tab:gcpsrs} and~\cite{gcpsrs, cr05}). 
Early highlights include the
double neutron star binary in M15~\cite{pakw91} and a low-mass binary
system with a $95 \mbox{-}\mathrm{min}$ orbital period in 47~Tucanae~\cite{clf00}, one of 22
millisecond pulsars currently known in this cluster alone~\cite{clf00, lcf03}.

On-going surveys of clusters continue to yield new
surprises~\cite{rgh01, dlm01} including the discovery of the most
eccentric binary pulsar so far -- J0514$-$4002 is a $4.99 \mathrm{\ ms}$ pulsar in
a highly eccentric ($e=0.89$) binary system in the globular cluster
NGC~1851~\cite{fgr04}. The latest sensation, however, is the
discovery~\cite{rhs05, ran05} of 28 millisecond pulsars in Terzan~5 with the
Green Bank Telescope~\cite{gbt}. This brings the total known in this
cluster to 26. The spin periods and orbital parameters of the new
pulsars reveal that, as a population, they are significantly different
to the pulsars of 47~Tucanae which have periods in the range
$2 \mbox{--} 8 \mathrm{\ ms}$~\cite{lcf03}. The spin periods of the new pulsars span a much
broader range ($1.67 \mbox{--} 80 \mathrm{\ ms}$) including the second and third
shortest spin periods of all pulsars currently known. The binary
pulsars include two systems with eccentric orbits and likely white
dwarf companions. No such systems
are known in 47~Tucanae. The difference between the two pulsar
populations may reflect the different evolutionary states and physical
conditions of the two clusters. In particular, the central stellar
density of Terzan~5 is about twice that of 47~Tucanae, suggesting that
the increased rate of stellar interactions might disrupt the recycling
process for the neutron stars in some binary systems and induce larger
eccentricities in others.


\subsubsection{Current and future surveys}

Motivated by the successes at Parkes, a multibeam receiver
has recently been installed at the Arecibo telescope~\cite{alfa}.
A preliminary survey using this instrument began
in 2004 and has so far discovered 10 pulsars~\cite{palfa}. 
The main virtue of this survey is the shorter integration times
employed and resulting higher sensitivity to accelerated systems
than the Parkes survey. Depending
on the stamina of the observers, up to 1000 pulsars
could be found in this survey over the next decade. 
Other surveys with the Green Bank Telescope~\cite{gbt} and with the Giant
Metre Wave Radio Telescope~\cite{gmrt} are also expected to make 
major contributions.

All surveys that have so far been conducted, or will be carried
out in the next few years, are likely to be surpassed by
the Square Kilometre Array~\cite{ska} -- the next generation
radio telescope, planned to come online around 2020. Simulations
suggest~\cite{kbc04} that at least $10^4$ pulsars, including $10^3$ millisecond
pulsars, could be detected in our Galaxy.


\subsection{Going further}

In this brief review of pulsar astronomy, we have only skimmed the
surface of many topics which are covered more thoroughly elsewhere. 
Two recent textbooks, {\it Pulsar
Astronomy}~\cite{ls05} and {\it Handbook of Pulsar
Astronomy}~\cite{lk05}, cover the literature and techniques and provide
excellent further reading. The morphological properties of pulsars
have recently been comprehensively discussed in two recent
reviews~\cite{smi03, sw04}. Those wishing to approach the subject from
a more theoretical viewpoint are advised to read {\it The Theory of
Neutron Star Magnetospheres}~\cite{mic91} and {\it The Physics of the
Pulsar Magnetosphere}~\cite{bgi93}. Our summary of evolutionary
aspects serves merely as a primer to the vast body of literature
available. Further insights can be found from more detailed
reviews~\cite{bv91, pk94, sta04}.

Pulsar resources available on the {\it Internet} are continually
becoming more extensive and useful. A good starting point for
pulsar-surfers is the {\it Handbook of Pulsar Astronomy} 
website~\cite{handbook}, as well as the
pages maintained at Arecibo~\cite{aopsr}, 
Berkeley~\cite{bkypsr}, Bonn~\cite{mpipsr}, Cagliari~\cite{cagpsr},
Jodrell Bank~\cite{jodpsr},
Princeton~\cite{pripsr}, Swinburne~\cite{swinpsr}, UBC~\cite{ubcpsr}
and Sydney~\cite{mbeampsr}.

\newpage


\section{Pulsar Statistics and Demography}
\label{sec:gal}

The observed pulsar sample is heavily biased towards the brighter
objects that are the easiest to detect. What we observe represents
only the tip of the iceberg of a much larger underlying
population~\cite{go70}. The extent to which the sample is incomplete
is well demonstrated by the projection of pulsars onto the Galactic
plane and their cumulative number distribution as a function of
distance shown in Figure~\ref{fig:incomplete}. The clustering of sources
around the Sun seen in the left panel of Figure~\ref{fig:incomplete} is
clearly at variance with the distribution of other stellar populations
which show a radial distribution about the Galactic centre.

The extent to which the pulsar sample is incomplete can be seen from
the cumulative number of pulsars as a function of the projected
distance from the Sun. In Figure~\ref{fig:incomplete} the observed
distribution is compared to the expected distribution for a simple
model population with no selection effects. The observed number
distribution becomes strongly deficient beyond a few kpc.

\epubtkImage{xy.png}{
  \begin{figure}[htbp]
    \def\epsfsize#1#2{0.95#1}
    \centerline{\epsfbox{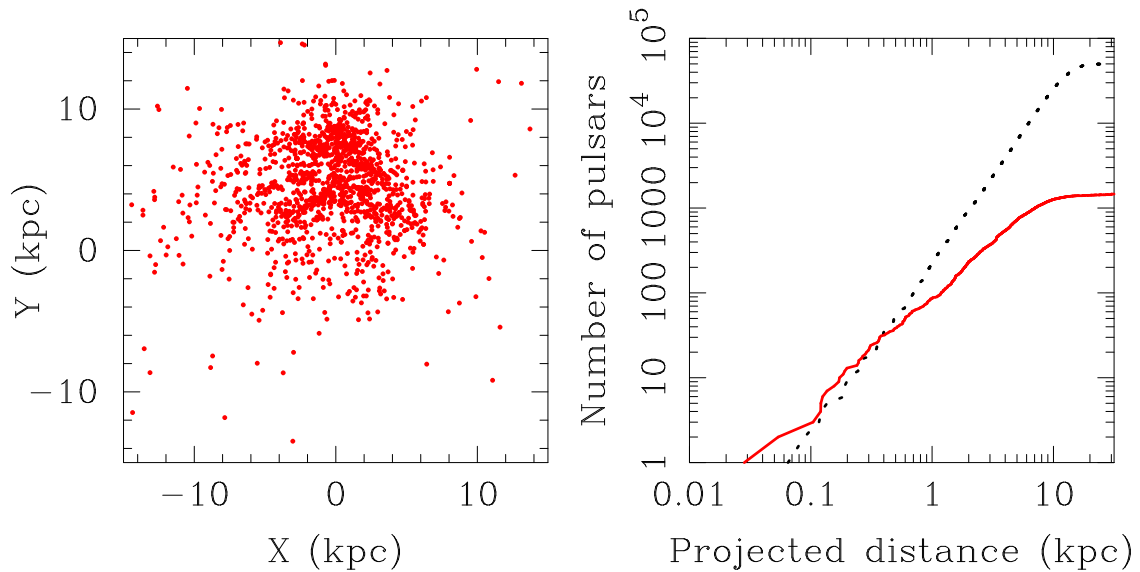}}
    \caption{\it Left panel: The current sample of all known radio
      pulsars projected onto the Galactic plane. The Galactic centre
      is at the origin and the Sun is at $ (0, 8.5) \mathrm{\ kpc} $.
      Note the spiral-arm structure seen in the distribution which
      is now required by the electron density
      model~\cite{cl02a, cl02b}. Right panel: Cumulative number of
      pulsars as a function of projected distance from the Sun. The
      solid line shows the observed sample while the dotted line shows
      a model population free from selection effects.}
    \label{fig:incomplete}
  \end{figure}}


\subsection{Selection effects in pulsar searches}
\label{sec:selfx}


\subsubsection{The inverse square law and survey thresholds}
\label{sec:invsq}

The most prominent selection effect is the inverse square law, i.e.\
for a given intrinsic luminosity\epubtkFootnote{Pulsar astronomers usually
define the luminosity $L = S d^2$, where $S$ is the mean flux density
at $400 \mathrm{\ MHz}$ (a standard observing frequency) and $d$ is the distance
derived from the DM (see Section~\ref{sec:dist}). Since this ignores any
assumptions about beaming or geometrical factors, it is sometimes
referred to as a ``pseudoluminosity''~\cite{acc02}.}, the observed
flux density falls off as the inverse square of the distance. This
results in the observed sample being dominated by nearby and/or high
luminosity
objects. Beyond distances of a few kpc from the Sun, the apparent
flux density falls below the detection thresholds $S_\mathrm{min}$ of most
surveys. Following~\cite{dss84}, we express this threshold as follows:
\begin{equation}
  S_\mathrm{min} =
  \frac{\mathrm{S/N}_\mathrm{min}}{\eta \sqrt{n_\mathrm{pol}}}
  \left( \frac{T_\mathrm{rec} + T_\mathrm{sky}}{\mathrm{K}} \right) 
  \left( \frac{G}{\mathrm{K\ Jy}^{-1}} \right)^{-1}
  \left( \frac{\Delta \nu}{\mathrm{MHz}} \right)^{-1/2}  
  \left( \frac{t_\mathrm{int}}{\mathrm{s}} \right)^{-1/2} 
  \left( \frac{W}{P-W} \right)^{1/2} \mathrm{\ mJy},
  \label{equ:defsmin}
\end{equation}
where S/N$_\mathrm{min}$ is the threshold signal-to-noise
ratio, $\eta$ is a generic fudge factor ($\lesssim 1$) which accounts
for losses in sensitivity (e.g., due to sampling and digitization noise),
$n_\mathrm{pol}$ is the number of polarizations recorded (either 1 or 2),
$T_\mathrm{rec}$ and $T_\mathrm{sky}$ are the receiver and sky noise
temperatures, $G$ is the gain of the antenna, $\Delta \nu$ is the
observing bandwidth, $t_\mathrm{int}$ is the integration time, $W$ is the
detected pulse width and $P$ is the pulse period.


\subsubsection{Interstellar pulse dispersion and multipath scattering}
\label{sec:dispandscatt}

It follows from Equation~(\ref{equ:defsmin}) that the sensitivity
decreases as $W/(P-W)$ and hence $W$
increases. Also note that if $W \gtrsim P$, the
pulsed signal is smeared into the background emission and is no longer
detectable, regardless of how luminous the source may be. The detected
pulse width $W$ may be broader than the intrinsic value largely as a
result of pulse dispersion and multipath scattering by free electrons in the
interstellar medium. The dispersive smearing scales as $\Delta
\nu/\nu^3$, where $\nu$ is the observing frequency. This can largely
be removed by dividing the pass-band into a number of channels and
applying successively longer time delays to higher frequency channels
\emph{before} summing over all channels to produce a sharp profile.
This process is known as incoherent dedispersion.

The smearing across the individual frequency channels, however, still
remains and becomes significant at high dispersions when searching for
short-period pulsars. Multipath scattering from electron density
irregularities results in a one-sided broadening due to the delay in
arrival times. A simple scattering model is shown in
Figure~\ref{fig:scatt} in which the scattering electrons are assumed to
lie in a thin screen between the pulsar and the observer~\cite{sch68}.
The timescale of this effect varies roughly as $\nu^{-4}$, which can
not be removed by instrumental means.

\epubtkImage{scatt.png}{
  \begin{figure}[htbp]
    \def\epsfsize#1#2{0.32#1}
    \centerline{\epsfbox{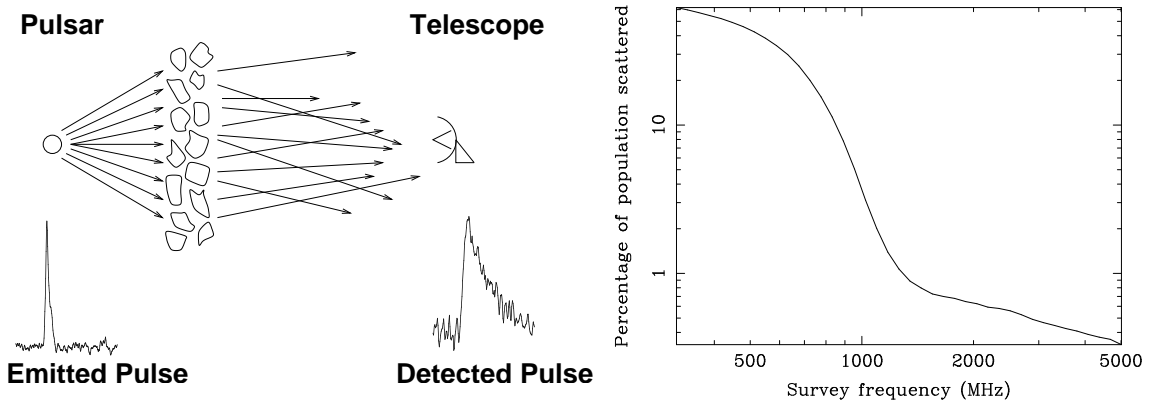}}
    \caption{\it Left panel: Pulse scattering caused by irregularities
      in the interstellar medium. The different path lengths and
      travel times of the scattered rays result in a ``scattering
      tail'' in the observed pulse profile which lowers its
      signal-to-noise ratio. Right panel: A simulation showing the
      percentage of Galactic pulsars that are likely to be
      undetectable due to scattering as a function of observing
      frequency. Low-frequency ($\lesssim 1 \mathrm{\ GHz} $) surveys
      clearly miss a large percentage of the population due to this
      effect.}
    \label{fig:scatt}
  \end{figure}}

Dispersion and scattering are most severe for distant pulsars in
the inner Galaxy where the number of free electrons along the line of
sight becomes large. The strong frequency dependence of both effects
means that they are considerably less of a problem for surveys at
observing frequencies $\gtrsim 1400 \mathrm{\ MHz}$~\cite{clj92, jlm92} compared 
to the $400 \mbox{-}\mathrm{MHz}$ search frequency used in early surveys.
An added bonus for such
observations is the reduction in $T_\mathrm{sky}$, since the spectral
index of the non-thermal Galactic emission is about
$-2.8$~\cite{lmop87}. Pulsars themselves have typical
spectral indices of $-1.6$~\cite{lylg95}, so that flux
densities are roughly an order of magnitude lower at $1400 \mathrm{\ MHz}$ compared
to $400 \mathrm{\ MHz}$. Fortunately, this can be at least partially compensated for by the 
use of larger receiver bandwidths at higher radio frequencies. For
example, the $1370 \mbox{-}\mathrm{MHz}$ system at Parkes has a bandwidth of $288 \mathrm{\ MHz}$~\cite{lcm00}
compared to the $430 \mbox{-}\mathrm{MHz}$ system, where nominally
$32 \mathrm{\ MHz}$ is available~\cite{mld96}.


\subsubsection{Orbital acceleration}
\label{sec:accn}

Standard pulsar searches use Fourier techniques~\cite{lk05} to search
for \emph{a-priori} unknown periodic signals and usually assume that
the apparent pulse period remains constant throughout the
observation. For searches with integration times much greater than a
few minutes, this assumption is only valid for solitary pulsars or
binary systems with orbital periods longer than
about a day. For shorter-period binary systems, the Doppler-shifting
of the period results in a spreading of the signal power over a number
of frequency bins in the Fourier domain, leading to a reduction in
S/N~\cite{jk91}. An observer will perceive the frequency of a pulsar
to shift by an amount $aT/(Pc)$, where $a$ is the (assumed constant)
line-of-sight acceleration during the observation of length $T$, $P$
is the (constant) pulsar period in its rest frame and $c$ is the speed
of light. Given that the width of a frequency bin in the Fourier
domain is $1/T$, we see that the signal will drift into more than one
spectral bin if $aT^2/(Pc)>1$. Survey sensitivities to
rapidly-spinning pulsars in tight orbits are therefore significantly
compromised when the integration times are large.

\epubtkImage{1913acsearch.png}{
  \begin{figure}[htbp]
    \def\epsfsize#1#2{0.6#1}
    \centerline{\epsfbox{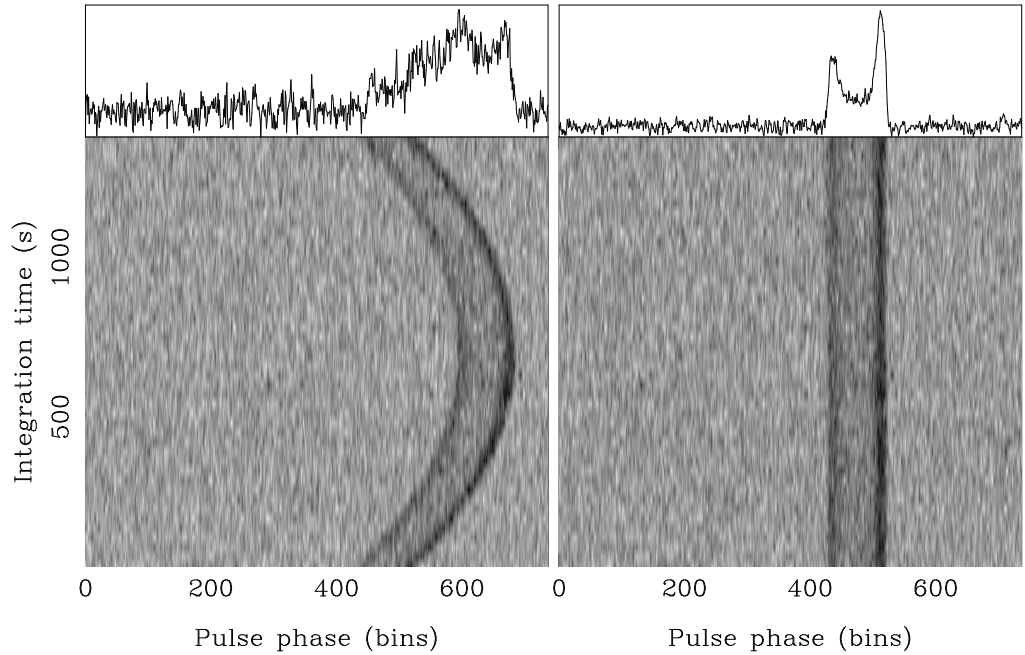}}
    \caption{\it Left panel: A $22.5 \mbox{-}\mathrm{min}$ Arecibo observation of the
      binary pulsar B1913+16. The assumption that the pulsar has a
      constant period during this time is clearly inappropriate given
      the quadratic drifting in phase of the pulse during the
      observation (linear grey scale plot). Right panel: The same
      observation after applying an acceleration search. This shows
      the effective recovery of the pulse shape and a significant
      improvement in the signal-to-noise ratio.}
    \label{fig:1913acc}
  \end{figure}}

As an example of this effect, as seen in the time domain,
Figure~\ref{fig:1913acc} shows a $22.5 \mbox{-}\mathrm{min}$ search mode observation of the
binary pulsar~B1913+16~\cite{ht75a, tw82, tw89}. Although this
observation covers only about 5\% of the orbit (7.75 hr), the severe effects
of the Doppler smearing on the pulse signal are very apparent. While
the standard search code nominally detects the pulsar with $\mathrm{S/N}=9.5$
for this observation, it is clear that this value is significantly
reduced due to the Doppler shifting of the
pulse period seen in the individual sub-integrations.

It is clearly desirable to employ a technique to recover the loss in
sensitivity due to Doppler smearing. One such technique, the so-called
``acceleration search''~\cite{mk84}, assumes the pulsar has a constant
acceleration during the observation. Each time series can then be
re-sampled to refer it to the frame of an inertial observer using the
Doppler formula to relate a time interval $\tau$ in the pulsar frame
to that in the observed frame at time $t$, as $\tau(t) \propto ( 1 +
at/c )$. Searching over a range of accelerations is desirable to find
the time series for which the trial acceleration most closely matches
the true value. In the ideal case, a time series is produced with a
signal of constant period for which full sensitivity is recovered (see
right panel of Figure~\ref{fig:1913acc}). This technique was first
used to find PSR B2127+11C~\cite{agk90}, a double neutron star binary
in M15 which has parameters similar to B1913+16. More recently, its
application to 47~Tucanae~\cite{clf00} resulted in the discovery of
nine binary millisecond pulsars, including one in a $96 \mbox{-}\mathrm{min}$ orbit
around a low-mass ($0.15\,M_{\odot}$) companion. This is
currently the shortest binary period for any known radio pulsar.

For intermediate orbital periods, in the range $30 \mathrm{\ min}$--several hours,
another promising technique is the dynamic power spectrum search.
Here the time series is split into a number of smaller contiguous
segments which are Fourier-transformed separately. The individual
spectra are displayed as a two-dimensional (frequency versus time)
image. Orbitally modulated pulsar signals appear as sinusoidal signals
in this plane as shown in Figure~\ref{fig:dps}.

\epubtkImage{dps.png}{
  \begin{figure}[htbp]
    \def\epsfsize#1#2{0.9#1}
    \centerline{\epsfbox{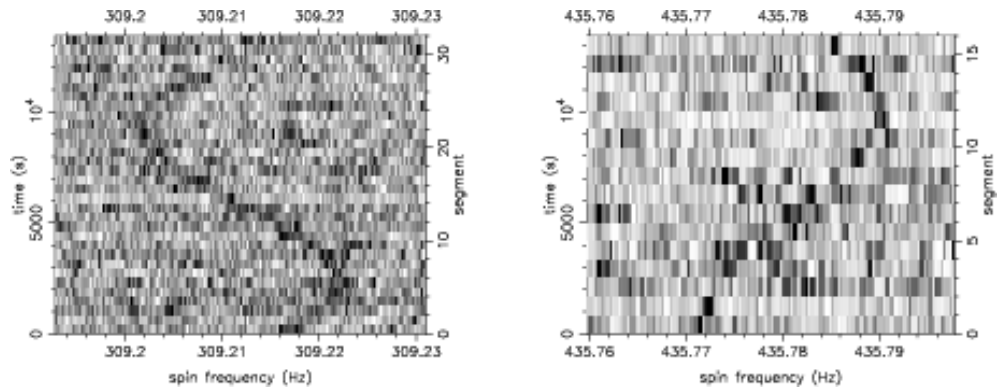}}
    \caption{\it Dynamic power spectra showing two recent pulsar
      discoveries in the globular cluster M62 showing fluctuation
      frequency as a function of time. Figure provided by Adam
      Chandler.}
    \label{fig:dps}
  \end{figure}}

This technique has been used by various groups where spectra are
inspected visually~\cite{lmbm00}. Much of
the human intervention can be removed using a hierarchical scheme for
selecting significant events~\cite{cha03}. This approach was recently
applied to a search of the globular cluster M62 resulting in the
discovery of three new pulsars. One of the new discoveries -- M62F, a
faint $2.3 \mbox{-}\mathrm{ms}$ pulsar in a $4.8 \mbox{-}\mathrm{hr}$ orbit -- was detectable only using the
dynamic power spectrum technique.

For the shortest orbital periods, the assumption of a constant
acceleration during the observation clearly breaks down. In this case,
a particularly efficient algorithm has been
developed~\cite{ran01, jrs02, rce03} which is optimised to finding
binaries with periods so short that many orbits can take place during
an observation. This ``phase modulation'' technique exploits the fact
that the Fourier components are modulated by the orbit to create a
family of periodic sidebands around the nominal spin frequency of the
pulsar. While this technique has so far not resulted in any new
discoveries, the existence of short period binaries in
47~Tucanae~\cite{clf00}, Terzan~5~\cite{rhs05} and the $11 \mbox{-}\mathrm{min}$ X-ray
binary X1820$-$303 in NGC~6624~\cite{spw87}, suggests that there may be
more ultra-compact binary pulsars that await discovery.


\subsection{Correcting the observed pulsar sample}
\label{sec:corsamp}

In the following, we review common techniques to account for
the various selection effects and form a less biased 
picture of the true pulsar population.


\subsubsection{Scale factor determination}
\label{sec:sfacts}

A very useful technique~\cite{pb81, vn81}, is to
define a scaling factor $\xi$ as the ratio of the total Galactic
volume weighted by pulsar density to the volume in which a pulsar
is detectable:
\begin{equation}
  \xi(P, L) = \frac{\int \int_\mathrm{G} \Sigma(R, z) R \, dR \, dz}
  {\int \int_{P, L} \Sigma(R, z) R \, dR \, dz}.
  \label{equ:sfac}
\end{equation}
Here, $\Sigma(R, z)$ is the assumed pulsar space density distribution
in terms of galactocentric radius $R$ and height above the Galactic
plane $z$. Note that $\xi$ is primarily a function of period $P$ and
luminosity $L$ such that short-period/low-luminosity pulsars have
smaller detectable volumes and therefore higher $\xi$ values than
their long-period/high-luminosity counterparts. 

In practice, $\xi$ is calculated for each pulsar separately using a Monte Carlo
simulation to model the volume of the Galaxy probed by the major
surveys~\cite{nar87}. For a sample of $N_\mathrm{obs}$ observed pulsars
above a minimum luminosity $L_\mathrm{min}$, the total number of
pulsars in the Galaxy is
\begin{equation}
  N_\mathrm{G} = \sum_{i=1}^{N_\mathrm{obs}} \frac{\xi_i}{f_i},
  \label{equ:ngal}
\end{equation}
where $f$ is the model-dependent ``beaming fraction'' discussed below
in Section~\ref{sec:beaming}. Note that this estimate 
applies to those pulsars with
luminosities $\gtrsim L_\mathrm{min}$. Monte Carlo simulations have shown
this method to be reliable, as long as $N_\mathrm{obs}$ is reasonably
large~\cite{lbdh93}.


\subsubsection{The small-number bias}
\label{sec:smallnumber}

For small samples of observationally-selected objects, the detected
sources are likely to be those with larger-than-average luminosities.
The sum of the scale factors~(\ref{equ:ngal}), therefore,
will tend to underestimate the true size of the population. This
``small-number bias'' was first pointed out~\cite{kal00, knst01} for
the sample of double neutron star binaries where we know of only four
systems relevant for calculations of the merging rate
(see Section~\ref{sec:nsns}). Only when $N_\mathrm{obs} \gtrsim 10$ does the sum of
the scale factors become a good indicator of the true population size.

\epubtkImage{smallnumber.png}{
  \begin{figure}[htbp]
    \def\epsfsize#1#2{0.9#1}
    \centerline{\epsfbox{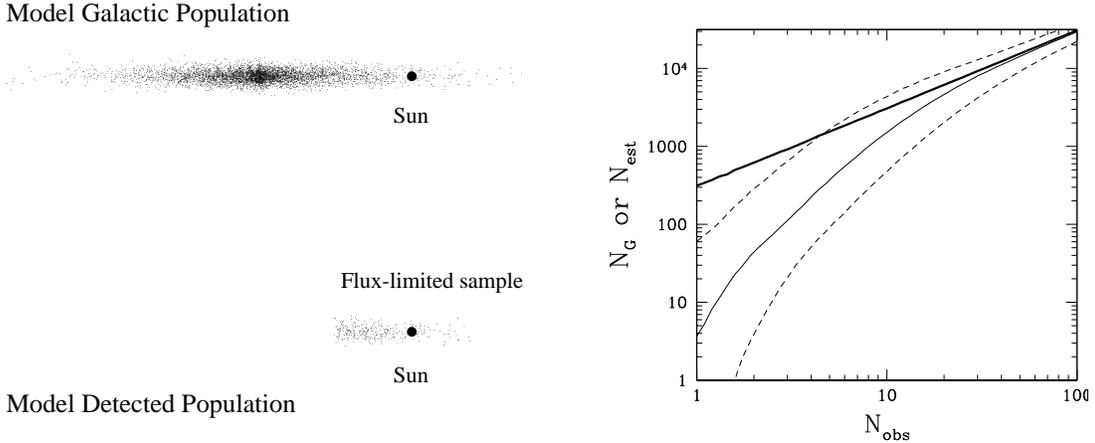}}
    \caption{\it Small-number bias of the scale factor estimates
      derived from a synthetic population of sources where the true
      number of sources is known. Left panel: An edge-on view of a
      model Galactic source population. Right panel: The thick line
      shows $N_\mathrm{G}$, the true number of objects in the model
      Galaxy, plotted against $N_\mathrm{observed}$, the number
      detected by a flux-limited survey. The thin solid line shows
      $N_\mathrm{est}$, the median sum of the scale factors, as a
      function of $N_\mathrm{obs}$ from a large number of Monte Carlo
      trials. Dashed lines show 25 and 75\% percentiles of the
      $N_\mathrm{est}$ distribution.}
    \label{fig:smallnumber}
  \end{figure}}

Despite a limited sample size, recent work~\cite{kkl03} has
demonstrated that rigorous confidence
intervals of $N_\mathrm{G}$ can be derived. Monte Carlo simulations
verify that the simulated number of detected objects
$N_\mathrm{detected}$
closely follows a Poisson distribution and that $N_\mathrm{detected} = \alpha
N_\mathrm{G}$, where $\alpha$ is a constant. By varying the
value of $N_\mathrm{G}$ in the simulations, the mean of
this Poisson distribution can be measured.
Using a Bayesian analysis it was shown~\cite{kkl03}
that, for a single object, the probability density function
of the total population is
\begin{equation}
  P(N_\mathrm{G}) = \alpha^2 N_\mathrm{G} \exp(-\alpha N_\mathrm{G}).
\end{equation}
Adopting the necessary assumptions required in the Monte
Carlo population about the underlying pulsar distribution,
this technique can be used to place interesting constraints
on the size and, as we shall see later, birth rate of the 
underlying population.


\subsubsection{The beaming correction}
\label{sec:beaming}

The ``beaming fraction'' $f$ in Equation~(\ref{equ:ngal}) is the
fraction of $4\pi$ steradians swept out by a pulsar's radio beam
during one rotation. Thus $f$ is the probability that the beam cuts
the line-of-sight of an arbitrarily positioned observer. A na\"{\i}ve
estimate for $f$ of roughly 20\% assumes a circular beam of width $\sim 10^{\circ}$
and a randomly distributed inclination angle between the spin and
magnetic axes~\cite{tm77}. Observational evidence suggests that
shorter period pulsars have wider beams and therefore larger beaming
fractions than their long-period
counterparts~\cite{nv83, lm88, big90b, tm98}. As can be seen in
Figure~\ref{fig:bfracts}, however, a consensus on the beaming
fraction-period relation has yet to be reached.

\epubtkImage{beam.png}{
  \begin{figure}[htbp]
    \def\epsfsize#1#2{0.5#1}
    \centerline{\epsfbox{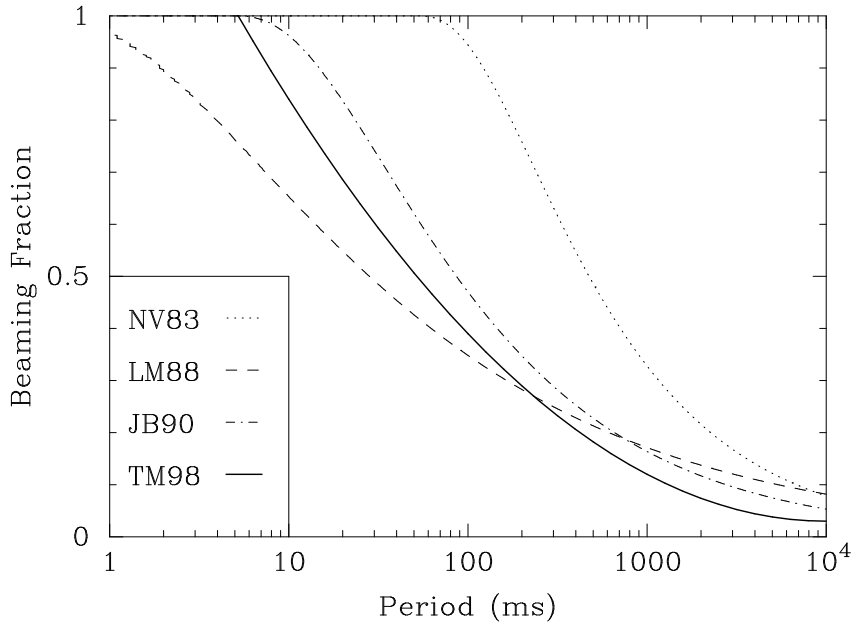}}
    \caption{\it Beaming fraction plotted against pulse period for
      four different beaming models: Narayan \& Vivekanand 1983 (NV83)
      \cite{nv83}, Lyne \& Manchester 1988 (LM88) \cite{lm88}, Biggs
      1990 (JDB90) \cite{big90b} and Tauris \& Manchester 1998 (TM88)
      \cite{tm98}.}
    \label{fig:bfracts}
  \end{figure}}

When most of these beaming models were originally proposed, the sample of
millisecond pulsars was $\lesssim 5 $ and hence their predictions about
the beaming fractions of short-period pulsars relied largely on
extrapolations from the normal pulsars. An analysis of a large
sample of millisecond pulsar profiles~\cite{kxl98} 
suggests that their beaming fraction lies between 50 and 100\%. 
The large beaming fraction and narrow pulses often observed
strongly suggests a fan beam model for millisecond pulsars~\cite{mic91}.


\subsection{The population of normal and millisecond pulsars}
\label{sec:nmsppop}


\subsubsection{Luminosity distributions and local number estimates}
\label{sec:lumfuns}

Based on a number of all-sky surveys carried out in the 1990s, the
scale factor approach has been used to derive the characteristics of
the true normal and millisecond pulsar populations and is based on the
sample of pulsars within $1.5 \mathrm{\ kpc}$ of the Sun~\cite{lml98}. Within this
region, the selection effects are well understood and easier to
quantify than in the rest of the Galaxy. These calculations
should therefore give a reliable \emph{local pulsar population} estimate.

\epubtkImage{lumfuns.png}{
  \begin{figure}[htbp]
    \def\epsfsize#1#2{0.32#1}
    \centerline{\epsfbox{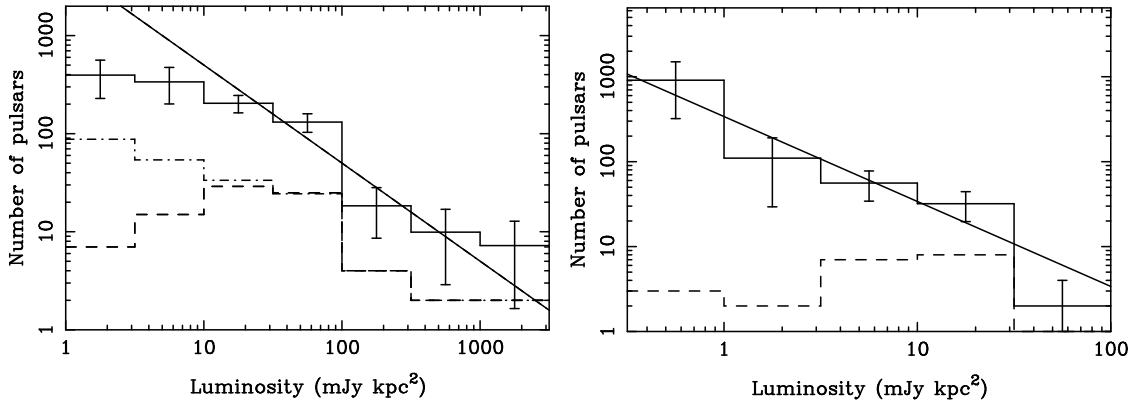}}
    \caption{\it Left panel: The corrected luminosity distribution
      (solid histogram with error bars) for normal pulsars. The
      corrected distribution \emph{before} the beaming model has been
      applied is shown by the dot-dashed line. Right panel: The
      corresponding distribution for millisecond pulsars. In both cases,
      the observed distribution is shown by the dashed line and the
      thick solid line is a power law with a slope of $-1$. The
      difference between the observed and corrected distributions
      highlights the severe under-sampling of low-luminosity pulsars.}
    \label{fig:lumfuns}
  \end{figure}}

The luminosity distributions obtained from this analysis are shown in
Figure~\ref{fig:lumfuns}. For the normal pulsars, integrating the
corrected distribution above $1 \mathrm{\ mJy\ kpc}^2$ and dividing by $\pi
\times (1.5)^2 \mathrm{\ kpc}^2$ yields a local surface density, assuming
a beaming model~\cite{big90b}, of $156 \pm 31 \mathrm{pulsars\
  kpc}^{-2}$ for luminosities above $1 \mathrm{\ mJy\ kpc}^2$. The same
analysis for the millisecond pulsars, assuming a mean beaming fraction
of 75\%~\cite{kxl98}, leads to a local surface density of $38 \pm 16
\mathrm{\ pulsars\ kpc}^{-2}$ also for luminosities above $1 \mathrm{\ mJy\ kpc}^2$.


\subsubsection{Galactic population and birth-rates}
\label{sec:psrpop}

Integrating the local surface densities of pulsars over the whole
Galaxy requires a knowledge of the presently rather uncertain
Galactocentric radial distribution~\cite{joh94, lor04}. One approach is
to assume that pulsars have a radial distribution similar to that of
other stellar populations and to scale the local number density with 
this distribution in order to estimate the total Galactic population. 
The corresponding local-to-Galactic scaling is
$ 1000 \pm 250 \mathrm{\ kpc}^2$~\cite{rv89}. This implies a population of
$\sim 160,000$ active normal pulsars and $\sim 40,000$
millisecond pulsars in the Galaxy. 

Based on these estimates, we are in a position to deduce the
corresponding rate of formation or birth-rate required to sustain
the observed population. From the
$P \mbox{--} \dot{P}$ diagram in Figure~\ref{fig:ppdot}, we infer a typical
lifetime for normal pulsars of $\sim 10^{7} \mathrm{\ yr}$, corresponding to a
Galactic birth rate of $\sim 1$ per $60 \mathrm{\ yr}$ -- consistent with the rate
of supernovae~\cite{vt91}. As noted in Section~\ref{sec:nms}, the
millisecond pulsars are much older, with ages close to that of the
Universe $\tau_\mathrm{u}$ (we assume here $\tau_\mathrm{u}=13.8
\mathrm{\ Gyr}$~\cite{yjb05}). Taking the maximum age of the millisecond pulsars
to be $\tau_\mathrm{u}$, we infer a mean birth rate of at least 1 per
$345,000 \mathrm{\ yr}$. This is consistent, within the uncertainties,
with the birth-rate of low-mass X-ray binaries~\cite{lnl95}.


\subsection{The population of relativistic binaries}
\label{sec:relpop}

Although no radio pulsar has so far been observed in orbit around a
black hole companion, we now know of several double neutron star and
neutron star--white dwarf binaries which will merge due to
gravitational wave emission within a reasonable timescale. The
current sample of objects is shown as a function of orbital period and
eccentricity in Figure~\ref{fig:mplane}. Isochrones showing various
coalescence times $\tau_\mathrm{g}$ are calculated using the
expression
\begin{equation}
  \tau_\mathrm{g} \simeq 9.83 \times 10^6 \mathrm{\ yr}
  \left( \frac{P_\mathrm{b}}{\mathrm{hr}} \right)^{8/3} 
  \left( \frac{m_1 + m_2}{M_\odot} \right)^{-2/3} 
  \left( \frac{\mu}{M_\odot} \right)^{-1}
  \left( 1 - e^2 \right)^{7/2},
  \label{equ:tgw}
\end{equation}
where $m_1$ and $m_2$ are the masses of the two stars, 
$\mu=m_1m_2/(m_1+m_2)$ is the so-called ``reduced mass'', 
$P_\mathrm{b}$ is the binary period and $e$ is
the eccentricity. This formula is a good analytic approximation
(within a few percent) to the numerical solution of the exact
equations for $\tau_\mathrm{g}$~\cite{pet64, pm63}. 

\epubtkImage{mplane.png}{
  \begin{figure}[htbp]
    \def\epsfsize#1#2{0.5#1}
    \centerline{\epsfbox{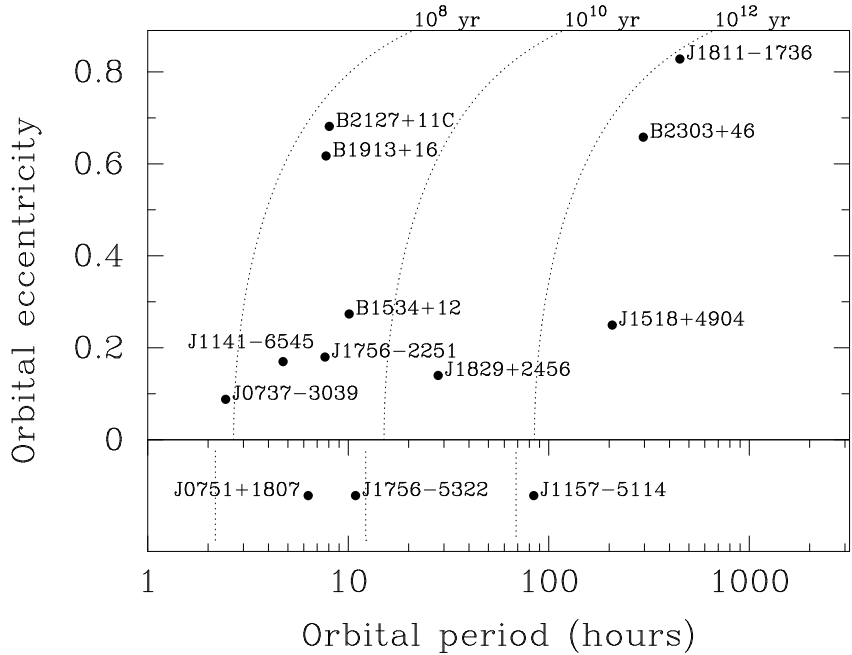}}
    \caption{\it The relativistic binary merging plane. Top: Orbital
      eccentricity versus period for eccentric binary systems
      involving neutron stars. Bottom: Orbital period distribution for
      the massive white dwarf--pulsar binaries.}
    \label{fig:mplane}
  \end{figure}}

In addition to testing general relativity through observations of
these systems (see Section~\ref{sec:tbin}), estimates of their Galactic
population and merger rate are of great interest as one of the prime
sources for current gravitational wave detectors such as
GEO600~\cite{geo600}, LIGO~\cite{ligo}, VIRGO~\cite{virgo} and
TAMA~\cite{tama}. In the following, we review recent empirical
determination of the population sizes and merging rates of binaries
where at least one component is visible as a radio pulsar.


\subsubsection{Double neutron star binaries}
\label{sec:nsns}

As discussed in Section~\ref{sec:nms}, double neutron star (DNS) binaries
are expected to be rare. This is certainly the case; despite extensive
searches, only five certain DNS binaries are currently known: PSRs
J0737$-$3039~\cite{bdp03}, B1534+12~\cite{wol91a}, J1756$-$2251~\cite{fkl05},
B1913+16~\cite{ht75a} and B2127+11C~\cite{pakw91}. Although we only
see both neutron stars as pulsars in J0737$-$3039~\cite{lbk04}, we are
``certain'' of the identification in the other four systems from 
precise component mass measurements from pulsar timing
observations (see Section~\ref{sec:tbin}). The spin and orbital
parameters, timescales and mass constraints for these systems
are listed in Table~\ref{tab:nsns}, along with three other
binaries with eccentric
orbits, mass functions and periastron advance measurements that are
\emph{consistent} with a DNS identification, but for which there is
presently not sufficient component mass information to confirm their nature.

\begin{table}[htbp]
  \centering
  \begin{tabular}{l|rrrr}
    \hline \hline
    \vsp & J0737$-$3039 & J1518+4904 & B1534+12 & J1756$-$2251 \\ [0.2 em]
    \hline
    $P$ [ms] & 22.7/2770 & 40.9 & 37.9 & 28.5 \\
    $P_\mathrm{b}$ [d] & 0.102 & 8.6 & 0.4 & 0.32 \\
    $e$ & 0.088 & 0.25 & 0.27 & 0.18 \\
    $\tau_\mathrm{c}$ [$10^8 \mathrm{\ yr}$] & 210/50& 200 & 2.5 & 4.4 \\
    $\tau_\mathrm{g}$ [$10^8 \mathrm{\ yr}$] & 87 & 24000 & 27 & 169 \\
    Masses measured? & Yes & No & Yes & Yes \\
    \hline \hline
    \vsp & J1811$-$1736 & J1829+2456 & B1913+16 & B2127+11C \\ [0.2 em]
    \hline
    $P$ [ms] & 104.2 & 41.0 & 59.0 & 30.5 \\
    $P_\mathrm{b}$ [d] & 18.8 & 1.18 & 0.3 & 0.3 \\
    $e$ & 0.83 & 0.14 & 0.62 & 0.68 \\
    $\tau_\mathrm{c}$ [$10^8 \mathrm{\ yr}$] & 9.2 & 130 & 1.1 & 0.97 \\
    $\tau_\mathrm{g}$ [$10^8 \mathrm{\ yr}$] & 10000 & 600 & 3.0 & 2.2 \\
    Masses measured? & No & No & Yes & Yes \\
    \hline \hline
  \end{tabular}
  \caption{\it Known and likely DNS binaries. Listed are the pulse
    period $P$, orbital period $P_\mathrm{b}$, orbital eccentricity
    $e$, characteristic age $\tau_\mathrm{c}$, and expected binary
    coalescence time-scale $\tau_\mathrm{g}$ due to gravitational wave
    emission calculated from Equation~(\ref{equ:tgw}). To distinguish
    between definite and candidate DNS systems, we also list whether
    the masses of both components have been determined via the
    measurement of two or more post Keplerian parameters as described
    in Section~\ref{sec:tbin}.}
  \label{tab:nsns}
\end{table}

Despite the uncertainties in identifying DNS binaries, for the
purposes of determining the Galactic merger rate, the
systems\epubtkFootnote{The likely DNS binary J1756$-$2251~\cite{fkl05} has a
merger time of $1.7 \mathrm{\ Gyr}$ and will make a small contribution to the
rate. It has so far not been incorporated into the latest
calculations.} for which $\tau_\mathrm{g}$ is less than $\tau_\mathrm{u}$
(i.e.\ PSRs J0737$-$3039, B1534+12, J1756$-$2251, B1913+16 and
B2127+11C) are primarily of interest. Of these PSR B2127+11C is in the
process of being ejected from the globular cluster
M15~\cite{pakw91, ps91} and is thought to make only a negligible
contribution to the merger rate~\cite{phi91}. The general approach
with the remaining systems is to derive scale factors for each
object, construct the probability density function of their total
population (as outlined in Section~\ref{sec:sfacts}) and then divide these
by a reasonable estimate for the lifetime. It is generally
accepted~\cite{knst01} that the \emph{observable lifetimes} for these
systems are determined by the timescale on which the current orbital
period is reduced by a factor of two~\cite{acw99}. Below this point,
the orbital smearing selection effect discussed in Section~\ref{sec:accn}
will render the binary undetectable by current surveys.

The results of the most recent study of this
kind~\cite{kkl04a, kkl04b}, which take into account the discovery of
PSR J0737$-$3039, are summarised in the left panel of
Figure~\ref{fig:rates}. The combined Galactic merger rate, dominated by
the double pulsar, is found to be $80_{-70}^{+210} \mathrm{\ Myr}^{-1}$, where
the uncertainties reflect the 95\% confidence level using the
techniques summarised in Section~\ref{sec:smallnumber}. Extrapolating this
number to include DNS binaries detectable by LIGO in other
galaxies~\cite{phi91}, the expected event rate is
$35^{+90}_{-30}\times 10^{-3} \mathrm{\ yr}^{-1}$ for initial LIGO and
$190^{+470}_{-150} \mathrm{\ yr}^{-1}$ for advanced LIGO. Future prospects for
detecting gravitational wave emission from binary neutron star
inspirals are therefore very encouraging. Since much of the
uncertainty in the rate estimates is due to our ignorance of the
underlying distribution of double neutron star systems, future
detection rates could ultimately constrain the properties of this
exciting binary species.

\epubtkImage{rates.png}{
  \begin{figure}[htbp]
    \def\epsfsize#1#2{0.37#1}
    \centerline{\epsfbox{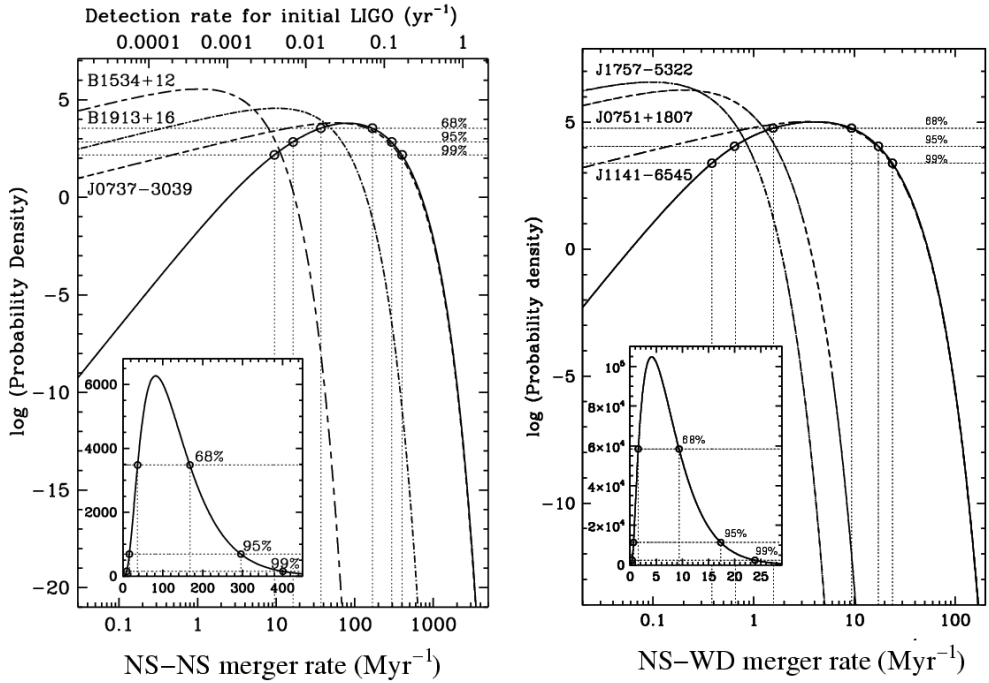}}
    \caption{\it The current best empirical estimates of the
      coalescence rates of relativistic binaries involving neutron
      stars. The individual contributions from each known binary
      system are shown as dashed lines, while the solid line shows the
      total probability density function on a logarithmic and (inset)
      linear scale. The left panel shows DNS
      binaries~\cite{kkl04a, kkl04b}, while the right panel shows the
      equivalent results for NS--WD binaries~\cite{kklw04}.}
    \label{fig:rates}
  \end{figure}}

Although the double pulsar system J0737$-$3039 will not be important
for ground-based detectors until its final coalescence in another
$85 \mathrm{\ Myr}$, it may be a useful calibration source for the future
space-based detector LISA~\cite{lisa}. Recent
calculations~\cite{kkl05b} show that a $1 \mbox{-}\mathrm{yr}$ observation with LISA would
detect (albeit with $\mathrm{S/N} \sim 2$) the continuous emission at a
frequency of $0.2 \mathrm{\ mHz}$ based on the current orbital parameters. Although
there is the prospect of using LISA to detect similar systems systems
through their continuous emission, current calculations~\cite{kkl05b}
suggest that significant ($\mathrm{S/N}>5$) detections are not likely. Despite
these limitations, it is likely that LISA observations will be able to
place independent constraints on the Galactic DNS binary population
after several years of operation.


\subsubsection{White dwarf--neutron star binaries}
\label{sec:nswd}

Although the population of white dwarf--neutron star (WDNS) binaries
in general is substantial, the fraction which will merge due to
gravitational wave emission is small. Like the DNS binaries, the
observed WDNS sample suffers from small-number statistics. From
Figure~\ref{fig:mplane}, we note that only three WDNS systems
are currently known that will merge within $\tau_\mathrm{u}$, PSRs
J0751+1807~\cite{lzc95}, J1757$-$5322~\cite{eb01} and 
J1141$-$6545~\cite{klm00}. Applying the same techniques as used
for the DNS population, the merging rate contributions of the
three systems can be calculated~\cite{kklw04} and are shown
in Figure~\ref{fig:rates}. The combined Galactic coalescence rate is
$4_{-3}^{+5} \mathrm{\ Myr}^{-1}$ (at 68\% confidence interval). Although
the orbital frequencies of these objects at coalescence are too 
low to be detected by LIGO, they do fall within the band planned
for LISA~\cite{lisa}. Unfortunately, an extrapolation of the
Galactic event rate out to distances at which such events would be detectable
by LISA does not suggest that these systems will be a major source
of detection~\cite{kklw04}. Similar conclusions were reached by
considering the statistics of low-mass X-ray binaries~\cite{coo04}.

\subsection{Going further}

Studies of pulsar population statistics represent a large proportion of 
the pulsar literature. Good starting points for further reading can be 
found in other review articles~\cite{mic91, bv91}. Our coverage of
compact object coalescence rates has concentrated on empirical methods.
An alternative approach is to undertake a full-blown Monte Carlo
simulation of the most likely evolutionary scenarios described in
Section~\ref{sec:evolution}. In this approach, a
population of primordial binaries is synthesized with a number of
underlying distribution functions: primary mass, binary mass ratio,
orbital period distribution etc. The evolution of both stars is then
followed to give a predicted sample of binary systems of all the
various types. Although selection effects are not always taken into account
in this approach, the final census is usually normalized to the
star formation rate.

Numerous population syntheses (most often to
populations of binaries where one or both members are NSs) can be
found in the literature~\cite{dc87, rom92, ty93, pv96, lpp96, bkb02}. A group
in Moscow has made a web interface to their code~\cite{scenario}.
Although extremely instructive, the uncertain assumptions about
initial conditions, the physics of mass transfer and the kicks applied
to the compact object at birth result in a wide range of predicted
event rates which are currently broader than the empirical
methods~\cite{kkl04b, kklw04, kkl05}. Ultimately, the detection
statistics from the gravitational wave detectors could provide far
tighter constraints on the DNS merging rate than the pulsar surveys
from which these predictions are made.

\newpage


\section{Principles and Applications of Pulsar Timing}
\label{sec:pultim}

It became clear soon after their discovery that pulsars are excellent
celestial clocks. The period of the first pulsar~\cite{hbp68} was
found to be stable to one part in $10^7$ over a few months. Following
the discovery of the millisecond pulsar B1937+21~\cite{bkh82} it was
demonstrated that its period could be measured to one part in
$10^{13}$ or better~\cite{dtwb85}. This unrivaled stability leads to a
host of applications including uses as time keepers, probes of relativistic
gravity and natural gravitational wave detectors.


\subsection{Observing basics}
\label{sec:timobs}

Each pulsar is typically observed at least once or twice per month
over the course of a year to establish its basic
properties. Figure~\ref{fig:timing} summarises the essential steps
involved in a ``time-of-arrival'' (TOA) measurement. Pulses from the
neutron star traverse the interstellar medium before being received at
the radio telescope where they are de-dispersed and added to form a
mean pulse profile.

\epubtkImage{timing.png}{
  \begin{figure}[htbp]
    \def\epsfsize#1#2{0.6#1}
    \centerline{\epsfbox{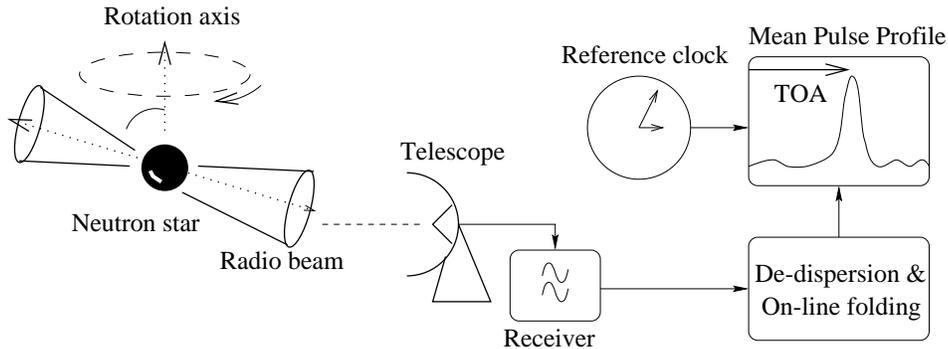}}
    \caption{\it Schematic showing the main stages involved in pulsar
      timing observations.}
    \label{fig:timing}
  \end{figure}}

During the observation, the data regularly receive a time stamp,
usually based on a caesium time standard or hydrogen maser at the
observatory plus a signal from the Global Positioning System of
satellites (GPS; see~\cite{gps}). The TOA is defined as the arrival time
of some fiducial point on the integrated profile with respect to either
the start or the midpoint of the observation. Since the profile has a
stable form at any given observing frequency (see Section~\ref{sec:profs}), the
TOA can be accurately determined by cross-correlation of the observed
profile with a high S/N ``template'' profile obtained from
the addition of many observations at the particular observing frequency.

Successful pulsar timing requires optimal TOA precision which
largely depends on the signal-to-noise ratio (S/N) of the 
pulse profile. Since the TOA uncertainty $\epsilon_\mathrm{TOA}$
is roughly the pulse width divided by the S/N, using 
Equation~(\ref{equ:defsmin}) we may write the fractional error as
\begin{equation}
  \frac{\epsilon_\mathrm{TOA}}{P} \simeq 
  \left( \frac{S_\mathrm{psr}}{\mathrm{mJy}} \right)^{-1}
  \left( \frac{T_\mathrm{rec} + T_\mathrm{sky}}{\mathrm{K}} \right)
  \left( \frac{G}{\mathrm{K\ Jy}^{-1}} \right)^{-1}
  \left( \frac{\Delta \nu}{\mathrm{MHz}} \right)^{-1/2}
  \left( \frac{t_\mathrm{int}}{\mathrm{s}} \right)^{-1/2}
  \left( \frac{W}{P} \right)^{3/2}\!\!\!\!\!\!\!.
  \label{equ:defsnr}
\end{equation}
Here, $S_\mathrm{psr}$ is the flux density of the pulsar, $T_\mathrm{rec}$
and $T_\mathrm{sky}$ are the receiver and sky noise temperatures, $G$ is
the antenna gain, $\Delta \nu$ is the observing bandwidth,
$t_\mathrm{int}$ is the integration time, $W$ is the pulse width and $P$ is the
pulse period (we assume $W \ll P$). Optimal results are thus obtained
for observations of short period pulsars with large flux densities and
small duty cycles ($W/P$) using large telescopes with low-noise
receivers and large observing bandwidths.

One of the main problems of employing large bandwidths is pulse
dispersion. As discussed in Section~\ref{sec:dist}, 
pulses emitted at lower radio frequencies travel
slower and arrive later than those emitted at higher
frequencies. This process has the effect of ``stretching'' the pulse
across a finite receiver bandwidth, increasing $W$
and therefore increasing $\epsilon_\mathrm{TOA}$.
For normal pulsars, dispersion can largely be compensated for
by the incoherent de-dispersion process outlined in Section~\ref{sec:selfx}.

The short periods of millisecond pulsars
offer the ultimate in timing precision. In order to fully exploit this,
a better method of dispersion removal is required. Technical
difficulties in building devices with very narrow channel bandwidths
require another dispersion removal technique. In the process of
coherent de-dispersion~\cite{han71, lk05} the incoming signals are
de-dispersed over the whole bandwidth using a filter which has the
inverse transfer function to that of the interstellar medium.
The signal processing can be done on-line either using finite
impulse response filter devices~\cite{bdz97} or completely in 
software~\cite{sta98, sst00}. 
Off-line data reduction, while disk-space limited,
allows for more flexible analysis schemes~\cite{bai03}.

The maximum time resolution obtainable via coherent dedispersion is the
inverse of the total receiver bandwidth. The current state of the art
is the detection~\cite{hkwe03}
of features on nanosecond timescales in pulses from
the $33 \mbox{-}\mathrm{ms}$ pulsar B0531+21 in the Crab nebula shown
in Figure~\ref{fig:crab}. Simple
light travel-time arguments can be made to show that, in the absence of
relativistic beaming effects~\cite{gm04}, these incredibly bright 
bursts originate from regions less than $1 \mathrm{\ m}$ in size.

\epubtkImage{crab.png}{
  \begin{figure}[htbp]
    \def\epsfsize#1#2{0.5#1}
    \centerline{\epsfbox{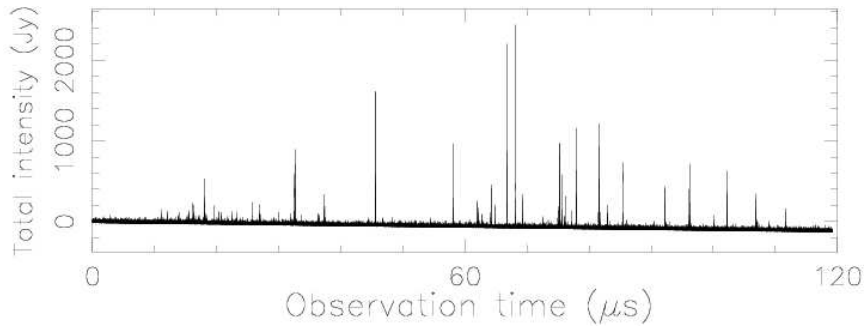}}
    \caption{\it A $120 \mathrm{\ \mu s}$ window centred on a
      coherently-dedispersed giant pulse from the Crab pulsar showing
      high-intensity nanosecond bursts. Figure provided by Tim
      Hankins~\cite{hkwe03}.}
    \label{fig:crab}
  \end{figure}}


\subsection{The timing model}
\label{sec:tmodel}

Ideally, in order to model the rotational behaviour of the neutron
star, we require TOAs measured by an inertial observer. An observatory
located on Earth experiences accelerations with respect to the neutron
star due to the Earth's rotation and orbital 
motion around the Sun and is therefore not in
an inertial frame. To a very good approximation, the solar system
centre-of-mass (barycentre) can be regarded as an
inertial frame. It is now standard practice~\cite{hun71} to transform the
observed TOAs to this frame using a planetary ephemeris such as the
JPL DE405~\cite{sn96}. The transformation is summarised as the
difference between barycentric ($\cal T$) and observed ($t$) TOAs:
\begin{equation}
  {\cal T} - t =
  \frac{\underline{r} . \hat{\underline{s}}}{c} +
  \frac{(\underline{r} . \hat{\underline{s}})^2 - |\underline{r}|^2}{2cd} +
  \Delta t_\mathrm{rel} - \Delta t_\mathrm{DM},
  \label{equ:bary}
\end{equation}
where $\underline{r}$ is the position of the observatory with respect to the
barycentre, $\hat{\underline{s}}$ is a unit vector in the direction
of the pulsar at a distance $d$ and $c$ is the speed of
light. The first term on the right hand side of
Equation~(\ref{equ:bary}) is the light travel time from the observatory to the
solar system barycentre. Incoming pulses from all but the nearest pulsars
can be approximated by plane wavefronts. The second term, which
represents the delay due to spherical wavefronts, yields the
parallax and hence $d$. This has so far only been measured
for five nearby millisecond pulsars~\cite{sbm97, tsb99, lkd04,
  sns05}. The term $\Delta t_\mathrm{rel}$ represents the Einstein and
Shapiro corrections due to general
relativistic time delays in the solar system~\cite{bh86}. Since
measurements can be carried out at different observing frequencies
with different dispersive delays, TOAs are generally referred to
the equivalent time that would be observed at infinite frequency.
This transformation is the term $\Delta t_\mathrm{DM}$ (see also
Equation~(\ref{equ:defdt})).

Following the accumulation of a number of TOAs, a surprisingly 
simple model is usually sufficient to account for the TOAs during
the time span of the observations and to predict the arrival times of
subsequent pulses. The model is a Taylor expansion of the
rotational frequency $\Omega = 2 \pi/P$ about a model value
$\Omega_0$ at some reference epoch ${\cal T}_0$. 
The model pulse phase is
\begin{equation}
  \phi({\cal T}) =
  \phi_0 + ({\cal T} - {\cal T}_0) \Omega_0 +
  \frac{1}{2} ({\cal T} - {\cal T}_0)^2 \dot{\Omega}_0 +
  \dots,
  \label{equ:phi}
\end{equation}
where ${\cal T}$ is the barycentric time and $\phi_0$ is the
pulse phase at ${\cal T}_0$. Based on this simple model, and
using initial estimates of the position, dispersion measure and pulse
period, a ``timing residual'' is calculated for each TOA as the
difference between the observed and predicted pulse phases.

A set of timing residuals for the nearby pulsar B1133+16 spanning
almost 10 years is shown in Figure~\ref{fig:1133}. Ideally, the
residuals should have a zero mean and be free from any systematic
trends (see Panel~a of Figure~\ref{fig:1133}). To reach this point, however, the model
needs to be refined in a bootstrap fashion. Early sets of residuals
will exhibit a number of trends indicating a systematic error in one
or more of the model parameters, or a parameter not incorporated into
the model.

\epubtkImage{1133.png}{
  \begin{figure}[htbp]
    \def\epsfsize#1#2{0.65#1}
    \centerline{\epsfbox{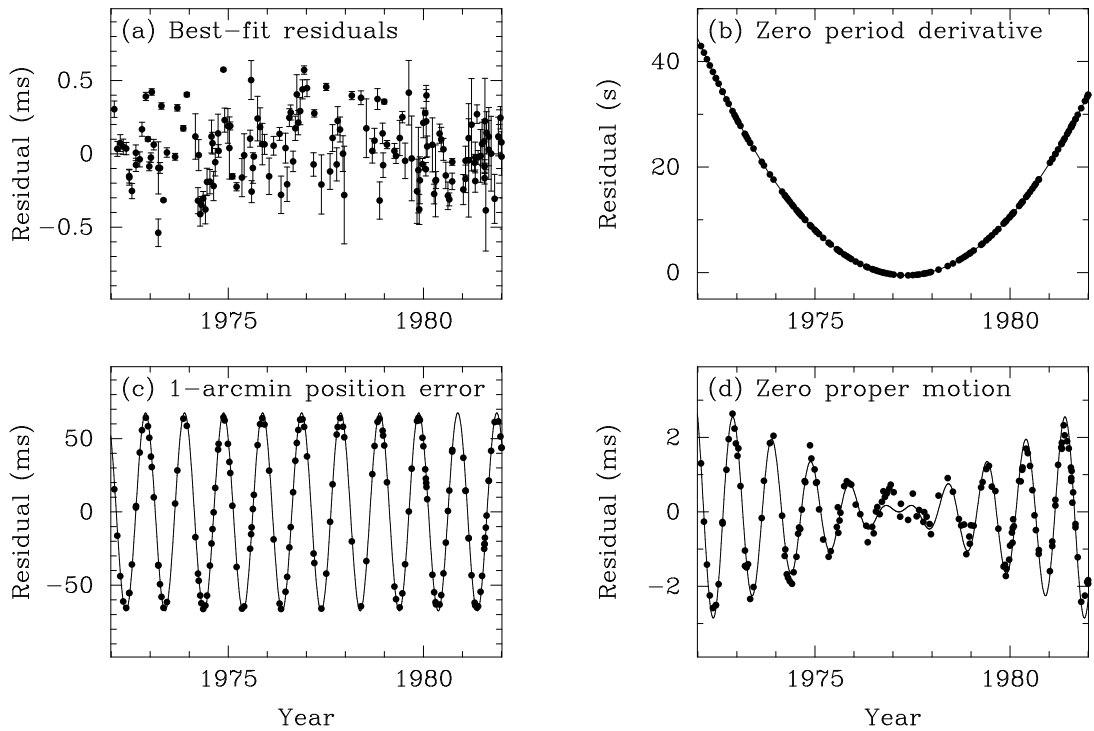}}
    \caption{\it Timing model residuals for PSR B1133+16. Panel~a:
      Residuals obtained from the best-fitting model which includes
      period, period derivative, position and proper motion. Panel~b:
      Residuals obtained when the period derivative term is set to
      zero. Panel~c: Residuals showing the effect of a $1 \mbox{-}\mathrm{arcmin}$
      position error. Panel~d: Residuals obtained neglecting the
      proper motion. The lines in Panels~b--d show the expected
      behaviour in the timing residuals for each effect. Data provided
      by Andrew Lyne.}
    \label{fig:1133}
  \end{figure}}

From Equation~(\ref{equ:phi}), an error in the assumed
$\Omega_0$ results in a linear slope with time. A parabolic
trend results from an error in $\dot{\Omega}_0$
(see Panel~b of Figure~\ref{fig:1133}). Additional effects will arise if the assumed
position of the pulsar (the unit vector $\hat{\underline{s}}$ in
Equation~(\ref{equ:bary})) used in the barycentric time calculation is
incorrect. A position error results in an annual sinusoid
(see Panel~c of Figure~\ref{fig:1133}). A proper motion produces an annual sinusoid
of linearly increasing magnitude (see Panel~d of Figure~\ref{fig:1133}).

After a number of iterations, and with the benefit of a modicum of
experience, it is possible to identify and account for each of these
various effects to produce a ``timing solution'' which is phase
coherent over the whole data span. The resulting model parameters
provide spin and astrometric information with a 
precision which improves as the length of the data span
increases. Timing observations of the original
millisecond pulsar B1937+21, spanning almost 9 years (exactly
165,711,423,279 rotations!), measure a period of
$1.5578064688197945\pm0.0000000000000004 \mathrm{\ ms}$~\cite{ktr94, kas94}
defined at midnight UT on
December 5 1988!  Astrometric measurements are no
less impressive, with position errors of $\sim 20 \mathrm{\ \mu arcsec}$ being
presently possible~\cite{van03}.


\subsection{Timing stability}
\label{sec:tstab}

Ideally, after correctly applying a timing model,
we would expect a set of uncorrelated timing residuals
with a zero mean and a Gaussian scatter with a standard deviation
consistent with the measurement uncertainties. As can be
seen in Figure~\ref{fig:tnoise}, this is not always the
case; the residuals of many pulsars exhibit a quasi-periodic
wandering with time.

\epubtkImage{tnoise.png}{
  \begin{figure}[htbp]
    \def\epsfsize#1#2{0.5#1}
    \centerline{\epsfbox{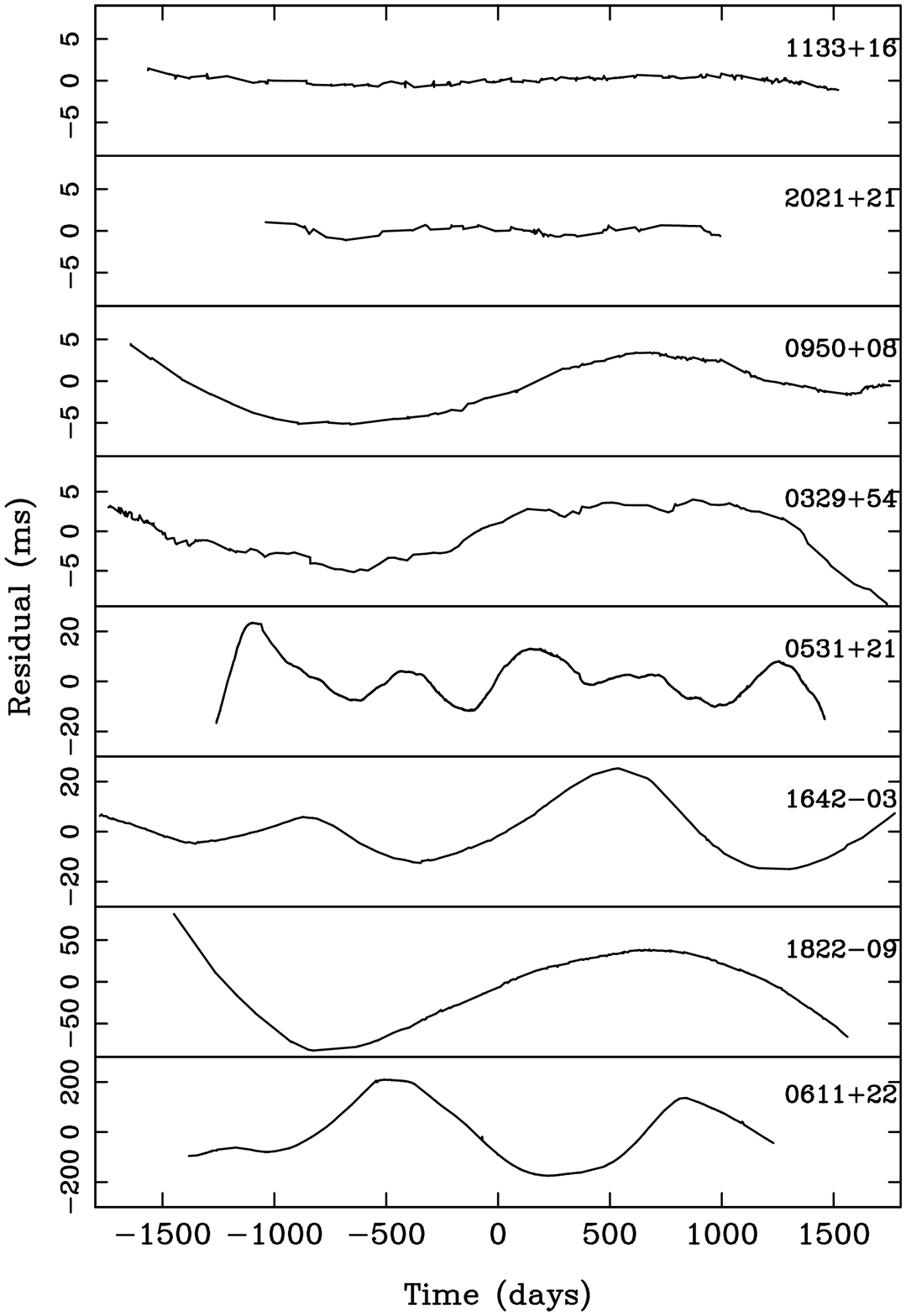}}
    \caption{\it Examples of timing residuals for a number of normal
      pulsars. Note the varying scale on the ordinate axis, the
      pulsars being ranked in increasing order of timing
      ``activity''. Data taken from the Jodrell Bank timing
      program~\cite{sl96, hlk04}. Figure provided by Andrew Lyne.}
    \label{fig:tnoise}
  \end{figure}}

Such ``timing noise'' is most prominent in the youngest of the normal
pulsars~\cite{mt74, ch80} and present at a lower level in the much older
millisecond pulsars~\cite{ktr94, antt94}. While the physical processes of
this phenomenon are not well understood, it seems likely that it
may be connected to superfluid processes and temperature changes
in the interior of the neutron star~\cite{anp86}, or to processes in the
magnetosphere~\cite{che87a, che87b}.

The relative dearth of timing noise for the older pulsars is a very
important finding. It implies that, presently, the measurement
precision depends primarily on the particular hardware constraints of
the observing system. Consequently, a large effort in hardware
development is presently being made to improve the precision of these
observations using, in particular, coherent dedispersion outlined in
Section~\ref{sec:timobs}. Much progress in this area has been made by
groups at Princeton~\cite{pripsr}, Berkeley~\cite{bkypsr},
Jodrell Bank~\cite{jodpsr},
UBC~\cite{ubcpsr}, Swinburne~\cite{swinpsr} and ATNF~\cite{atnfpsr}.
From high quality observations spanning over a
decade~\cite{rt91a, rt91b, ktr94}, these groups have demonstrated that
the timing stability of millisecond pulsars over such time-scales is
comparable to terrestrial atomic clocks.

This phenomenal stability is demonstrated in Figure~\ref{fig:sigmaz}
which shows $\sigma_z$~\cite{mtem97}, a parameter closely resembling
the Allan variance used by the clock community to estimate the
stability of atomic clocks~\cite{tay91, allan}. Both PSRs B1937+21 and
B1855+09 seem to be limited by a power law component which produces a
minimum in $\sigma_z$ after $2 \mathrm{\ yr}$ and $5 \mathrm{\ yr}$ respectively. This is most
likely a result of a small amount of intrinsic timing
noise~\cite{ktr94}. Although the baseline for the bright millisecond
pulsar J0437$-$4715 is shorter, its $\sigma_z$ is already an order of
magnitude smaller than the other two pulsars or the atomic clocks.
Timing observations of an array of millisecond pulsars in the context
of detecting gravitational waves from the Big Bang are discussed
further in Section~\ref{sec:array}.

\epubtkImage{sigmaz.png}{
  \begin{figure}[htbp]
    \def\epsfsize#1#2{0.5#1}
    \centerline{\epsfbox{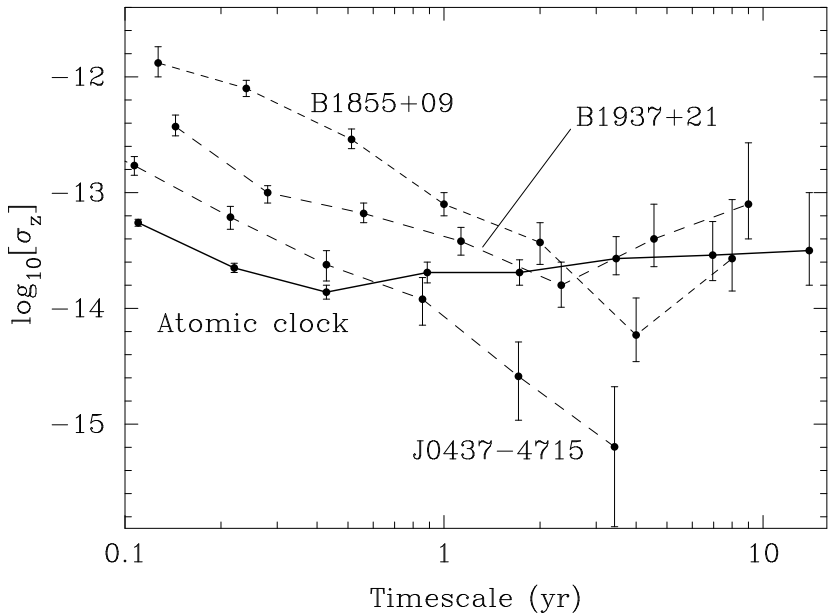}}
    \caption{\it The fractional stability of three millisecond pulsars
      compared to an atomic clock. Both PSRs B1855+09 and B1937+21 are
      comparable, or just slightly worse than, the atomic clock
      behaviour over timescales of a few years~\cite{mtem97}. More
      recent timing of the millisecond pulsar J0437$-$4715~\cite{hot04}
      indicates that it is more stable than the atomic clock.}
    \label{fig:sigmaz}
  \end{figure}}


\subsection{Timing binary pulsars}
\label{sec:tbin}

For binary pulsars, the simple timing model introduced in
Section~\ref{sec:tmodel} needs to be extended to incorporate the additional
motion of the pulsar as it orbits the common centre-of-mass of the
binary system. Treating the binary orbit using Kepler's laws to refer
the TOAs to the binary barycentre requires five additional model
parameters: the orbital period $P_\mathrm{b}$, projected semi-major
orbital axis $x$, orbital eccentricity
$e$, longitude of periastron $\omega$ and the epoch of periastron
passage $T_0$. This description, using five ``Keplerian
parameters'', is identical to that used for spectroscopic binary
stars. Analogous to the radial velocity curve in a spectroscopic
binary, for binary pulsars the orbit is described by the apparent
pulse period against time. An example of this is shown in Panel~a of
Figure~\ref{fig:orbits}. Alternatively, when radial accelerations can be 
measured, the orbit can also be visualised in a plot of acceleration
versus period as shown in Panel~b of Figure~\ref{fig:orbits}. This method is particularly
useful for determining binary pulsar
orbits from sparsely sampled data~\cite{fkl01}.

\epubtkImage{orbitcurves.png}{
  \begin{figure}[htbp]
    \def\epsfsize#1#2{0.65#1}
    \centerline{\epsfbox{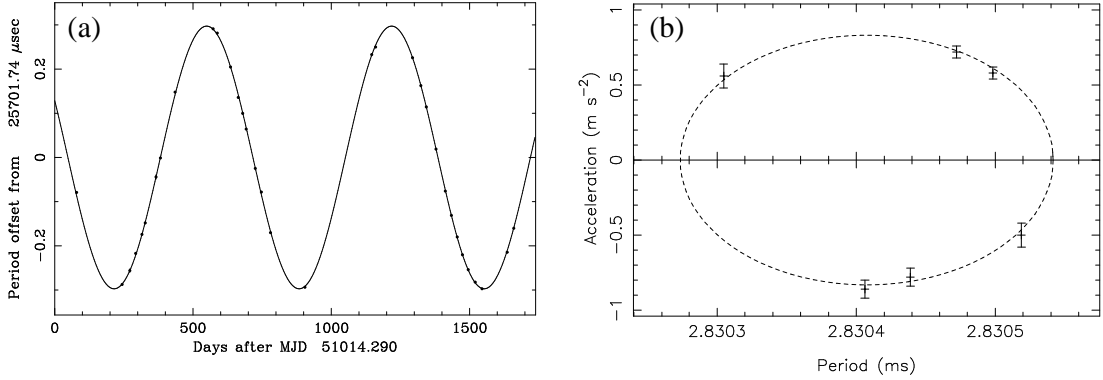}}
    \caption{\it Panel~a: Keplerian orbital fit to the $669 \mbox{-}\mathrm{day}$ binary
      pulsar J0407+1607~\cite{lxf05}. Panel~b: Orbital fit in the
      period-acceleration plane for the globular cluster pulsar
      47~Tuc~S~\cite{fkl01}.}
    \label{fig:orbits}
  \end{figure}}

Constraints on the masses of the pulsar $m_\mathrm{p}$ and the
orbiting companion $m_\mathrm{c}$ can be placed by combining $x$ and
$P_\mathrm{b}$ to obtain the mass function
\begin{equation}
  f_\mathrm{mass} = \frac{4\pi^2}{G} \frac{x^3}{P_\mathrm{b}^2} = 
  \frac{(m_\mathrm{c} \sin i)^3}{(m_\mathrm{p} + m_\mathrm{c})^2},
  \label{equ:massfn}
\end{equation}
where $G$ is Newton's gravitational constant and $i$ is the (initially
unknown) angle between the orbital plane and the plane of the sky
(i.e.\ an orbit viewed edge-on corresponds to $i=90^{\circ}$). In the
absence of further information, the standard practice is to consider a
random distribution of inclination angles. Since the probability that
$i$ is \emph{less} than some value $i_0$ is $p(<i_0) = 1 - \cos(i_0)$,
the 90\% confidence interval for $i$ is $26^{\circ} < i <
90^{\circ}$. For an assumed pulsar mass, the 90\% confidence interval
for $m_\mathrm{c}$ can be obtained by solving Equation~(\ref{equ:massfn})
for $i = 26^{\circ}$ and $90^{\circ}$. If the sum of the
masses $M=m_\mathrm{p}+m_\mathrm{c}$ can be determined (e.g., through a
measurement of relativistic periastron advance described below), then
the condition $\sin i<1$ sets a lower limit on the companion mass
$m_\mathrm{c}>(f_\mathrm{mass} M^2)^{1/3}$ and a corresponding upper limit
on the pulsar mass.

Although most of the presently known binary pulsar systems can be
adequately timed using Kepler's laws, there are a number which require
an additional set of ``post-Keplerian'' (PK) parameters which have
a distinct functional form for a given relativistic theory of
gravity. In general relativity (GR) the PK formalism gives the relativistic
advance of periastron
\begin{equation}
  \dot\omega = 3 \left( \frac{P_\mathrm{b}}{2\pi} \right)^{-5/3}
  (T_\odot M)^{2/3} (1 - e^2)^{-1},
  \label{equ:omdot}
\end{equation}
the time dilation and gravitational redshift parameter
\begin{equation}
  \gamma = e \left(\frac{P_\mathrm{b}}{2\pi}\right)^{1/3}
  T_\odot^{2/3}M^{-4/3} m_\mathrm{c} (m_\mathrm{p} + 2m_\mathrm{c}),
  \label{equ:gamma}
\end{equation}
the rate of orbital decay due to gravitational radiation
\begin{equation}
  \dot P_\mathrm{b} = -\frac{192\pi}{5}
  \left( \frac{P_\mathrm{b}}{2\pi} \right)^{-5/3}
  \left( 1 + \frac{73}{24} e^2 + \frac{37}{96} e^4 \right)
  \left( 1 - e^2 \right)^{-7/2}
  T_\odot^{5/3} m_\mathrm{p} m_\mathrm{c} M^{-1/3}
  \label{equ:pbdot}
\end{equation}
and the two Shapiro delay parameters
\begin{equation}
  r = T_\odot m_\mathrm{c}
  \label{equ:r}
\end{equation}
and
\begin{equation}
  s = x \left( \frac{P_\mathrm{b}}{2\pi} \right)^{-2/3}
  T_\odot^{-1/3} M^{2/3} m_\mathrm{c}^{-1}
  \label{equ:s}
\end{equation}
which describe the delay in the pulses around superior conjunction
where the pulsar radiation traverses the gravitational well of
its companion. In the above expressions, all masses are in solar units,
$M\equiv m_\mathrm{p}+m_\mathrm{c}$, $x\equiv
a_\mathrm{p} \sin i/c$,
$s\equiv\sin i$ and $T_\odot\equiv G M_\odot/c^3 = 4.925490947
\mathrm{\ \mu s}$.
Some combinations, or all, of the PK parameters have now been 
measured for a number of binary pulsar systems.
Further PK parameters due to aberration and relativistic
deformation~\cite{dd86} are not listed here but may soon
be important for the double pulsar~\cite{lk05}.

The key point in the PK definitions is that,
given the precisely measured Keplerian parameters, the only two unknowns
are the masses of the pulsar and its companion,
$m_\mathrm{p}$ and $m_\mathrm{c}$. Hence, from a measurement of just
two PK parameters (e.g., $\dot{\omega}$ and $\gamma$) one can
solve for the two masses and, using 
Equation~(\ref{equ:massfn}), find the orbital inclination angle $i$.
If three (or more) PK parameters are measured, the system is
``overdetermined'' and can be used to test GR (or, more
generally, any other
theory of gravity) by comparing
the third PK parameter with the predicted value based on
the masses determined from the other two.

\epubtkImage{1913.png}{
  \begin{figure}[htbp]
    \def\epsfsize#1#2{0.5#1}
    \centerline{\epsfbox{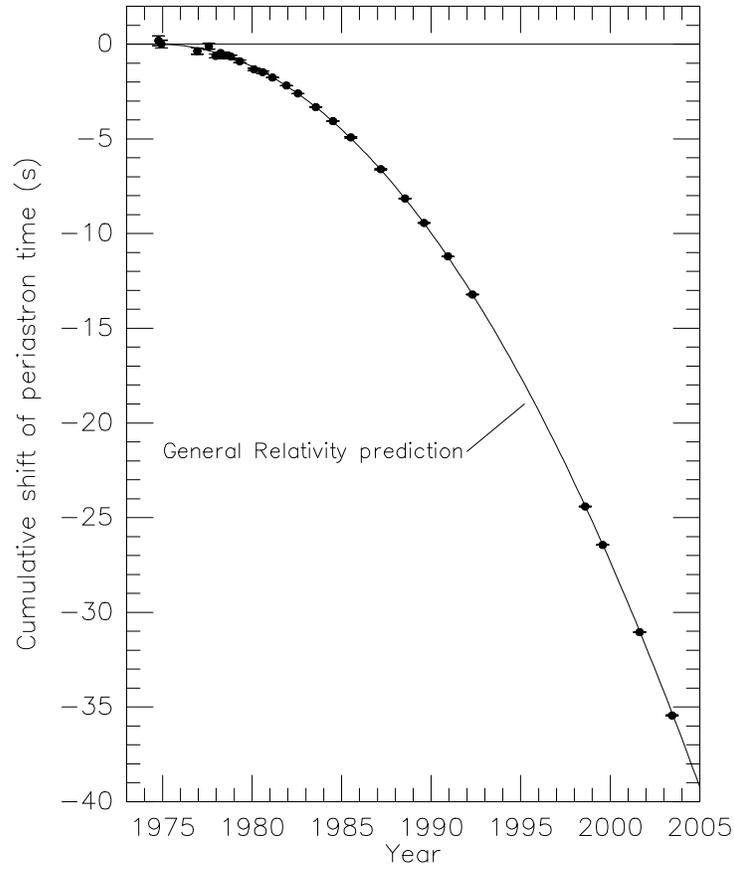}}
    \caption{\it Orbital decay in the binary pulsar B1913+16 system
      demonstrated as an increasing orbital phase shift for periastron
      passages with time. The GR prediction due entirely to the
      emission of gravitational radiation is shown by the
      parabola. Figure provided by Joel Weisberg~\cite{wt05}.}
    \label{fig:1913}
  \end{figure}}

The first binary pulsar used to test GR in this way
was PSR B1913+16 discovered by
Hulse \& Taylor in 1974~\cite{ht75a}. Measurements of three PK
parameters ($\dot{\omega}$, $\gamma$ and $\dot{P_\mathrm{b}}$) were
obtained from long-term timing observations at
Arecibo~\cite{tw82, tw89}. The measurement of orbital decay, which
corresponds to a shrinkage of about $3.2 \mathrm{\ mm}$ per orbit, is seen most
dramatically as the gradually increasing shift in orbital phase for
periastron passages with respect to a non-decaying orbit shown in
Figure~\ref{fig:1913}. This figure includes recent Arecibo data taken
since the upgrade of the telescope in the mid 1990s. The measurement
of orbital decay, now spanning a $30 \mbox{-}\mathrm{yr}$ baseline~\cite{wt05}, is
within 0.2\% of the GR prediction and
provided the first indirect evidence for the existence of gravitational
waves. Hulse and Taylor were awarded the 1993 Nobel Physics 
prize~\cite{nobpr1993, hul94, tay94} in recognition of their discovery of
this remarkable laboratory for testing GR.

More recently, five PK parameters have been
measured for PSRs B1534+12~\cite{sttw02} and
J0737$-$\hspace{0 em}3039A~\cite{kra05}. For PSR B1534+12, the test of GR comes from
measurements of $\dot{\omega}$, $\gamma$ and $s$, where the agreement
between theory and observation is within 0.7\%~\cite{sttw02}. This
test will improve in the future as the timing baseline extends and a
more significant measurement of $r$ can be made. Although a
significant measurement of $\dot{P_\mathrm{b}}$ exists, it is known to be
contaminated by kinematic effects which depend on the assumed distance
to the pulsar~\cite{sac98}. Assuming GR to be correct, the observed
and theoretical $\dot{P_\mathrm{b}}$ values can be reconciled to provide
a ``relativistic measurement'' of the distance $d = 1.04 \pm
0.03 \mathrm{\ kpc}$~\cite{sta05b}. Prospects for independent parallax
measurements of the distance to this pulsar using radio
interferometry await more sensitive telescopes~\cite{sta05}.

\epubtkImage{0737m1m2.png}{
  \begin{figure}[htbp]
    \def\epsfsize#1#2{0.6#1}
    \centerline{\epsfbox{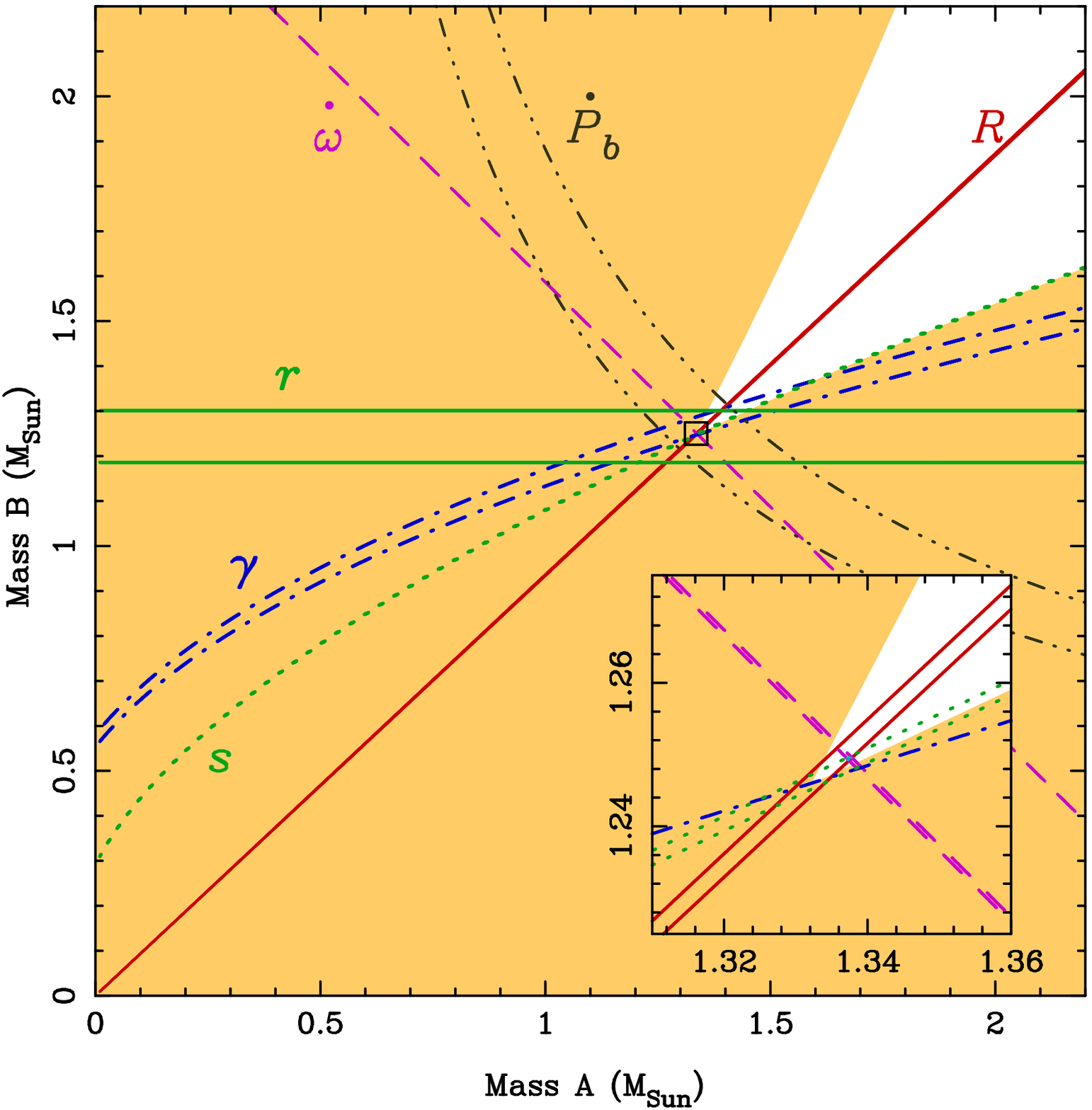}}
    \caption{\it `Mass--mass' diagram showing the observational
      constraints on the masses of the neutron stars in the double
      pulsar system J0737$-$3039. Inset is an enlarged view of the
      small square encompassing the intersection of the tightest
      constraints. Figure provided by Michael Kramer~\cite{kra05}.}
    \label{fig:0737m1m2}
  \end{figure}}

For PSR J0737$-$3039, where two independent pulsar clocks can be timed,
five PK parameters of the $22.7 \mbox{-}\mathrm{ms}$ pulsar ``A'' have been measured
as well as two additional constraints from the measured mass function
and projected semi-major axis of the $2.7 \mbox{-}\mathrm{s}$ pulsar ``B''. In terms of a
laboratory for GR, then, J0737$-$3039 promises to go well beyond the
results possible from PSRs B1913+16 and B1534+12. A useful means of
summarising the limits so far is Figure~\ref{fig:0737m1m2}
which shows the allowed regions of parameter space in terms of the
masses of the two pulsars. The shaded regions are excluded by the
requirement that $\sin i < 1$. Further
constraints are shown as pairs of lines enclosing permitted regions as
predicted by GR. The measurement of $\dot{\omega}=16.899 \pm
0.001 \mathrm{\ deg\ yr}^{-1}$ gives the total system mass $M=2.587
\pm 0.001\,M_\odot$. The measurement of the projected semi-major axes of both
orbits gives the mass ratio $R=1.071 \pm 0.001$. The mass ratio
measurement is unique to the double pulsar system and rests on
the basic assumption that momentum is conserved. This constraint
should apply to any reasonable theory of gravity.
The intersection between the lines for $\dot{\omega}$ and $R$
yield the masses of A and B as $m_\mathrm{A}=1.338\pm0.001\,M_\odot$
and $m_\mathrm{B}=1.249\pm0.001\,M_\odot$. From these values,
using Equations~(\ref{equ:gamma}--\ref{equ:s}) the expected values of
$\gamma$, $\dot{P}_\mathrm{b}$, $r$ and $s$ may be calculated and
compared with the observed values. These four tests of GR all agree
with the theory to within the uncertainties. Currently the tightest
constraint is the Shapiro delay parameter $s$ where the observed value
is in agreement with GR at 0.1\% level.

Less than two years after its discovery, the double pulsar system has
already surpassed the three decades of monitoring PSR B1913+16 and over a
decade of timing PSR B1534+12 as a precision test of GR.
On-going precision timing measurements
of the double pulsar system should soon provide even more stringent
and new tests of GR. Crucial to these measurements will be the timing
of the $2.7 \mbox{-}\mathrm{s}$ pulsar B, where the observed profile is significantly
affected by A's relativistic wind~\cite{lbk04, mkl04}. A careful
decoupling of these profile variations is required to accurately
measure TOAs for this pulsar and determine the extent to which the
theory-independent mass ratio $R$ can be measured.

The relativistic effects observed in the double pulsar system are so
large that corrections to higher post-Newtonian order may soon need to
be considered. For example, $\dot{\omega}$ may be measured precisely
enough to require terms of second post-Newtonian order to be included
in the computations~\cite{ds88}. In addition, in contrast to
Newtonian physics, GR predicts that the spins of the neutron stars
affect their orbital motion via spin-orbit coupling. This effect
would most clearly be visible as a contribution to the observed
$\dot{\omega}$ in a secular~\cite{bo75b} and periodic
fashion~\cite{wex95}. For the J0737$-$3039 system, the expected
contribution is about an order of magnitude larger than for PSR
B1913+16~\cite{lbk04}. As the exact value depends on the pulsars'
moment of inertia, a potential measurement of this effect allows the
moment of inertia of a neutron star to be determined for the first
time~\cite{ds88}. Such a measurement would be invaluable for studies
of the neutron star equation of state and our understanding of
matter at extreme pressure and densities~\cite{lp04}.

The systems discussed above are all double neutron star binaries. A
further self-consistency test of GR has recently been made in the $4.7 \mbox{-}\mathrm{hr}$
relativistic binary J1141$-$6545, where the measurement~\cite{bokh03}
of $\dot{\omega}$, $\gamma$ and $\dot{P}_\mathrm{b}$ yield a pulsar mass
of $1.30\pm0.02\,M_\odot$ and a companion mass of
$0.99\pm0.02\,M_\odot$. Since the mass of the companion
is some seven standard deviations below the mean neutron star
mass (see Figure~\ref{fig:masses}), it is most likely a white dwarf.
The observed $\dot{P}_\mathrm{b}=-(4 \pm 1)
\times 10^{-13}$ is consistent, albeit with limited precision,
with the predicted value from GR ($-3.8 \times 10^{-13}$).
Continued timing should reduce the relative error
in $\dot{P}_\mathrm{b}$ down to 1\% by 2010~\cite{bokh03}.

\epubtkImage{masses.png}{
  \begin{figure}[htbp]
    \def\epsfsize#1#2{0.6#1}
    \centerline{\epsfbox{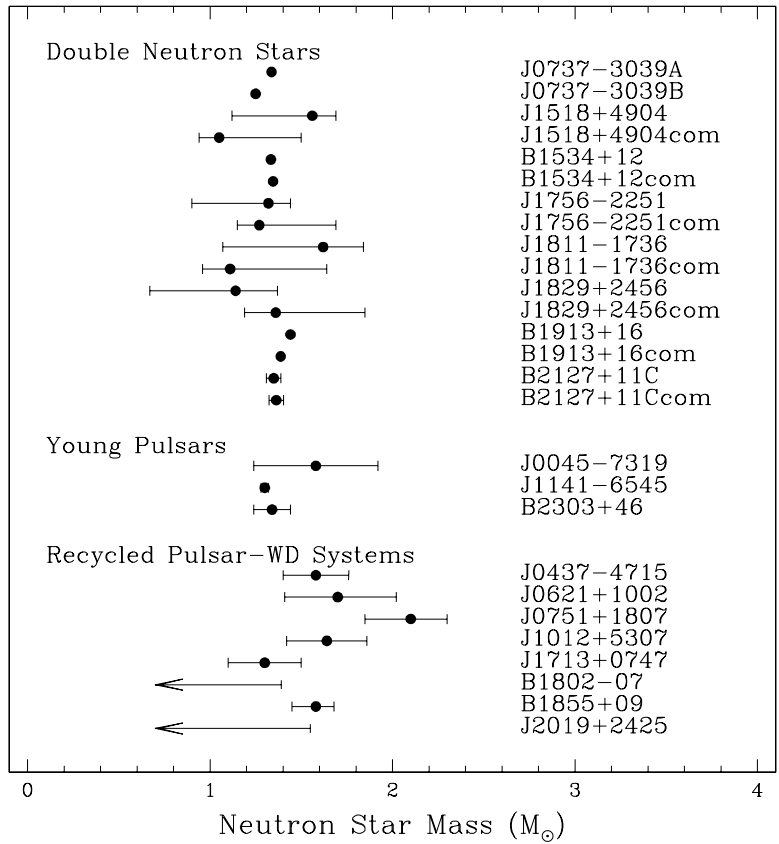}}
    \caption{\it Distribution of neutron star masses as inferred from
      timing observations of binary pulsars~\cite{sta04}. The vertical
      dotted line shows the canonical neutron star mass of
      $1.4\,M_\odot$. Figure provided by Ingrid Stairs.}
    \label{fig:masses}
  \end{figure}}

PK parameters have now been measured for a number of other binary
pulsars which provide interesting constraints on neutron star
masses~\cite{tc99, sta04}. Figure~\ref{fig:masses} shows the distribution
taken from a recent compilation~\cite{sta04}. While the young
pulsars and the double neutron star binaries are consistent
with, or just below, the canonical $1.4\,M_\odot$, we note
that the millisecond pulsars in binary systems have, on average,
significantly larger masses. This provides strong support for their
formation through an extended period of accretion in the past,
as discussed in Section~\ref{sec:evolution}.


\subsection{Pulsar timing and gravitational wave detection}
\label{sec:gwdet}

We have seen in the previous Section\ref{sec:tbin} how pulsar timing can be used to
provide indirect evidence for the existence of gravitational waves
from coalescing stellar-mass binaries. In this final section, we look
at how pulsar timing might soon be used to detect gravitational
radiation directly. The idea to use pulsars as natural gravitational
wave detectors was first explored in the late
1970s~\cite{saz78, det79}. The basic concept is to treat the solar
system barycentre and a distant pulsar as opposite ends of an
imaginary arm in space. The pulsar acts as the reference clock at one
end of the arm sending out regular signals which are monitored by an
observer on the Earth over some time-scale $T$. The effect of a
passing gravitational wave would be to perturb the local spacetime
metric and cause a change in the observed rotational frequency of the
pulsar. For regular pulsar timing observations with typical TOA
uncertainties of $\epsilon_\mathrm{TOA}$, this ``detector'' would be
sensitive to waves with dimensionless amplitudes
$h \gtrsim \epsilon_\mathrm{TOA}/T$ and frequencies
$f \sim 1/T$~\cite{bcr83, bnr84}.


\subsubsection{Probing the gravitational wave background}
\label{sec:uplim}

Many cosmological models predict that the Universe is presently filled
with a low-frequency
stochastic gravitational wave background (GWB) produced during
the big bang era~\cite{pee93}. A significant 
component~\cite{rr95, jb03} is the gravitational
radiation from massive black hole mergers.
In the ideal case, the change in the observed frequency
caused by the GWB should be detectable in the set of timing residuals
after the application of an appropriate model for the rotational,
astrometric and, where necessary, binary parameters of the pulsar. As
discussed in Section~\ref{sec:tmodel}, all other effects being negligible,
the rms scatter of these residuals $\sigma$ would be due to the
measurement uncertainties and intrinsic timing noise from the neutron star.

For a GWB with a flat energy spectrum in the frequency band $f \pm
f/2$ there is an additional contribution to the timing residuals
$\sigma_\mathrm{g}$~\cite{det79}. When $fT \gg 1$, the corresponding
wave energy density is
\begin{equation}
  \rho_\mathrm{g} = \frac{243 \pi^3 f^4 \sigma_\mathrm{g}^2}{208 G}.
\end{equation}
An upper limit to
$\rho_\mathrm{g}$ can be obtained from a set of timing residuals by assuming
the rms scatter is entirely due to this effect ($\sigma=\sigma_\mathrm{g}$).
These limits are commonly expressed as a fraction of the
energy density required to close the Universe
\begin{equation}
  \rho_\mathrm{c} = \frac{3 H_0^2}{8 \pi G} \simeq
  2 \times 10^{-29} h^2 \mathrm{\ g\ cm}^{-3},
\end{equation}
where the Hubble constant $H_0 = 100 \, h \mathrm{\ km\ s}^{-1} \mathrm{\ Mpc}$.

This technique was first applied~\cite{rt83} to a set of TOAs for PSR
B1237+25 obtained from regular observations over a period of 11 years
as part of the JPL pulsar timing programme~\cite{dr83}. This pulsar
was chosen on the basis of its relatively low level of timing activity
by comparison with the youngest pulsars, whose residuals are
ultimately plagued by timing noise (see Section~\ref{sec:tstab}). By ascribing
the rms scatter in the residuals ($\sigma = 240 \mathrm{\ ms}$) to the GWB, the
limit $\rho_\mathrm{g}/\rho_\mathrm{c} \lesssim 4 \times 10^{-3} h^{-2}$ for a
centre frequency $f = 7 \mathrm{\ nHz}$.

This limit, already well below the energy density required to close
the Universe, was further reduced following the long-term timing
measurements of millisecond pulsars at Arecibo
(see Section~\ref{sec:tstab}). In the intervening period, more
elaborate techniques had been devised~\cite{bcr83, bnr84, srtr90} to
look for the likely signature of a GWB in the frequency spectrum of
the timing residuals and to address the possibility of ``fitting
out'' the signal in the TOAs. Following~\cite{bcr83} it is convenient
to define
\begin{equation}
  \Omega_\mathrm{g} = \frac{1}{\rho_\mathrm{c}}
  \frac{\mathrm{d}\log \rho_\mathrm{g}}{\mathrm{d}\log f},
\end{equation}
i.e.\ the energy density of the GWB per logarithmic
frequency interval relative to $\rho_\mathrm{c}$. With this definition, the
power spectrum of the GWB~\cite{hr84, bnr84} is
\begin{equation}
  {\cal P}(f) = \frac{G \rho_\mathrm{g}}{3\pi^3 f^4} =
  \frac{H_0^2 \Omega_\mathrm{g}}{8\pi^4 f^5} =
  1.34 \times 10^4 \Omega_\mathrm{g} h^2
  f^{-5}_{\mathrm{yr}^{-1}} \mathrm{\ \mu s}^2 \mathrm{\ yr},
\end{equation}
where $f_\mathrm{yr^{-1}}$ is frequency in cycles per year. The long-term
timing stability of B1937+21, discussed in Section~\ref{sec:tstab}, limits its use
for periods $\gtrsim 2 \mathrm{\ yr}$. Using the more stable residuals for PSR B1855+09,
and a rigorous statistical analysis~\cite{td96}, the current
95\% confidence upper limit is $\Omega_\mathrm{g} h^2 < 10^{-8}$.
This limit is difficult to reconcile with most
cosmic string models for galaxy formation~\cite{ca92, td96}.

For binary pulsars, the orbital period provides an additional clock
for measuring the effects of gravitational waves. In this case, the
range of frequencies is not limited by the time span of the
observations, allowing the detection of waves with periods as large as
the light travel time to the binary system~\cite{bcr83}. The most
stringent results presently available are based on the B1855+09 limit
$\Omega_\mathrm{g} h^2 < 2.7 \times 10^{-4}$ in the frequency range
$10^{-11} < f < 4.4 \times 10^{-9} \mathrm{\ Hz}$~\cite{kop97}.


\subsubsection{Constraints on massive black hole binaries}

In addition to probing the GWB, pulsar timing is beginning to place
interesting constraints on the existence of massive black hole
binaries. Arecibo data for PSRs~B1937+21 and J1713+0747 already make
the existence of an equal-mass black hole binary in Sagittarius~$\mathrm{A}^*$
unlikely~\cite{lb01}. More recently, timing data from B1855+09
have been used to virtually rule out the existence of a proposed
supermassive black hole as the explanation for the periodic motion
seen at the centre of the radio galaxy 3C66B~\cite{simy03}.

A simulation of the expected modulations of the timing
residuals for the putative binary system, with a total mass of 
$5.4\times 10^{10}\,M_\odot$, is shown along with the
observed timing residuals in Figure~\ref{fig:3c66b}. Although
the exact signature depends on the orientation and eccentricity
of the binary system, Monte Carlo simulations show that the
existence of such a massive black hole binary is ruled out
with at least 95\% confidence~\cite{jllw04}.

\epubtkImage{3c66b.png}{
  \begin{figure}[htbp]
    \def\epsfsize#1#2{0.65#1}
    \centerline{\epsfbox{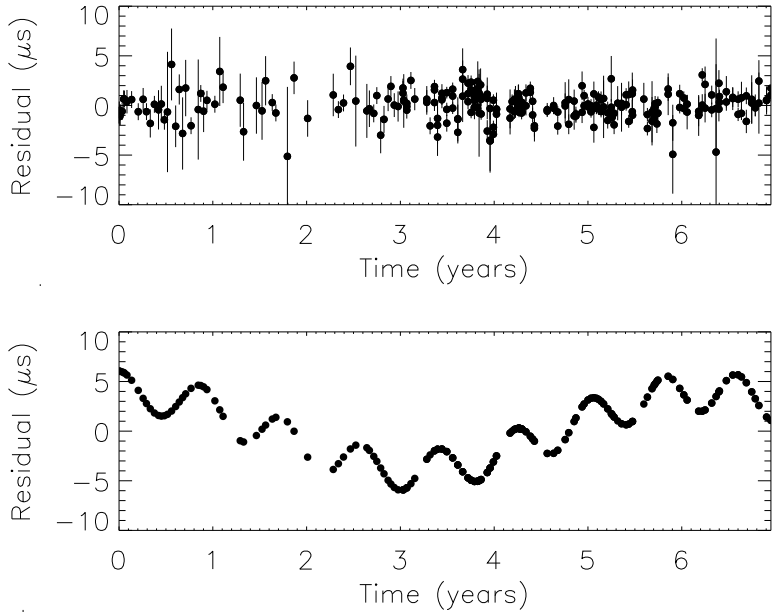}}
    \caption{\it Top panel: Observed timing residuals for
      PSR B1855+09. Bottom panel: Simulated timing residuals induced
      from a putative black hole binary in 3C66B. Figure provided by
      Rick Jenet~\cite{jllw04}.}
    \label{fig:3c66b}
  \end{figure}}


\subsubsection{A millisecond pulsar timing array}
\label{sec:array}

A natural extension of the single-arm detector concept discussed above
is the idea of using timing data for a number of pulsars distributed
over the whole sky to detect gravitational waves~\cite{hd83}. Such a
``timing array'' would have the advantage over a single arm
in that, through a cross-correlation analysis of the residuals for
pairs of pulsars distributed over the sky, it should be possible to
separate the timing noise of each pulsar from the signature of the
GWB common to all pulsars in the array. To illustrate this, consider
the fractional frequency shift of the $i$th pulsar in an array
\begin{equation}
  \frac{\delta \nu_i}{\nu_i} = \alpha_i {\cal A}(t) + {\cal N}_i(t),
\end{equation}
where $\alpha_i$ is a geometric factor dependent on
the line-of-sight direction to the pulsar and the propagation
and polarisation vectors of the gravitational wave of dimensionless
amplitude ${\cal A}$. The timing noise intrinsic to the pulsar
is characterised by the function ${\cal N}_i$. The result of a
cross-correlation between pulsars $i$ and $j$ is then
\begin{equation}
  \alpha_i \alpha_j \langle {\cal A}^2 \rangle +
  \alpha_i \langle {\cal A}{\cal N}_j \rangle +
  \alpha_j \langle {\cal A}{\cal N}_i \rangle +
  \langle {\cal N}_i{\cal N}_j \rangle,
\end{equation}
where the bracketed terms indicate cross-correlations. Since the wave
function and the noise contributions from the two pulsars are
independent, the cross correlation tends to $\alpha_i
\alpha_j \langle{\cal A}^2\rangle$ as the number of residuals becomes
large. Summing the cross-correlation functions over a large number of
pulsar pairs provides additional information on this term as a
function of the angle on the sky~\cite{hel90} and allows
the separation of the effects of clock and 
ephemeris errors from the GWB~\cite{fb90}.

\epubtkImage{gwspec.png}{
  \begin{figure}[htbp]
    \def\epsfsize#1#2{0.7#1}
    \centerline{\epsfbox{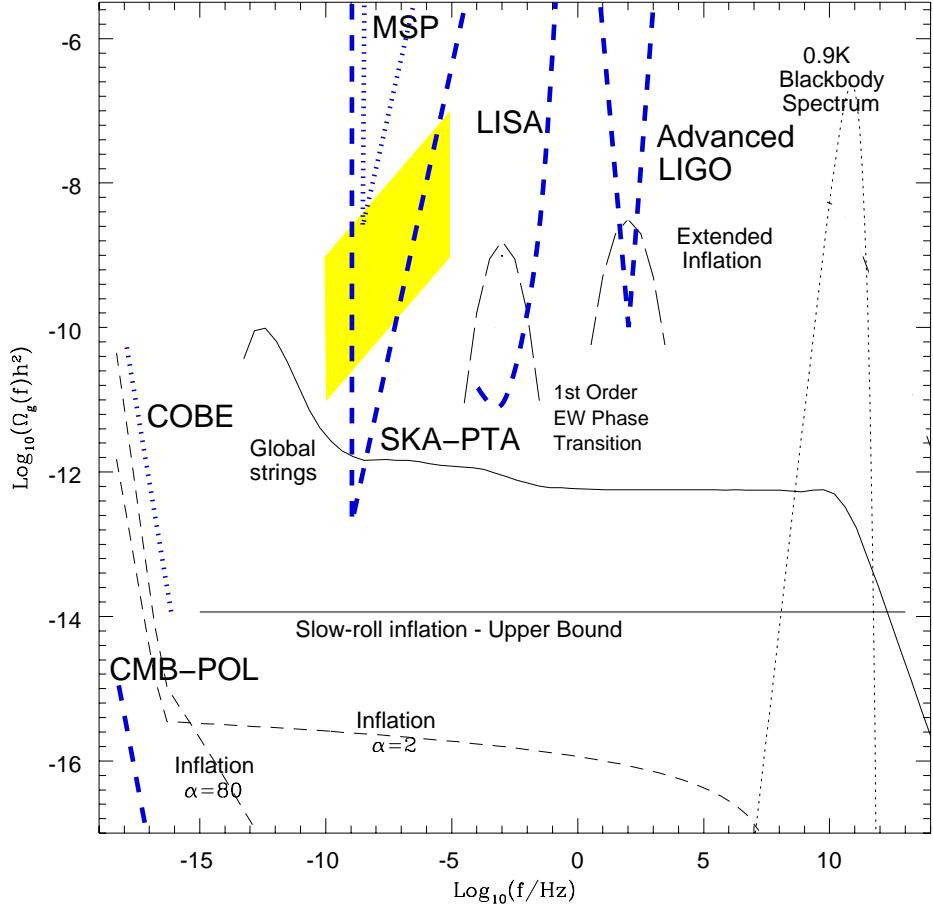}}
    \caption{\it Summary of the gravitational wave spectrum showing
      the location in phase space of the pulsar timing array (PTA) and
      its extension with the Square Kilometre Array (SKA). Figure
      updated by Michael Kramer~\cite{kbc04} from an original design
      by Richard Battye.}
    \label{fig:gwspec}
  \end{figure}}

A preliminary analysis
applying the timing array concept to Arecibo data for three
millisecond pulsars (B1937+21, B1855+09 and J1737$+$0747) now
spanning 17 years has reduced the energy density
limit to $\Omega_\mathrm{g} h^2 < 2 \times
10^{-9}$~\cite{lom02}\epubtkFootnote{It is likely, however, that a more
  detailed statistical analysis of the data are required to confirm or
  refute this result~\cite{dv04}.}.
This limit, and the region of the gravitational wave
energy density spectrum probed by the current pulsar timing
array is shown in Figure~\ref{fig:gwspec} where it can be
seen that the pulsar timing regime is complementary to the
higher frequency bands of LISA and LIGO.

A number of long-term timing projects are now underway to make a
large-scale pulsar timing array a reality. At Arecibo and Green Bank,
regular timing of a dozen or more millisecond pulsars are now carried
out using on-line data acquisition systems by groups at
Berkeley~\cite{bkypsr}, Princeton~\cite{pripsr} and
UBC~\cite{ubcpsr}. A similar system is also being commissioned at
Jodrell Bank~\cite{jlk03}. The Berkeley group have installed identical
sets of data taking equipment at Effelsberg and Nan\c{c}ay which has
enabled these telescopes to make regular high-quality millisecond
pulsar timing observations~\cite{lkd04, cb04}. In the southern
hemisphere, a high-precision timing program at Parkes~\cite{van03} has
been in existence since the discovery bright nearby millisecond pusar
J0437$-$4715~\cite{jlh93}. Since February 2004, weekly Parkes
observations of $\sim$20 millisecond pulsars are already achieving a
limit of $\Omega_\mathrm{g} h^2 < 10^{-4}$, with the ultimate goal being
to reach a limits of $\Omega_\mathrm{g} h^2 < 10^{-10}$ before 2010~\cite{hob05}.

Looking further ahead, the increase in sensitivity provided by
the Square Kilometre Array~\cite{ska, kbc04} should further
improve the limits of the spectrum probed by pulsar timing. 
As Figure~\ref{fig:gwspec} shows, the SKA could provide
up to two orders of magnitude improvement over current capabilities.


\subsection{Going further}
\label{sec:tfurther}

For further details on the techniques and prospects of pulsar timing,
a number of excellent reviews are
available~\cite{bh86, tay91, bel98, sta03, lk05}. Two audio files and
slides from lectures at the centennial meeting of the American
Physical Society are also available
on-line~\cite{backertalk, willtalk}. The tests of general relativity
discussed in Section~\ref{sec:tbin}, along with further effects, are
reviewed elsewhere in this journal by Will~\cite{wil01} and
Stairs~\cite{sta03}. Further discussions on the realities of using
pulsars as gravity wave detectors can be found in other review
articles~\cite{rom89, bac96, hob05}.

\newpage


\section{Summary and Future Prospects}
\label{sec:future}

The main aim of this article was to review some of the many
astrophysical applications provided by the present sample of binary
and millisecond radio pulsars. The topics covered here, along with the
bibliography and associated tables of observational parameters, should
be useful to those wishing to delve deeper into the vast body of literature
that exists.

Through an understanding of the Galactic population of radio pulsars
summarised in Section~\ref{sec:gal} it is possible to predict the detection
statistics of terrestrial gravitational wave detectors to nearby
rapidly spinning neutron stars (see Section~\ref{sec:nmsppop}), as well as
coalescing relativistic binaries at cosmic distances
(see Section~\ref{sec:relpop}). Continued improvements in gravitational wave
detector sensitivities should result in a number of interesting
developments and contributions in this area. These developments and
contributions might include the detection of presently known radio
pulsars, as well as a population of coalescing binary systems which
have not yet been detected as radio pulsars.

The phenomenal timing stability of radio pulsars leads naturally to a
large number of applications, including their use as laboratories for
relativistic gravity (see Section~\ref{sec:tbin}) and as natural detectors
of gravitational radiation (see Section~\ref{sec:gwdet}). Long-term timing
experiments of the present sample of millisecond and binary pulsars
currently underway appear to have tremendous potential in these areas
and perhaps detect the gravitational wave background (if it exists) 
within the next decade.

These applications will benefit greatly from the continuing discovery
of new systems by the present generation of radio pulsar searches
which continue to probe new areas of parameter space. Based on the
results presented in Section~\ref{sec:nmsppop}, it is clear that we are
aware of only about 1\% of the total active pulsar population in our
Galaxy. It is therefore likely that we have not seen all of the pulsar
zoo. More sensitive surveys using a multibeam system on the Arecibo
telescope~\cite{alfa} are now being carried out. Future
surveys with the Square Kilometre Array~\cite{ska} will ultimately
provide a far more complete census of the Galactic pulsar
population. With the double pulsar now checked off the list of
pulsar milestones given in the last version of this review~\cite{lor01}, 
we look forward to new discoveries. Two possible ``holy
grails'' of pulsar astronomy which may soon be found are the following
ones:
\begin{description}
\item[A pulsar--black hole binary system:]
  Following the wide variety of science possible from the new double
  pulsar J0737$-$3039 (see Sections~\ref{sec:evolution}, \ref{sec:nsns}
  and~\ref{sec:tbin}), a radio pulsar with a black-hole companion would
  without doubt be a fantastic gravitational laboratory. Excellent places
  to look for such systems are globular clusters and the Galactic
  centre~\cite{pl04}.
\end{description}
\begin{description}
\item[A sub-millisecond pulsar:]
  With the new discoveries in Terzan~5, it now appears that the
  original millisecond pulsar, B1937+21 is no longer the most rapidly
  rotating neutron star known~\cite{ran05}. Do neutron stars with kHz
  spin rates exist? Searches now have sensitivity to such
  objects~\cite{bd97} and a discovery of even one would constrain the
  equation of state of matter at high densities. As mentioned in
  Section~\ref{sec:nms}, the R-mode instability may prevent neutron
  stars from reaching such high spin rates~\cite{bil98, cmm03,
    ajks00}.
\end{description}
Pulsar astronomy remains an extremely active area of modern
astrophysics and the next decade will undoubtedly continue to produce
new results from currently known objects as well as new surprises. In
my opinion, pulsar research is currently limited by a shortage of
researchers, and not necessarily resources. Keen graduate students
more than ever are needed to help shape the future of this
exciting and continually evolving field.

\newpage


\section{Acknowledgements}

Many thanks to Maura McLaughlin, Michael Kramer and Ingrid Stairs who
read and provided very useful comments on this revised review, as well
as Paul Ray and Matthew Bailes for spotting errors in the previous
version. I am indebted to a number of colleagues who, as noted in many
of the figure captions, kindly gave permission to use their figures in
this article. Frequent use was made of NASA's magnificent Astrophysics
Data System~\cite{nasaads}, the arXiv e-Print service~\cite{lanl} and
of course {\it Google} during the literature searches. The
Royal Society supports my research through their University Research
Fellowship scheme. Finally, I'd like to thank the {\it Living Reviews}
editors for their support and patience during the completion of this
long-overdue update.

\newpage

\appendix


\section{Tables of Binary and Millisecond Pulsars}
\label{appendix}

\begin{table}[htbp]
  \centering
  \begin{tabular}{l|rrrrrr}
    \hline \hline
    \vsp PSR &
    \multicolumn{1}{c}{$P$} &
    \multicolumn{1}{c}{$\log \tau_\mathrm{c}$} &
    \multicolumn{1}{c}{$\log B$} &
    \multicolumn{1}{c}{$d$} &
    \multicolumn{1}{c}{$v_\mathrm{t}$} &
    \multicolumn{1}{c}{Refs.} \\
    &
    \multicolumn{1}{c}{[ms]} &
    \multicolumn{1}{c}{[$\log(\mathrm{yr})$]} &
    \multicolumn{1}{c}{[$\log(\mathrm{G})$]} &
    \multicolumn{1}{c}{[kpc]} &
    \multicolumn{1}{c}{[km/s]} \\ [0.4 em]
    \hline \hline
    J0030+0451   &  4.865 &  9.9 & 8.3\z & 0.23  & $<65$ & \cite{lzb01} \\
    J0609+2130   & 55.698 &  9.6 &  9.56 & 1.20  &     ? & \cite{lma04, lxf05} \\
    J0711$-$6830 &  5.491 & 10.4 & 8.2\z & 1.04  &   139 & \cite{bjb97, tsb99} \\
    J1024$-$0719 &  5.162 &  9.7 & 8.5\z & 0.35  &    45 & \cite{bjb97, tsb99} \\
    J1629$-$6902 &  6.000 & 10.0 & 8.4\z & 1.36  &     ? & \cite{eb01b} \\ [1.0 em]
    J1721$-$2457 &  3.497 & 10.0 & 8.2\z & 1.56  &     ? & \cite{eb01b} \\
    J1730$-$2304 &  8.123 &  9.9 & 8.6\z & 0.51  &    53 & \cite{lnl95, tsb99} \\
    J1744$-$1134 &  4.075 &  9.9 & 8.3\z & 0.17  &    20 & \cite{bjb97, tsb99} \\
    J1801$-$1417 &  3.625 & 10.0 & 8.1\z & 1.80  &     ? & \cite{fsk04} \\
    J1843$-$1113 &  1.846 &  9.5 & 8.1\z & 1.97  &     ? & \cite{hfs04} \\ [1.0 em]
    B1937+21     &  1.558 &  8.4 & 8.6\z & 9.65  &    22 & \cite{bkh82, ktr94} \\
    J1944+0907   &  5.185 &  9.7 & 8.5\z & 1.8\z &     ? & \cite{mla05} \\
    J2124$-$3358 &  4.931 &  9.9 & 8.4\z & 0.25  &    67 & \cite{bjb97, tsb99} \\
    J2235+1506   & 59.767 & 10.0 & 9.4\z & 1.15  &    98 & \cite{cnt93} \\
    J2322+2057   &  4.808 & 10.2 & 8.2\z & 0.78  &    89 &\cite{nt95, cnt96} \\
    \hline \hline
  \end{tabular}
  \caption{\it Parameters for the 15 isolated millisecond pulsars
    currently known in the Galactic disk. Two of the pulsars
    (J0609+2130 and J2235+1506) are thought to be the descendants of
    high-mass binary systems. Listed are the spin period $P$, the
    base-10 logarithms of the characteristic age $\tau_\mathrm{c}$ and
    surface magnetic field strength $B$ (see Section~\ref{sec:nms}),
    the distance $d$ derived from the Cordes \& Lasio electron density
    model~\cite{cl02a, cl02b} or independently (when available), and
    the transverse speed $v_\mathrm{t}$ inferred from $d$ and a proper
    motion measurement (when available). Key publications for each
    pulsar are referenced to the bibliography.}
  \label{tab:imsps}
\end{table}

\begin{landscape}

\begin{table}
  \centering
  \begin{tabular}{l|rrrrrrrrrr}
    \hline \hline
    \vsp PSR &
    \multicolumn{1}{c}{$P$} &
    \multicolumn{1}{c}{$\log \tau_\mathrm{c}$} &
    \multicolumn{1}{c}{$\log B$} &
    \multicolumn{1}{c}{$d$} &
    \multicolumn{1}{c}{$v_\mathrm{t}$} &
    \multicolumn{1}{c}{$P_\mathrm{b}$} &
    \multicolumn{1}{c}{$x$} &
    \multicolumn{1}{c}{$e$} &
    \multicolumn{1}{c}{$m_2$} &
    \multicolumn{1}{c}{Refs.} \\
    &
    \multicolumn{1}{c}{[ms]} &
    \multicolumn{1}{c}{[$\log(\mathrm{yr})$]} &
    \multicolumn{1}{c}{[$\log(\mathrm{G})$]} &
    \multicolumn{1}{c}{[kpc]} &
    \multicolumn{1}{c}{[km/s]} &
    \multicolumn{1}{c}{[days]} &
    \multicolumn{1}{c}{[s]} & &
    \multicolumn{1}{c}{[$M_\odot$]} & \\ [0.4 em]
    \hline \hline
    J0045$-$7319  &   926.276 &  6.5 & 12.3 &  57.00 &   ? &   51.2\z &  174.3\z &  0.81\z &    8.8\z & \cite{bbs95} \\
    J0737$-$3039A &    22.699 &  8.3 &  9.8 &   0.57 &   ? &    0.1\z &    1.4\z &   0.088 &     1.25 & \cite{bdp03, lbk04} \\
    J0737$-$3039B & 2773.46\z &  7.7 & 12.2 &   0.57 &   ? &    0.1\z &    1.5\z &   0.088 &     1.34 & \cite{bdp03, lbk04} \\
    J1141$-$6545  &   393.898 &  6.2 & 12.1 &   3.20 & 115 &     0.20 &     1.86 &  0.17\z &   1.0\z  & \cite{klm00, obv02} \\
    B1259$-$63    &    47.762 &  5.5 & 11.5 &   4.60 &   ? & 1237\dzz & 1296\dzz &  0.87\z &   10.0\z & \cite{jlm92, jml92} \\ [1.0 em]
    J1518+4904    &    40.935 & 10.3 &  9.0 &   0.70 &  27 &     8.63 &    20.04 &  0.25\z &    1.3\z & \cite{nst96, nst99} \\ 
    B1534+12      &     7.904 &  8.4 & 10.0 &   0.68 &  80 &     0.42 &     3.73 &  0.27\z &    1.3\z & \cite{wol91a, sac98} \\
    J1638$-$4715  &  764\dzzz &    ? &    ? &      ? &   ? & 1830\dzz &        ? &  0.9\zz & $>4.5$\z & \cite{lyn05} \\
    J1740$-$3052  &   570.309 &  5.5 & 12.6 & 10.8\z &   ? &  231\dzz &  756.9\z &  0.58\z &   16\dzz & \cite{sml01, sml03} \\
    J1756$-$2251  &    28.462 &  8.6 &  9.7 &  2.5\z &   ? &     0.32 &    2.8\z &  0.18\z &    1.2\z & \cite{fkl05} \\ [1.0 em]
    J1811$-$1736  &   104.182 &  8.9 & 10.1 &   5.94 &   ? &    18.77 &    34.78 &  0.83\z &    0.7\z & \cite{lcm00} \\
    B1820$-$11    &   279.828 &  6.5 & 11.8 &   6.26 &   ? &  357.8\z &  200.7\z &  0.79\z &    0.7\z & \cite{lm89, pv91} \\
    J1829+2456    &    41.010 & 10.1 &  9.2 &   0.75 &   ? &     1.18 &     7.24 &  0.14\z &    1.3\z & \cite{clm04, cha05} \\
    B1913+16      &    59.030 &  8.0 & 10.4 &   7.13 & 100 &     0.32 &     2.34 &  0.62\z &    1.4\z & \cite{tw82, tw89} \\
    B2303+46      &  1066.371 &  7.5 & 11.9 &   4.35 &   ? &    12.34 &    32.69 &  0.66\z &    1.2\z & \cite{lb90, vk99} \\
    \hline \hline
  \end{tabular}
  \caption{\it Parameters for the 15 eccentric ($e>0.05$) binary
    pulsars currently known that are not in globular clusters. In
    addition to the parameters listed in Table~\ref{tab:imsps}, we
    also give the binary period $P_\mathrm{b}$, the projected
    semi-major axis of the orbit $x$ in units of light seconds, the
    orbital eccentricity $e$, and the companion mass $m_2$ evaluated
    from the mass function assuming a pulsar mass of $1.4\,M_\odot$
    and an inclination angle of 60 degrees (see
    Section~\ref{sec:tbin}) or (when known) from independent
    measurements.}
  \label{tab:ebpsrs}
\end{table}

\end{landscape}

\begin{landscape}

\begin{longtable}{l|rrrrrrrrrr}
  \hline \hline
  \vsp PSR &
  \multicolumn{1}{c}{$P$} &
  \multicolumn{1}{c}{$\log \tau_\mathrm{c}$} &
  \multicolumn{1}{c}{$\log B$} &
  \multicolumn{1}{c}{$d$} &
  \multicolumn{1}{c}{$v_\mathrm{t}$} &
  \multicolumn{1}{c}{$P_\mathrm{b}$} &
  \multicolumn{1}{c}{$x$} &
  \multicolumn{1}{c}{$e$} &
  \multicolumn{1}{c}{$m_2$} &
  \multicolumn{1}{c}{Refs.} \\
  &
  \multicolumn{1}{c}{[ms]} &
  \multicolumn{1}{c}{[$\log(\mathrm{yr})$]} &
  \multicolumn{1}{c}{[$\log(\mathrm{G})$]} &
  \multicolumn{1}{c}{[kpc]} &
  \multicolumn{1}{c}{[km/s]} &
  \multicolumn{1}{c}{[days]} &
  \multicolumn{1}{c}{[s]} & &
  \multicolumn{1}{c}{[$M_\odot$]} & \\ [0.4 em]
  \hline \hline
  \endhead
  J0034$-$0534 &    1.877 &  9.9 &  7.9\z &  0.98 &    71\dz &    1.59 &    1.44 & $<0.00002$\zzz & 0.1\z & \cite{bhl94} \\
  J0218+4232   &    2.323 &  8.7 &  8.6\z &  5.85 &     ?\dz &    2.03 &    1.98 & $<0.00002$\zzz & 0.2\z & \cite{nbf95} \\
  J0407+1607   & 25.702\z &  9.7 &  9.2\z &  1.30 &    35\dz & 669.1\z & 106.5\z &    0.000937\zz & 0.2\z & \cite{lxf05} \\
  J0437$-$4715 &    5.757 &  9.7 &  8.5\z &  0.18 &   121\dz &    5.74 &    3.37 &    0.000019\zz & 0.1\z & \cite{jlh93} \\
  J0613$-$0200 &    3.062 &  9.7 &  8.2\z &  2.19 &    77\dz &    1.20 &    1.09 &    0.000007\zz & 0.1\z & \cite{lnl95, tsb99} \\ [1.0 em]
  J0621+1002   &   28.854 & 10.1 &  9.0\z &  1.88 &     ?\dz &    8.31 &   12.03 &    0.0025\zzzz & 0.5\z & \cite{cnst96} \\
  B0655+64     &  195.671 &  9.7 & 10.1\z &  0.48 &    32\dz &    1.03 &    4.13 &    0.000008\zz & 0.7\z & \cite{jl88, vk95} \\
  J0751+1807   &    3.479 &  9.8 &  8.2\z &  2.02 &     ?\dz &    0.26 &    0.40 & $>0.0$\zzzzzzz & 0.1\z & \cite{lzc95} \\
  B0820+02     &  864.873 &  8.1 & 11.5\z &  1.43 &    35\dz & 1232.47 &  162.15 &    0.011868\zz & 0.2\z & \cite{mncl80, vk95} \\
  J1012+5307   &    5.256 &  9.8 &  8.4\z &  0.52 &   102\dz &    0.60 &    0.58 & $<0.0000008$\z & 0.1\z & \cite{nll95, lcw01} \\ [1.0 em]
  J1022+1001   &   16.453 &  9.8 &  8.9\z &  0.60 & $>50$\dz &    7.81 &   16.77 &    0.000098\zz & 0.7\z & \cite{cnst96, kxc99} \\
  J1045$-$4509 &    7.474 & 10.0 &  8.5\z &  3.25 &    52\dz &    4.08 &    3.02 &    0.000024\zz & 0.2\z & \cite{lnl95, tsb99} \\
  J1157$-$5114 &   43.589 &  9.7 &  9.4\z &  1.88 &     ?\dz &    3.51 &   14.29 &    0.00040\zzz & 1.2\z & \cite{eb01} \\
  J1232$-$6501 &   88.282 &  9.2 &  9.9\z & 10.00 &     ?\dz &    1.86 &    1.61 &    0.00011\zzz & 0.1\z & \cite{clm01} \\
  B1257+12     &    6.219 &  9.5 &  8.6\z &  0.62 &   281\dz & \multicolumn{4}{c}{planetary system} & \cite{wf92, wol94} \\ [1.0 em]
  J1420$-$5625 &   34.117 &  9.9 &  9.2\z &  1.74 &     ?\dz &  40.3\z &  29.5\z &    0.0035\zzzz & 0.4\z & \cite{hfs04} \\
  J1435$-$6100 &    9.348 &  9.8 &  8.7\z &  3.25 &     ?\dz &    1.35 &    6.18 &    0.00001\zzz & 0.9\z & \cite{clm01} \\
  J1454$-$5846 &   45.249 &  9.0 &  9.8\z &  3.32 &     ?\dz &   12.42 &   26.52 &    0.0019\zzzz & 0.9\z & \cite{clm01} \\
  J1455$-$3330 &    7.987 &  9.9 &  8.5\z &  0.74 &   100\dz &   76.17 &   32.36 &    0.000170\zz & 0.3\z & \cite{lnl95, tsb99} \\
  J1603$-$7202 &   14.842 & 10.2 &  8.7\z &  1.64 &    27\dz &    6.31 &    6.88 & $>0.0$\zzzzzzz & 0.3\z & \cite{llb96, tsb99} \\ [1.0 em]
  J1618$-$39   &   11.987 &    ? &      ? &  4.76 &     ?\dz &  22.8\z &  10.2\z &    0.0\zzzzzzz & 0.2\z & \cite{eb01b} \\
  J1640+2224   &    3.163 & 10.2 &  8.0\z &  1.18 &    76\dz &  175.46 &   55.33 &    0.0008\zzzz & 0.3\z & \cite{wdk00} \\
  J1643$-$1224 &    4.622 &  9.7 &  8.4\z &  4.86 &   159\dz &  147.02 &   25.07 &    0.000506\zz & 0.1\z & \cite{lnl95, tsb99} \\
  J1709+2313   &    4.631 & 10.3 &  8.1\z &  1.83 &    89\dz &  22.7\z &  15.3\z &    0.0000187\z & 0.3\z & \cite{lwf04} \\
  J1713+0747   &    4.570 & 10.0 &  8.3\z &  0.89 &    27\dz &   67.83 &   32.34 &    0.000075\zz & 0.3\z & \cite{fwc93, cfw94} \\ [1.0 em]
  \newpage
  J1732$-$5049 &    5.313 &  9.8 &   8.43 &  1.81 &     ?\dz &    5.26 &    3.98 &    0.0000098\z & 0.2\z & \cite{eb01b} \\
  J1745$-$0952 &   19.376 &  9.5 &  9.1\z &  2.38 &     ?\dz &    4.94 &    2.38 &    0.0000180\z & 0.1\z & \cite{eb01b} \\
  J1751$-$2857 &    3.915 &  9.7 &  8.3\z &  1.40 &     ?\dz &  110.75 &   32.53 &    0.000128\zz & 0.2\z & \cite{sfl05} \\
  J1757$-$5322 &    8.870 &  9.7 &  8.7\z &  1.36 &     ?\dz &    0.45 &    2.09 &    0.000004\zz & 0.6\z & \cite{eb01} \\
  B1800$-$27   &  334.415 &  8.5 & 10.9\z &  3.62 &     ?\dz &  406.78 &   58.94 &    0.000507\zz & 0.1\z & \cite{jlm92} \\ [1.0 em]
  J1804$-$2717 &    9.343 &  9.5 &  8.8\z &  1.17 &     ?\dz &   11.13 &    7.28 &    0.000035\zz & 0.2\z & \cite{llb96} \\
  J1810$-$2005 &   32.822 &  9.6 &  9.3\z &  4.04 &     ?\dz &   15.01 &   11.98 &    0.000025\zz & 0.3\z & \cite{clm01} \\
  B1831$-$00   &  520.954 &  8.8 & 10.9\z &  2.63 &     ?\dz &    1.81 &    0.72 & $>0.0$\zzzzzzz & 0.1\z & \cite{dmr86} \\
  J1853+1303   &    4.092 &  9.9 &  8.3\z &  1.60 &     ?\dz &  115.65 &  40.8\z &    0.0000237\z & 0.3\z & \cite{sfl05} \\
  B1855+09     &    5.362 &  9.7 &  8.5\z &  1.00 &     29.2 &   12.33 &    9.23 &    0.000022\zz & 0.2\z & \cite{srs86, ktr94} \\ [1.0 em]
  J1904+0412   &   71.095 & 10.1 &  9.4\z &  4.01 &     ?\dz &   14.93 &    9.63 &    0.00022\zzz & 0.2\z & \cite{clm01} \\
  J1909$-$3744 &    2.947 &  9.5 &  8.3\z &  0.82 &   143\dz &    1.53 &    1.89 &     0.00000026 & 0.2\z & \cite{jbv03} \\
  J1910+1256   &    4.984 &  9.9 &  8.3\z &  1.90 &     ?\dz &   58.47 &  21.1\z &    0.0002302\z & 0.2\z & \cite{sfl05} \\
  J1911$-$1114 &    3.626 &  0.0 &  0.0\z &  1.59 &   183\dz &    2.72 &    1.76 & $>0.0$\zzzzzzz & 0.1\z & \cite{llb96, tsb99} \\
  J1918$-$0642 &    7.646 & $5.05 \times 10^9$ & $4.33 \times 10^8$ & 1.40 & ?\dz & 10.91 & 8.35 & 0.0000222\z & & \cite{eb01b} \\ [1.0 em]
  B1953+29     &    6.133 &  9.5 &  8.6\z &  5.39 &    98\dz &  117.35 &   31.41 &    0.00033\zzz & 0.2\z & \cite{bbf83, wdk00} \\
  B1957+20     &    1.607 &  9.4 &  8.1\z &  1.53 &   190\dz &    0.38 &    0.09 & $>0.0$\zzzzzzz &  0.02 & \cite{fst88, aft94} \\
  J2016+1948   &   64.940 &  9.4 &  9.7\z &  1.84 &     ?\dz & 635.0\z & 150.7\z &    0.00128\zzz &       & \cite{naf03, lf05} \\
  J2019+2425   &    3.935 & 10.4 &  8.0\z &  0.91 &    83\dz &   76.51 &   38.77 &    0.000111\zz & 0.3\z & \cite{nt95, nss01} \\
  J2033+1734   &    5.949 &  9.9 &  8.4\z &  1.38 &     ?\dz &   56.31 &   20.16 &    0.00013\zzz & 0.2\z & \cite{rtj96} \\ [1.0 em]
  J2051$-$0827 &    4.509 &  9.7 &  8.4\z &  1.28 &    14\dz &    0.10 &    0.05 & $>0.0$\zzzzzzz &  0.03 & \cite{sbl96, tsb99} \\
  J2129$-$5721 &    3.726 &  9.5 &  8.4\z &  2.55 &    56\dz &    6.63 &    3.50 & $>0.0$\zzzzzzz & 0.1\z & \cite{llb96, tsb99} \\
  J2145$-$0750 &   16.052 & 10.3 &  8.6\z &  0.50 &    38\dz &    6.84 &   10.16 &    0.000019\zz & 0.4\z & \cite{bhl94, tsb99} \\
  J2229+2643   &    2.978 & 10.4 &  7.9\z &  1.43 &   113\dz &   93.02 &   18.91 &    0.00026\zzz & 0.1\z & \cite{wdk00} \\
  J2317+1439   &    3.445 & 10.6 &  7.9\z &  1.89 &    68\dz &    2.46 &    2.31 & $>0.0$\zzzzzzz & 0.2\z & \cite{cnt93, cam95a} \\
  \hline \hline
  \multicolumn{1}{c}{} \\
  \caption{\it Parameters for 50 low-eccentricity binary pulsars
    currently known in the Galactic disk. The parameters listed are
    defined in Tables~\ref{tab:imsps} and~\ref{tab:ebpsrs}.}
  \label{tab:bmsps}
\end{longtable}

\end{landscape}

\begin{landscape}

\begin{longtable}{l|rrrrrrrr}
  \hline \hline
  \vsp PSR &
  \multicolumn{1}{c}{$P$} &
  \multicolumn{1}{c}{$\dot{P}$} &
  \multicolumn{1}{c}{Cluster} &
  \multicolumn{1}{c}{$P_\mathrm{b}$} &
  \multicolumn{1}{c}{$x$} &
  \multicolumn{1}{c}{$e$} &
  \multicolumn{1}{c}{$m_2$} &
  \multicolumn{1}{c}{Refs.} \\
  &
  \multicolumn{1}{c}{[ms]} &
  \multicolumn{1}{c}{[$\times 10^{-20}$]} & &
  \multicolumn{1}{c}{[days]} &
  \multicolumn{1}{c}{[s]} & &
  \multicolumn{1}{c}{[$M_\odot$]} & \\ [0.4 em]
  \hline \hline
  \endhead
  J0023$-$7204C &     5.756 &       $-4.98$ &   47~Tuc &           &         &                 &         & \cite{fck03} \\
  J0024$-$7204D &     5.357 &       $-0.34$ &   47~Tuc &           &         &                 &         & \cite{fck03} \\
  J0024$-$7205E &     3.536 &       $+9.85$ &   47~Tuc &     2.256 &   1.981 &        0.000315 & 0.15\zz & \cite{fck03} \\
  J0024$-$7204F &     2.623 &       $+6.45$ &   47~Tuc &           &         &                 &         & \cite{fck03} \\
  J0024$-$7204G &     4.040 &       $-4.21$ &   47~Tuc &           &         &                 &         & \cite{fck03} \\ [1.0 em]
  J0024$-$7204H &     3.210 &       $-0.18$ &   47~Tuc &     2.357 &   2.152 &       0.07056\z & 0.16\zz & \cite{fck03} \\
  J0024$-$7204I &     3.484 &       $-4.58$ &   47~Tuc &     0.229 &   0.038 &    $<0.0004$\zz & 0.013\z & \cite{fck03} \\
  J0023$-$7203J &     2.100 &       $-0.97$ &   47~Tuc &     0.120 &   0.040 &    $<0.00004$\z & 0.021\z & \cite{fck03} \\
  J0024$-$7204L &     4.346 &     $-12.2$\z &   47~Tuc &           &         &                 &         & \cite{fck03} \\
  J0023$-$7205M &     3.676 &       $-3.84$ &   47~Tuc &           &         &                 &         & \cite{fck03} \\ [1.0 em]
  J0024$-$7204N &     3.053 &       $-2.18$ &   47~Tuc &           &         &                 &         & \cite{fck03} \\
  J0024$-$7204O &     2.643 &       $+3.03$ &   47~Tuc &     0.135 &   0.045 &    $<0.00016$\z & 0.022\z & \cite{fck03} \\
  P             &     3.643 &               &   47~Tuc &     0.147 &   0.038 &                 & 0.017\z & \cite{clf00} \\
  J0024$-$7204Q &     4.033 &       $+3.40$ &   47~Tuc &     1.189 &   1.462 &       0.00008\z & 0.17\zz & \cite{fck03} \\
  R             &     3.480 &               &   47~Tuc &     0.066 &   0.033 &                 & 0.026\z & \cite{fck03} \\ [1.0 em]
  J0024$-$7204S &     2.830 &     $-12.0$\z &   47~Tuc &     1.201 &   0.766 &       0.00039\z & 0.088\z & \cite{fck03} \\
  J0024$-$7204T &     7.588 &     $+29.3$\z &   47~Tuc &     1.126 &   1.338 &       0.0004\zz & 0.16\zz & \cite{fck03} \\
  J0024$-$7203U &     4.342 &       $+9.52$ &   47~Tuc &     0.429 &   0.526 &       0.00014\z & 0.12\zz & \cite{fck03} \\
  V             &     4.810 &               &   47~Tuc &    0.2\zz &  0.8\zz &                 & 0.34\zz & \cite{clf00} \\
  J0024$-$7204W &     2.352 &               &   47~Tuc &     0.133 &   0.243 &                 & 0.12\zz & \cite{fck03} \\ [1.0 em]
  X             &     4.771 &               &   47~Tuc &         ? &         &                 &         & \cite{lcf03} \\
  Y             &     2.196 &               &   47~Tuc &     0.521 &   0.671 &                 & 0.13\zz & \cite{lcf03} \\
  J0514$-$4002  &     4.990 &               & NGC~1851 &    18.785 & 36\dzzz &       0.88\zzzz & 0.89\zz & \cite{fgr04} \\
  B1310+18A     &    33.163 &               &      M53 &  255\dzzz & 84\dzzz &    $<0.01$\zzzz & 0.29\zz & \cite{kapw91} \\ 
  J1342+23A     &     2.545 &               &       M3 &         ? &         &                 &         & \cite{rhs05b} \\ [1.0 em]
  \newpage
  B             &     2.389 &               &       M3 &    1.42\z &  1.9\zz &                 & 0.20\zz & \cite{rhs05b} \\
  C             &     2.166 &               &       M3 &         ? &         &                 &         & \cite{rhs05b} \\
  D             &     5.443 &               &       M3 &         ? &         &                 &         & \cite{rhs05b} \\
  B1516+02A     &     5.553 &       $+4.12$ &       M5 &           &         &                 &         &  \cite{awkp97} \\
  B1516+02B     &     7.946 &      $-0.3$\z &       M5 &     6.858 &   3.048 &       0.1378\zz & 0.11\zz & \cite{awkp97} \\ [1.0 em]
  C             &     2.484 &               &       M5 &     0.087 &   0.057 &                 & 0.037\z & \cite{rhs05b} \\
  D             &     2.988 &               &       M5 &    1.22\z &  1.6\zz &                 & 0.19\zz & \cite{rhs05b} \\
  E             &     3.182 &               &       M5 &    1.10\z &  1.2\zz &                 & 0.15\zz & \cite{rhs05b} \\
  B1620$-$26    &    11.075 &       $-5.46$ &       M4 &   191.442 &  64.809 &        0.025315 & 0.27\zz & \cite{srh03} \\
  B1639+36A     &    10.377 &      $<4.5$\z &      M13 &           &         &                 &         & \cite{kapw91} \\ [1.0 em]
  B1639+36B     &     3.528 &               &      M13 &     1.259 &  1.38\z    & $<0.001$\zzz & 0.15\zz & \cite{and92} \\
  C             &     3.722 &               &      M13 &           &         &                 &         & \cite{rhs05b} \\
  D             &     3.118 &               &      M13 &     0.591 &  0.92\z &                 & 0.17\zz & \cite{rhs05b} \\
  E             &     2.487 &               &      M13 &     0.213 &  0.17\z &                 & 0.061\z & \cite{rhs05b} \\
  J1701$-$3006A &     5.241 &      $-13.19$ & NGC~6266 &     3.805 &   3.483 &     $<0.000004$ & 0.19\zz & \cite{pdm03} \\ [1.0 em]
  J1701$-$3006B &     3.593 &      $-34.97$ & NGC~6266 &     0.144 &   0.252 &    $<0.00007$\z & 0.12\zz & \cite{pdm03} \\
  J1701$-$3006C &     3.806 &       $-3.18$ & NGC~6266 &     0.215 &   0.192 &    $<0.00006$\z & 0.069\z & \cite{pdm03} \\
  D             &     3.418 &               & NGC~6266 &    1.12\z &  0.98\z &                 & 0.12\zz & \cite{cha03} \\
  E             &     3.234 &               & NGC~6266 &    0.16\z &  0.07\z &                 & 0.030\z & \cite{cha03} \\
  F             &     2.295 &               & NGC~6266 &    0.20\z &  0.05\z &                 & 0.018\z & \cite{cha03} \\ [1.0 em]
  B1718$-$19    & 1004\dzzz & $+150000$\dzz & NGC~6342 &     0.258 &   0.352 &    $<0.005$\zzz & 0.11\zz & \cite{lbhb93} \\
  J1740$-$5340  &     3.650 &     $+16$\dzz & NGC~6397 &     1.354 &   1.652 &    $<0.0001$\zz & 0.18\zz & \cite{dpm01}\\
  B1745$-$20    &  288.60\z &  $+40000$\dzz & NGC~6440 &           &         &                 &         & \cite{lmd96} \\
  J1748$-$2446A &    11.563 &      $-3.4$\z & Terzan~5 &    0.075  &   0.119 &    $<0.0012$\zz & 0.087\z & \cite{lmd90} \\
  J1748$-$2446C &     8.436 &     $-60$\dzz & Terzan~5 &           &         &                 &         & \cite{lmbm00} \\ [1.0 em]
  \newpage
  D             &     4.713 &               & Terzan~5 &           &         &                 &         & \cite{ran01} \\
  E             &     2.197 &               & Terzan~5 &   60.06\z & 23.6\zz & $\sim0.02$\zzzz & 0.21\zz & \cite{rhs05} \\
  F             &     5.540 &               & Terzan~5 &           &         &                 &         & \cite{rhs05} \\
  G             &    21.671 &               & Terzan~5 &           &         &                 &         & \cite{rhs05} \\
  H             &     4.925 &               & Terzan~5 &           &         &                 &         & \cite{rhs05} \\ [1.0 em]
  I             &     9.570 &               & Terzan~5 &     1.328 &   1.818 &       0.428\zzz & 0.20\zz & \cite{rhs05} \\
  J             &    80.337 &               & Terzan~5 &     1.102 &   2.454 &       0.350\zzz & 0.33\zz & \cite{rhs05} \\
  K             &     2.969 &               & Terzan~5 &           &         &                 &         & \cite{rhs05} \\
  L             &     2.244 &               & Terzan~5 &           &         &                 &         & \cite{rhs05} \\
  M             &     3.569 &               & Terzan~5 &     0.443 &   0.596 &                 & 0.13\zz & \cite{rhs05} \\ [1.0 em]
  N             &     8.666 &               & Terzan~5 &     0.385 &   1.619 &        0.000045 & 0.46\zz & \cite{rhs05} \\
  O             &     1.676 &               & Terzan~5 &     0.259 &   0.112 &                 & 0.035\z & \cite{rhs05} \\
  P             &     1.728 &               & Terzan~5 &     0.362 &   1.272 &                 & 0.36\zz & \cite{rhs05} \\
  Q             &     2.812 &               & Terzan~5 & $>1$\dzz? &         &                 &         & \cite{rhs05} \\
  R             &     5.028 &               & Terzan~5 &           &         &                 &         & \cite{rhs05} \\ [1.0 em]
  S             &     6.116 &               & Terzan~5 &           &         &                 &         & \cite{rhs05} \\
  T             &     7.084 &               & Terzan~5 &           &         &                 &         & \cite{rhs05} \\
  U             &     3.289 &               & Terzan~5 & $>1$\dzz? &         &                 &         & \cite{rhs05} \\
  V             &     2.072 &               & Terzan~5 &     0.503 &   0.567 &                 & 0.11\zz & \cite{rhs05} \\
  W             &     4.205 &               & Terzan~5 &     4.877 &   5.869 &       0.015\zzz & 0.29\zz & \cite{rhs05} \\ [1.0 em]
  \newpage
  X             &     2.999 &               & Terzan~5 & $>1$\dzz? &         &                 &         & \cite{rhs05} \\
  Y             &     2.048 &               & Terzan~5 &    1.17\z &  1.16\z &                 & 0.13\zz & \cite{rhs05} \\
  Z             &     2.462 &               & Terzan~5 &         ? &       ? &                 &         & \cite{ran05} \\
  AA            &     5.788 &               & Terzan~5 &         ? &       ? &                 &         & \cite{ran05} \\
  AB            &    5.12\z &               & Terzan~5 &         ? &       ? &                 &         & \cite{ran05} \\
  AC            &    5.09\z &               & Terzan~5 &         ? &       ? &                 &         & \cite{ran05} \\
  AD            &    1.40\z &               & Terzan~5 &         ? &       ? &                 &         & \cite{ran05} \\
  AE            &    3.66\z &               & Terzan~5 &         ? &       ? &                 &         & \cite{ran05} \\
  AF            &    3.30\z &               & Terzan~5 &         ? &       ? &                 &         & \cite{ran05} \\
  J1750$-$37    &  111.61\z &               & NGC~6441 &   17.3\zz & 24.4\zz &       0.71\zzzz & 0.57\zz & \cite{pdc05} \\ [1.0 em]
  B1802$-$07    &    23.100 &     $+47$\dzz & NGC~6539 &     2.616 &   3.920 &       0.212\zzz & 0.29\zz & \cite{dbl93, tamt93} \\
  J1803$-$30    &     7.101 &               & NGC~6522 &           &         &                 &         & \cite{pdc05} \\
  J1807$-$24A   &     3.059 &               & NGC~6544 &     0.071 &   0.012 &                 &  0.0089 & \cite{rgh01} \\
  B             &     4.186 &               & NGC~6544 &         ? &         &                 &         & \cite{cha03} \\
  B1820$-$30A   &     5.440 &    $+338$\dzz & NGC~6624 &           &         &                 &         & \cite{bbl94} \\ [1.0 em]
  B1820$-$30B   &  378.6\zz &   $+3150$\dzz & NGC~6624 &           &         &                 &         & \cite{bbl94} \\
  C             &  405.9\zz &               & NGC~6624 &           &         &                 &         & \cite{cha03} \\
  B1821$-$24    &     3.054 &    $+161$\dzz &      M28 &           &         &                 &         & \cite{lbm87} \\
  J1905+01A     &     3.193 &               & NGC~6749 &           &         &                 &         & \cite{rhs05b} \\ 
  B             &     4.968 &               & NGC~6749 &           &         &                 &         & \cite{rhs05b} \\ [1.0 em] 
  J1911$-$5958A &     3.266 &       $+0.30$ & NGC~6752 &     0.837 &   1.206 &    $<0.00001$\z & 0.18\zz & \cite{dpf02} \\
  J1910$-$5959B &     8.357 &     $-79$\dzz & NGC~6752 &           &         &                 &         & \cite{dpf02} \\
  J1911$-$6000C &     5.277 &      $+0.2$\z & NGC~6752 &           &         &                 &         & \cite{dpf02} \\
  J1910$-$5959D &     9.035 &     $+96$\dzz & NGC~6752 &           &         &                 &         & \cite{dpf02} \\
  J1910$-$5959E &     4.571 &     $-43$\dzz & NGC~6752 &           &         &                 &         & \cite{dpf02} \\ [1.0 em]
  \newpage
  J1911+0102A   &     3.618 &       $-0.65$ & NGC~6760 &     0.140 &   0.037 &    $<0.00013$\z & 0.017\z & \cite{fhn05} \\
  J1911+0101B   &     5.384 &      $-0.2$\z & NGC~6760 &           &         &                 &         & \cite{fhn05} \\
  J1953+18A     &     4.888 &               &      M71 &     0.176 &   0.078 &                 & 0.032\z & \cite{rhs05b} \\
  B2127+11A     &   110.664 &   $-2107$\dzz &      M15 &           &         &                 &         & \cite{and92} \\
  B2127+11B     &    56.133 &    $+956$\dzz &      M15 &           &         &                 &         & \cite{and92} \\ [1.0 em]
  B2127+11C     &    30.529 &    $+499$\dzz &      M15 &     0.335 &   2.518 &       0.681\zzz & 0.92\zz & \cite{and92} \\
  B2127+11D     &     4.802 &    $-107$\dzz &      M15 &           &         &                 &         & \cite{and92} \\
  B2127+11E     &     4.651 &     $+17$\dzz &      M15 &           &         &                 &         & \cite{and92} \\
  B2127+11F     &     4.027 &      $+3$\dzz &      M15 &           &         &                 &         & \cite{and92} \\
  B2127+11G     &    37.660 &    $+195$\dzz &      M15 &           &         &                 &         & \cite{and92} \\ [1.0 em]
  B2127+11H     &     6.743 &      $+2$\dzz &      M15 &           &         &                 &         & \cite{and92} \\
  J2140$-$23A   &    11.019 &       $-5.18$ &      M30 &    0.234  &         &    $<0.00012$\z &    0.10 & \cite{rsb04} \\
  B             &   13.0\zz &               &      M30 & $>0.1$\zz &         &    $>0.52$\zzzz &         & \cite{rsb04} \\
  \hline \hline
  \multicolumn{1}{c}{} \\
  \caption{\it Parameters for the 108 pulsars currently known in
    globular clusters. In addition to the parameters defined in
    Tables~\ref{tab:imsps} and~\ref{tab:ebpsrs}, we also list the
    measured spin period derivative $\dot{P}$ and name of associated
    cluster.}
  \label{tab:gcpsrs}
\end{longtable}

\end{landscape}

\newpage


\bibliography{refs}

\begin{thebibliography}{100}

\bibitem{allan}
Allan, D.W., ``The Allan Variance'', personal homepage, Allan's Time Interval
  Metrology Enterprises, (1999). URL (cited on 7 December 2000):
  \newline\url{http://www.allanstime.com/AllanVariance/}.
  \epubtkKeywords{Pulsar timing, Fundamental physics}

\bibitem{acrs82}
Alpar, M.A., Cheng, A.F., Ruderman, M.A., and Shaham, J., ``A new class of
  radio pulsars'', {\em Nature}, {\bf 300}, 728--730, (1982).
  \epubtkKeywords{Accretion, Binary systems, Neutron stars, Pulsars}

\bibitem{anp86}
Alpar, M.A., Nandkumar, R., and Pines, D., ``Vortex Creep and the Internal
  Temperature of Neutron Stars: Timing Noise of Pulsars'', {\em Astrophys. J.},
  {\bf 311}, 197--213, (1986). \epubtkKeywords{Fundamental physics, Neutron
  stars}

\bibitem{and92}
Anderson, S.B., {\em A study of recycled pulsars in globular clusters}, Ph.D.
  Thesis, (California Institute of Technology, Pasadena, U.S.A., 1992).
  \epubtkKeywords{Radio astronomy, Pulsars, Neutron stars, Data analysis}

\bibitem{agk90}
Anderson, S.B., Gorham, P.W., Kulkarni, S.R., Prince, T.A., and Wolszczan, A.,
  ``Discovery of two radio pulsars in the globular cluster M15'', {\em Nature},
  {\bf 346}, 42--44, (1990). \epubtkKeywords{Radio astronomy, Pulsars}

\bibitem{awkp97}
Anderson, S.B., Wolszczan, A., Kulkarni, S.R., and Prince, T.A., ``Observations
  of Two Millisecond Pulsars in the Globular Cluster NGC 5904'', {\em
  Astrophys. J.}, {\bf 482}, 870--873, (1997). \epubtkKeywords{Pulsars, Neutron
  stars}

\bibitem{ajks00}
Andersson, N., Jones, D.I., Kokkotas, K., and Stergioulas, N., ``r-Mode runaway
  and rapidly rotating neutron stars'', {\em Astrophys. J. Lett.}, {\bf 534},
  L75--L78, (2000). \epubtkKeywords{X-ray binaries, Gravitational waves,
  Neutron stars}

\bibitem{acc02}
Arzoumanian, Z., Chernoff, D.~F., and Cordes, J.~M., ``The Velocity
  Distribution of Isolated Radio Pulsars'', {\em Astrophys. J.}, {\bf 568},
  289--301, (2002). \epubtkKeywords{Pulsars, Pulsar velocities}

\bibitem{acw99}
Arzoumanian, Z., Cordes, J.M., and Wasserman, I., ``Pulsar Spin Evolution,
  Kinematics, and the Birthrate of Neutron Star Binaries'', {\em Astrophys.
  J.}, {\bf 520}, 696--705, (1999). \epubtkKeywords{Neutron stars, Binary
  systems, Gravitational wave sources}

\bibitem{aft94}
Arzoumanian, Z., Fruchter, A.S., and Taylor, J.H., ``Orbital Variability in the
  Eclipsing Pulsar Binary PSR B1957+20'', {\em Astrophys. J. Lett.}, {\bf 426},
  L85--L88, (1994). \epubtkKeywords{Pulsars, Neutron stars, Binary systems}

\bibitem{antt94}
Arzoumanian, Z., Nice, D.J., Taylor, J.H., and Thorsett, S.E., ``Timing
  Behavior of 96 Radio Pulsars'', {\em Astrophys. J.}, {\bf 422}, 671--680,
  (1994). \epubtkKeywords{Pulsar timing}

\bibitem{atnfpsr}
Australia Telescope National Facility, ``ATNF Pulsar Group'', project homepage,
  (2005). URL (cited on 22 January 2005):
  \newline\url{http://www.atnf.csiro.au/research/pulsar}. \epubtkKeywords{Radio
  astronomy, Pulsars}

\bibitem{psrcat}
Australia Telescope National Facility, ``The Pulsar Catalogue'', web interface
  to database, (2005). URL (cited on 17 January 2005):
  \newline\url{http://www.atnf.csiro.au/research/pulsar/psrcat}.
  \epubtkKeywords{Radio astronomy, Pulsars}

\bibitem{bac96}
Backer, D.C., ``Timing of Millisecond Pulsars'', in van Paradijs, J., van~del
  Heuvel, E.P.J., and Kuulkers, E., eds., {\em Compact Stars in Binaries (IAU
  Symposium 165)}, Proceedings of the 165th Symposium of the International
  Astronomical Union, held in the Hague, The Netherlands, August 15--19, 1994,
  197--211, (Kluwer, Dordrecht, Netherlands; Boston, U.S.A., 1996).
  \epubtkKeywords{Radio astronomy, Pulsars, Pulsar timing, Cosmic gravitational
  wave background}

\bibitem{backertalk}
Backer, D.C., ``Pulsars - Nature's Best Clocks'', lecture notes, American
  Physical Society, (1999). URL (cited on 8 December 2000):
  \newline\url{http://www.apscenttalks.org/pres_masterpage.cfm?nameID=65}.
  \epubtkKeywords{Radio astronomy, Pulsars, Fundamental physics}

\bibitem{bdz97}
Backer, D.C., Dexter, M.R., Zepka, A.F., Ng, D., Werthimer, D.J., Ray, P.S.,
  and Foster, R.S., ``A Programmable 36 MHz Digital Filter Bank for Radio
  Science'', {\em Publ. Astron. Soc. Pac.}, {\bf 109}, 61--68, (1997).
  \epubtkKeywords{Radio astronomy, Signal processing, Pulsars}

\bibitem{bfs93}
Backer, D.C., Foster, R.F., and Sallmen, S., ``A second companion of the
  millisecond pulsar 1620--26'', {\em Nature}, {\bf 365}, 817--819, (1993).
  \epubtkKeywords{Radio astronomy, Pulsars, Fundamental physics}

\bibitem{bh86}
Backer, D.C., and Hellings, R.W., ``Pulsar Timing and General Relativity'',
  {\em Annu. Rev. Astron. Astrophys.}, {\bf 24}, 537--575, (1986).
  \epubtkKeywords{Radio astronomy, Pulsars, Pulsar timing, General relativity,
  Cosmology, Fundamental physics}

\bibitem{bkh82}
Backer, D.C., Kulkarni, S.R., Heiles, C., Davis, M.M., and Goss, W.M., ``A
  millisecond pulsar'', {\em Nature}, {\bf 300}, 615--618, (1982).
  \epubtkKeywords{Radio astronomy, Pulsars, Neutron stars, Fundamental physics}

\bibitem{bai89}
Bailes, M., ``The Origin of pulsar velocities and the velocity-magnetic moment
  correlation.'', {\em Astrophys. J.}, {\bf 342}, 917--927, (1989).
  \epubtkKeywords{Magnetic fields, Binary systems, Neutron stars, Pulsars}

\bibitem{bai03}
Bailes, M., ``Precision timing at the Parkes 64-m radio telescope'', in Bailes,
  M., Nice, D.J., and Thorsett, S.E., eds., {\em Radio Pulsars}, Proceedings of
  meeting held at Mediterranean Agonomic Institute of Chania, Crete, Greece,
  26--29 August 2002, vol. 302 of ASP Conference Series,  57--64, (Astronomical
  Society of the Pacific, San Fransisco, U.S.A., 2003). \epubtkKeywords{Radio
  astronomy, Pulsars}

\bibitem{bhl94}
Bailes, M., Harrison, P.A., Lorimer, D.R., Johnston, S., Lyne, A.G.,
  Manchester, R.N., D'Amico, N., Nicastro, L., Tauris, T.M., and Robinson, C.,
  ``Discovery of Three Galactic Binary Millisecond Pulsars'', {\em Astrophys.
  J. Lett.}, {\bf 425}, L41--L44, (1994). \epubtkKeywords{Pulsars, Neutron
  stars, White dwarfs, Binary systems}

\bibitem{bjb97}
Bailes, M., Johnston, S., Bell, J.F., Lorimer, D.R., Stappers, B.W.,
  Manchester, R.N., Lyne, A.G., D'Amico, N., and Gaensler, B.M., ``Discovery of
  four isolated millisecond pulsars'', {\em Astrophys. J.}, {\bf 481},
  386--391, (1997). \epubtkKeywords{Radio astronomy, Pulsars}

\bibitem{bmk90b}
Bailes, M., Manchester, R.N., Kesteven, M.J., Norris, R.P., and Reynolds, J.E.,
  ``The proper motion of six southern radio pulsars'', {\em Mon. Not. R.
  Astron. Soc.}, {\bf 247}, 322--326, (1990). \epubtkKeywords{Pulsars, Galactic
  astronomy}

\bibitem{bokh03}
Bailes, M., Ord, S.M., Knight, H.S., and Hotan, A.W., ``Self-Consistency of
  Relativistic Observables with General Relativity in the White Dwarf-Neutron
  Star Binary PSR J1141-6545'', {\em Astrophys. J. Lett.}, {\bf 595}, L49--L52,
  (2003). \epubtkKeywords{Pulsars, Binary systems, General relativity}

\bibitem{bo75b}
Barker, B.M., and O'Connell, R.F., ``Gravitational two-body problem with
  arbitrary masses, spins, and quadrupole moments'', {\em Phys. Rev. D}, {\bf
  12}, 329--335, (1975). \epubtkKeywords{Binary systems, Neutron stars, General
  relativity}

\bibitem{bkb02}
Belczynski, K., Kalogera, V.M., and Bulik, T., ``A Comprehensive Study of
  Binary Compact Objects as Gravitational Wave Sources: Evolutionary Channels,
  Rates, and Physical Properties'', {\em Astrophys. J.}, {\bf 572}, 407--431,
  (2002). \epubtkKeywords{Binary pulsars, Pulsar evolution}

\bibitem{bel98}
Bell, J.F., ``Radio Pulsar Timing'', {\em Adv. Space Res.}, {\bf 21}, 131--147,
  (1998). \epubtkKeywords{Pulsars, Pulsar timing, Observational tests of
  relativity theory}

\bibitem{bbm97}
Bell, J.F., Bailes, M., Manchester, R.N., Lyne, A.G., Camilo, F., and Sandhu,
  J.S., ``Timing measurements and their implications for four binary
  millisecond pulsars'', {\em Mon. Not. R. Astron. Soc.}, {\bf 286}, 463--469,
  (1997). \epubtkKeywords{Radio astronomy, Pulsars}

\bibitem{bbs95}
Bell, J.F., Bessell, M.S., Stappers, B.W., Bailes, M., and Kaspi, V.M., ``PSR
  J0045-7319: A dual-line binary radio pulsar'', {\em Astrophys. J. Lett.},
  {\bf 447}, L117--L119, (1995). \epubtkKeywords{Binary pulsars}

\bibitem{bcr83}
Bertotti, B., Carr, B.J., and Rees, M.J., ``Limits from the timing of pulsars
  on the cosmic gravitational wave background'', {\em Mon. Not. R. Astron.
  Soc.}, {\bf 203}, 945--954, (1983). \epubtkKeywords{Cosmic gravitational wave
  background, Pulsars, Pulsar timing}

\bibitem{bgi93}
Beskin, V.S., Gurevich, A.V., and Istomin, Y.N., {\em Physics of the Pulsar
  Magnetosphere}, (Cambridge University Press, Cambridge, U.K.; New York,
  U.S.A., 1993). \epubtkKeywords{Neutron stars, Fundamental physics}

\bibitem{bv91}
Bhattacharya, D., and van~den Heuvel, E.P.J., ``Formation and evolution of
  binary and millisecond radio pulsars'', {\em Phys. Rep.}, {\bf 203}, 1--124,
  (1991). \epubtkKeywords{Accretion, Binary systems, Neutron stars, Pulsars}

\bibitem{big90b}
Biggs, J.D., ``Meridional compression of radio pulsar beams'', {\em Mon. Not.
  R. Astron. Soc.}, {\bf 245}, 514--521, (1990). \epubtkKeywords{Neutron stars,
  Pulsars}

\bibitem{bbl94}
Biggs, J.D., Bailes, M., Lyne, A.G., Goss, W.M., and Fruchter, A.S., ``Two
  radio pulsars in the globular cluster NGC 6624'', {\em Mon. Not. R. Astron.
  Soc.}, {\bf 267}, 125--128, (1994). \epubtkKeywords{Pulsars, Neutron stars}

\bibitem{bil98}
Bildsten, L., ``Gravitational radiation and rotation of accreting neutron
  stars'', {\em Astrophys. J. Lett.}, {\bf 501}, L89--L93, (1998).
  \epubtkKeywords{X-ray binaries, Gravitational waves, Neutron stars}

\bibitem{bk74}
Bisnovatyi-Kogan, G.S., and Komberg, B.V., ``Pulsars and Close Binary
  Systems'', {\em Sov. Astron.}, {\bf 18}, 217--221, (1974).
  \epubtkKeywords{Accretion, Binary systems, Neutron stars, Pulsars}

\bibitem{bla61}
Blaauw, A., ``On the origin of the O- and B-type stars with high velocities
  (the `Run-away' stars), and related problems'', {\em Bull. Astron. Inst.
  Neth.}, {\bf 15}, 265--290, (1961). \epubtkKeywords{Binary systems}

\bibitem{bnr84}
Blandford, R.D., Narayan, R., and Romani, R.W., ``Arrival-time analysis for a
  millisecond pulsar'', {\em J. Astrophys. Astron.}, {\bf 5}, 369--388, (1984).
  \epubtkKeywords{Cosmic gravitational wave background, Pulsars, Pulsar timing}

\bibitem{bbf83}
Boriakoff, V., Buccheri, R., and Fauci, F., ``Discovery of a 6.1-ms binary
  pulsar, PSR 1953+29'', {\em Nature}, {\bf 304}, 417--419, (1983).
  \epubtkKeywords{Pulsars, Neutron stars, Binary systems}

\bibitem{bd97}
Burderi, L., and D'Amico, N., ``Probing the Equation of State of Ultradense
  Matter with a Submillisecond Pulsar Search Experiment'', {\em Astrophys. J.},
  {\bf 490}, 343--352, (1997). \epubtkKeywords{Pulsars, Neutron stars,
  Fundamental physics}

\bibitem{bur04}
Burgay, M., {\em The Parkes High-Latitude Pulsar Survey and the Discovery of
  the first Double Pulsar}, Ph.D. Thesis, (Bologna University, Bologna, Italy,
  2004). \epubtkKeywords{Radio astronomy, Pulsars, Data analysis}

\bibitem{bbp03}
Burgay, M., Burderi, L., Possenti, A., D'Amico, N., Manchester, R.N., Lyne,
  A.G., Camilo, F., and Campana, S., ``A Search for Pulsars in Quiescent Soft
  X-Ray Transients. I.'', {\em Astrophys. J.}, {\bf 589}, 902--910, (2003).
  \epubtkKeywords{Radio astronomy, X-ray binaries, Pulsar searches}

\bibitem{bdp03}
Burgay, M., D'Amico, N., Possenti, A., Manchester, R.N., Lyne, A.G., Joshi,
  B.C., McLaughlin, M.A., Kramer, M., Sarkissian, J.M., Camilo, F., Kalogera,
  V., Kim, C., and Lorimer, D.R., ``An increased estimate of the merger rate of
  double neutron stars from observations of a highly relativistic system'',
  {\em Nature}, {\bf 426}, 531--533, (2003). \epubtkKeywords{Radio astronomy,
  Pulsars, Tests of relativistic gravity}

\bibitem{ca92}
Caldwell, R.R., and Allen, B., ``Cosmological constraints on cosmic-string
  gravitational radiation'', {\em Phys. Rev. D}, {\bf 45}, 3447--3468, (1992).
  \epubtkKeywords{Cosmology}

\bibitem{ligo}
Caltech/MIT, ``Laser Interferometric Gravitational Wave Observatory'', project
  homepage, (2005). URL (cited on 29 January 2005):
  \newline\url{http://www.ligo.caltech.edu}. \epubtkKeywords{Gravitational wave
  detectors}

\bibitem{cam95}
Camilo, F., ``Millisecond Pulsar Searches'', in Alpar, A., Kizilo\u{g}lu, \"U.,
  and van Paradis~J., eds., {\em The Lives of the Neutron Stars}, Proceedings
  of the NATO Advanced Study Institute, Kemer, Turkey, August 29 - September
  12, 1993, vol. 450 of NATO ASI Series, Series C,  243--257, (Kluwer,
  Dordrecht, Netherlands; Boston, U.S.A., 1995). \epubtkKeywords{Radio
  astronomy, Pulsars, Binary systems}

\bibitem{cam95a}
Camilo, F., {\em A Search for Millisecond Pulsars}, Ph.D. Thesis, (Princeton
  University, Princeton, U.S.A., 1995). \epubtkKeywords{Pulsars, Neutron stars,
  Radio astronomy}

\bibitem{cam96c}
Camilo, F., ``Intermediate-Mass Binary Pulsars: a New Class of Objects?'', in
  Johnston, S., Walker, M.A., and Bailes, M., eds., {\em Pulsars: Problems and
  Progress (IAU Colloquium 160)}, Proceedings of the 160th Colloquium of the
  IAU, held at the Research Centre for Theoretical Astrophysics, University of
  Sydney, Australia, 8-12 January 1996, vol. 105 of ASP Conference Series,
  539--540, (Astronomical Society of the Pacific, San Francisco, U.S.A., 1996).
  \epubtkKeywords{Neutron stars, Pulsars, White dwarfs, Binary systems}

\bibitem{cam97}
Camilo, F., ``Present and future pulsar searches'', in Jackson, N., and Davis,
  R.J., eds., {\em High-Sensitivity Radio Astronomy}, Proceedings of a meeting
  held at Jodrell Bank, University of Manchester, January 22--26, 1996,
  Cambridge Contemporary Astrophysics,  14--22, (Cambridge University Press,
  Cambridge, U.K.; New York, U.S.A., 1997). \epubtkKeywords{Radio astronomy,
  Pulsars, Binary systems}

\bibitem{cam98}
Camilo, F., ``Pulsar Searches in the Northern Hemisphere'', in Arzoumanian, Z.,
  van~der Hooft, F., and van~den Heuvel, E.P.J., eds., {\em Pulsar Timing,
  General Relativity, and the Internal Structure of Neutron Stars}, Proceedings
  of the colloquium, Amsterdam, 24--27 September 1996,  115, (Royal Netherlands
  Academy of Arts and Sciences, Amsterdam, Netherlands, 1999).
  \epubtkKeywords{Radio astronomy, Pulsars, Binary systems}

\bibitem{cfw94}
Camilo, F., Foster, R.S., and Wolszczan, A., ``High-precision timing of PSR
  J1713+0747: Shapiro delay'', {\em Astrophys. J. Lett.}, {\bf 437}, L39--L42,
  (1994). \epubtkKeywords{Pulsar timing, Radio astronomy, General relativity}

\bibitem{clf00}
Camilo, F., Lorimer, D.R., Freire, P.C.C., Lyne, A.G., and Manchester, R.N.,
  ``Observations of 20 Millisecond Pulsars in 47 Tucanae at 20 Centimeters'',
  {\em Astrophys. J.}, {\bf 535}, 975--990, (2000). \epubtkKeywords{Radio
  astronomy, Pulsars, Binary systems}

\bibitem{clm01}
Camilo, F., Lyne, A.G., Manchester, R.N., Bell, J.F., Stairs, I.H., D'Amico,
  N., Kaspi, V.M., Possenti, A., Crawford, F., and McKay, N.P.F., ``Discovery
  of Five Binary Radio Pulsars'', {\em Astrophys. J. Lett.}, {\bf 548},
  L187--L191, (2001). \epubtkKeywords{Neutron stars, Pulsars, White dwarfs,
  Binary systems}

\bibitem{cnt93}
Camilo, F., Nice, D.J., and Taylor, J.H., ``Discovery of two fast-rotating
  pulsars'', {\em Astrophys. J. Lett.}, {\bf 412}, L37--L40, (1993).
  \epubtkKeywords{Pulsars, Neutron stars, Binary systems}

\bibitem{cnt96}
Camilo, F., Nice, D.J., and Taylor, J.H., ``A search for millisecond pulsars at
  galactic latitudes $-50^\circ < b < -20^\circ$'', {\em Astrophys. J.}, {\bf
  461}, 812--819, (1996). \epubtkKeywords{Pulsars, Neutron stars, Radio
  astronomy, Observational tests of relativity theory}

\bibitem{cnst96}
Camilo, F., Nice, D.J., Taylor, J.H., and Shrauner, J.A., ``Princeton-Arecibo
  Declination-Strip Survey for Millisecond Pulsars. I.'', {\em Astrophys. J.},
  {\bf 469}, 819--827, (1996). \epubtkKeywords{Pulsars, Neutron stars, Binary
  systems, White dwarfs}

\bibitem{cr05}
Camilo, F., and Rasio, F.A., ``Radio Pulsars in Globular Clusters'', in Rasio,
  F.A., and Stairs, I.H., eds., {\em Binary Radio Pulsars}, Proceedings of
  meeting held at the Aspen Center for Physics, USA, 12 Janaury - 16 January
  2004, vol. 328 of ASP Conference Series, (Astronomical Society of the
  Pacific, San Francisco, U.S.A., 2005). \epubtkKeywords{Radio astronomy,
  Pulsars, Globular clusters}

\bibitem{cm98}
Chakrabarty, D., and Morgan, E.H., ``The 2 hour orbit of a binary millisecond
  X-ray pulsar'', {\em Nature}, {\bf 394}, 346--348, (1998).
  \epubtkKeywords{Accretion, Neutron stars, Pulsars, Binary systems}

\bibitem{cmm03}
Chakrabarty, D., Morgan, E.H., Muno, M.P., Galloway, D.K., Wijnands, R.,
  van~der Klis, M., and Markwardt, C.B., ``Nuclear-powered millisecond pulsars
  and the maximum spin frequency of neutron stars'', {\em Nature}, {\bf 424},
  42--44, (2003). \epubtkKeywords{X-ray binaries, Gravitational waves, Neutron
  stars}

\bibitem{cha05}
Champion, D.J., ``Latest timing results for PSR J1829+2456'', personal
  communication. \epubtkKeywords{Pulsars, Tests of relativistic gravity}

\bibitem{clm04}
Champion, D.J., Lorimer, D.R., McLaughlin, M.A., Arzoumanian, Z., Cordes, J.M.,
  Weisberg, J., and Taylor, J.H., ``PSR J1829+2456: a new relativistic binary
  system'', {\em Mon. Not. R. Astron. Soc.}, {\bf 350}, L61--L65, (2004).
  \epubtkKeywords{Radio astronomy, Pulsars, Relativistic binary systems}

\bibitem{cha03}
Chandler, A.M., {\em Pulsar Searches: From Radio to Gamma-Rays}, Ph.D. Thesis,
  (California Institute of Technology, Pasadena, U.S.A., 2003).
  \epubtkKeywords{Radio astronomy, Pulsars, Data analysis}

\bibitem{cr93}
Chen, K., and Ruderman, M., ``Pulsar death rates and death valley'', {\em
  Astrophys. J.}, {\bf 402}, 264--270, (1993). \epubtkKeywords{Pulsar
  statistics, Radio emission}

\bibitem{che87b}
Cheng, K.S., ``Could Glitches Inducing Magnetospheric Fluctuations Produce Low
  Frequency Timing Noise?'', {\em Astrophys. J.}, {\bf 321}, 805--812, (1987).
  \epubtkKeywords{Pulsar timing, Neutron stars, Fundamental physics}

\bibitem{che87a}
Cheng, K.S., ``Outer Magnetospheric Fluctuations and Pulsar Timing Noise'',
  {\em Astrophys. J.}, {\bf 321}, 799--804, (1987). \epubtkKeywords{Pulsar
  timing, Neutron stars, Fundamental physics}

\bibitem{clj92}
Clifton, T.R., Lyne, A.G., Jones, A.W., McKenna, J., and Ashworth, M., ``A high
  frequency survey of the Galactic plane for young and distant pulsars'', {\em
  Mon. Not. R. Astron. Soc.}, {\bf 254}, 177--184, (1992).
  \epubtkKeywords{Radio astronomy, Pulsars}

\bibitem{cb04}
Cognard, I., and Backer, D.C., ``A Microglitch in the Millisecond Pulsar PSR
  B1821-24 in M28'', {\em Astrophys. J. Lett.}, {\bf 612}, L125--L127, (2004).
  \epubtkKeywords{Radio astronomy, Pulsars, Pulsar timing, Pulsar glitches}

\bibitem{coo04}
Cooray, A., ``Gravitational wave background of neutron star--white dwarf
  binaries'', {\em Mon. Not. R. Astron. Soc.}, {\bf 354}, 25--30, (2005).
  \epubtkKeywords{Gravitational wave sources}

\bibitem{cc97}
Cordes, J.M., and Chernoff, D.F., ``Neutron Star Population Dynamics. I.
  Millisecond Pulsars'', {\em Astrophys. J.}, {\bf 482}, 971--992, (1997).
  \epubtkKeywords{Neutron stars, Pulsars, Monte Carlo methods, Galactic
  astronomy}

\bibitem{cc98}
Cordes, J.M., and Chernoff, D.F., ``Neutron Star Population Dynamics. II.
  Three-dimensional Space Velocities of Young Pulsars'', {\em Astrophys. J.},
  {\bf 505}, 315--338, (1998). \epubtkKeywords{Neutron stars, Pulsars, Galactic
  astronomy}

\bibitem{ch80}
Cordes, J.M., and Helfand, D.J., ``Pulsar timing III. Timing Noise of 50
  Pulsars'', {\em Astrophys. J. Lett.}, {\bf 239}, 640--650, (1980).
  \epubtkKeywords{Radio astronomy, Pulsars, Pulsar timing}

\bibitem{cl02a}
Cordes, J.M., and Lazio, T.J.W., ``NE2001. I. A New Model for the Galactic
  Distribution of Free Electrons and its Fluctuations'', (2002). URL (cited on
  22 January 2005): \newline\url{http://arXiv.org/abs/astro-ph/0207156}.
  \epubtkKeywords{Radio astronomy, Pulsars, Interstellar medium}

\bibitem{cl02b}
Cordes, J.M., and Lazio, T.J.W., ``NE2001. II. Using Radio Propagation Data to
  Construct a Model for the Galactic Distribution of Free Electrons'', (2003).
  URL (cited on 22 January 2005):
  \newline\url{http://arXiv.org/abs/astro-ph/0301598}. \epubtkKeywords{Radio
  astronomy, Pulsars, Interstellar medium}

\bibitem{lanl}
Cornell University, ``arXiv.org e-Print archive'', web interface to database,
  (2001). URL (cited on 19 March 2001): \newline\url{http://arXiv.org}.
  \epubtkKeywords{Astronomy, Astronomical observations}

\bibitem{dbl93}
D'Amico, N., Bailes, M., Lyne, A.G., Manchester, R.N., Johnston, S., Fruchter,
  A.S., and Goss, W.M., ``PSR B1802-07: A globular cluster pulsar in an
  eccentric binary system'', {\em Mon. Not. R. Astron. Soc.}, {\bf 260},
  L7--L10, (1993). \epubtkKeywords{Pulsars, Neutron stars, Binary systems}

\bibitem{dlm01}
D'Amico, N., Lyne, A.G., Manchester, R.N., Possenti, A., and Camilo, F.,
  ``Discovery of Short-Period Binary Millisecond Pulsars in Four Globular
  Clusters'', {\em Astrophys. J. Lett.}, {\bf 548}, L171--L174, (2001).
  \epubtkKeywords{Radio astronomy, Pulsars, Binary systems}

\bibitem{dpf02}
D'Amico, N., Possenti, A., Fici, L., Manchester, R.N., Lyne, A.G., Camilo, F.,
  and Sarkissian, J., ``Timing of Millisecond Pulsars in NGC 6752: Evidence for
  a High Mass-to-Light Ratio in the Cluster Core'', {\em Astrophys. J. Lett.},
  {\bf 570}, L89--L92, (2002). \epubtkKeywords{Radio astronomy, Pulsars, Binary
  systems}

\bibitem{dpm01}
D'Amico, N., Possenti, A., Manchester, R.N., Sarkissian, J., Lyne, A.G., and
  Camilo, F., ``An Eclipsing Millisecond Pulsar with a Possible Main-Sequence
  Companion in NGC 6397'', {\em Astrophys. J. Lett.}, {\bf 561}, L89--L92,
  (2001). \epubtkKeywords{Radio astronomy, Pulsars, Binary systems}

\bibitem{dd86}
Damour, T., and Deruelle, N., ``General Relativistic Celestial Mechanics of
  Binary Systems. II. The Post-Newtonian Timing Formula'', {\em Ann. Inst.
  Henri Poincare A}, {\bf 44}, 263--292, (1986). \epubtkKeywords{Pulsar timing,
  Pulsars, General relativity, Observational tests of relativity theory}

\bibitem{ds88}
Damour, T., and Sch{\"a}fer, G., ``Higher-Order Relativistic Periastron
  Advances and Binary Pulsars'', {\em Nuovo Cimento}, {\bf 101}, 127, (1988).
  \epubtkKeywords{Binary systems, Neutron stars, General relativity}

\bibitem{dv04}
Damour, T., and Vilenkin, A., ``Gravitational radiation from cosmic
  (super)strings: Bursts, stochastic background, and observational windows'',
  {\em Phys. Rev. D}, {\bf 71}, 063510--1--13, (2005). Related online version
  (cited on 30 January 2005):
  \newline\url{http://arXiv.org/abs/hep-th/0410222}. \epubtkKeywords{Pulsars,
  Cosmic gravitational wave background}

\bibitem{gps}
Dana, P.H., ``The Global Positioning System'', project homepage, Department of
  Geography, University of Colorado at Boulder, (2000). URL (cited on 19 March
  2001):
  \newline\url{http://www.colorado.edu/geography/gcraft/notes/gps/gps_f.html}.
  \epubtkKeywords{Pulsar timing, Fundamental physics}

\bibitem{dtwb85}
Davis, M.M., Taylor, J.H., Weisberg, J.M., and Backer, D.C., ``High-precision
  timing observations of the millisecond pulsar PSR 1937+21'', {\em Nature},
  {\bf 315}, 547--550, (1985). \epubtkKeywords{Pulsars, Neutron stars, Pulsar
  timing, Fundamental physics}

\bibitem{det79}
Detweiler, S.L., ``Pulsar timing measurements and the search for gravitational
  waves'', {\em Astrophys. J.}, {\bf 234}, 1100--1104, (1979).
  \epubtkKeywords{Cosmic gravitational wave background, Pulsars, Pulsar timing}

\bibitem{dc87}
Dewey, R.J., and Cordes, J.M., ``Monte Carlo Simulations of Radio Pulsars and
  their Progenitors'', {\em Astrophys. J.}, {\bf 321}, 780--798, (1987).
  \epubtkKeywords{Neutron stars, White dwarfs, Binary systems, Monte Carlo
  methods}

\bibitem{dmr86}
Dewey, R.J., Maguire, C.M., Rawley, L.A., Stokes, G.H., and Taylor, J.H.,
  ``Binary pulsar with a very small mass function'', {\em Nature}, {\bf 322},
  712--714, (1986). \epubtkKeywords{Pulsars, Neutron stars, Binary systems}

\bibitem{dss84}
Dewey, R.J., Stokes, G.H., Segelstein, D.J., Taylor, J.H., and Weisberg, J.M.,
  ``The Period Distribution of Pulsars'', in Reynolds, S.P., and Stinebring,
  D.R., eds., {\em Birth and Evolution of Neutron Stars: Issues Raised by
  Millisecond Pulsars}, Proceedings of a workshop, held at the National Radio
  Astronomy Observatory, Green Bank, West Virginia, June 6--8, 1984, vol.~8 of
  NRAO Workshop,  234--240, (National Radio Astronomy Observatory, Green Bank,
  U.S.A., 1984). \epubtkKeywords{Pulsars, Pulsar statistics, Neutron stars}

\bibitem{dr83}
Downs, G.S., and Reichley, P.E., ``JPL pulsar timing observations. II.
  Geocentric arrival times'', {\em Astrophys. J. Suppl. Ser.}, {\bf 53},
  169--240, (1983). \epubtkKeywords{Radio astronomy, Pulsar timing}

\bibitem{eb01}
Edwards, R., and Bailes, M., ``Discovery of two relativistic neutron star-white
  dwarf binaries'', {\em Astrophys. J. Lett.}, {\bf 547}, L37--L40, (2001).
  \epubtkKeywords{Pulsars, Binary systems, Gravitational wave sources}

\bibitem{eb01b}
Edwards, R., and Bailes, M., ``Recycled Pulsars Discovered at High Radio
  Frequency'', {\em Astrophys. J.}, {\bf 553}, 801--808, (2001).
  \epubtkKeywords{Pulsars, Binary systems, Gravitational wave sources}

\bibitem{edw00}
Edwards, R.T., ``Discovery of Eight Recycled Pulsars -- The Swinburne
  Intermediate Latitude Pulsar Survey'', in Kramer, M., Wex, N., and
  Wielebinski, R., eds., {\em Pulsar Astronomy: 2000 and Beyond (IAU Colloquium
  177)}, Proceedings of the 177th Colloquium of the IAU, held at the
  Max-Planck-Institut f\"ur Radioastronomie, Bonn, Germany, 30 August - 3
  September 1999, vol. 202 of ASP Conference Series,  33--34, (Astronomical
  Society of the Pacific, San Francisco, U.S.A., 2000). \epubtkKeywords{Radio
  astronomy, Pulsars, Binary systems}

\bibitem{fkl05}
Faulkner, A.J., Kramer, M., Lyne, A.G., Manchester, R.N., McLaughlin, M.A.,
  Stairs, I.H., Hobbs, G.B., Possenti, A., Lorimer, D.R., D'Amico, N., Camilo,
  F., and Burgay, M., ``PSR J1756-2251: A New Relativistic Double Neutron Star
  System'', {\em Astrophys. J. Lett.}, {\bf 618}, L119--L122, (2005).
  \epubtkKeywords{Radio astronomy, Pulsars, Relativistic binary systems}

\bibitem{fsk04}
Faulkner, A.J., Stairs, I.H., Kramer, M., Lyne, A.G., Hobbs, G.B., Possenti,
  A., Lorimer, D.R., Manchester, R.N., McLaughlin, M.A., D'Amico, N., Camilo,
  F., and Burgay, M., ``The Parkes Multibeam Pulsar Survey - V. Finding binary
  and millisecond pulsars'', {\em Mon. Not. R. Astron. Soc.}, {\bf 355},
  147--158, (2004). \epubtkKeywords{Radio astronomy, Pulsars, Neutron stars,
  Binary systems}

\bibitem{fv75}
Flannery, B.P., and van~den Heuvel, E.P.J., ``On the origin of the binary
  pulsar PSR 1913+16'', {\em Astron. Astrophys.}, {\bf 39}, 61--67, (1975).
  \epubtkKeywords{Accretion, Binary systems, Neutron stars, Pulsars}

\bibitem{fgl92}
Fomalont, E.B., Goss, W.M., Lyne, A.G., Manchester, R.N., and Justtanont, K.,
  ``Positions and Proper Motions of Pulsars'', {\em Mon. Not. R. Astron. Soc.},
  {\bf 258}, 497--510, (1992). \epubtkKeywords{Pulsars, Galactic astronomy}

\bibitem{fb90}
Foster, R.S., and Backer, D.C., ``Constructing a Pulsar Timing Array'', {\em
  Astrophys. J.}, {\bf 361}, 300--308, (1990). \epubtkKeywords{Radio astronomy,
  Pulsars, Pulsar timing, Cosmic gravitational wave background}

\bibitem{fwc93}
Foster, R.S., Wolszczan, A., and Camilo, F., ``A new binary millisecond
  pulsar'', {\em Astrophys. J. Lett.}, {\bf 410}, L91--L94, (1993).
  \epubtkKeywords{Pulsars, White dwarfs, Pulsar timing, Neutron stars}

\bibitem{fck03}
Freire, P.C., Camilo, F., Kramer, M., Lorimer, D.R., Lyne, A.G., Manchester,
  R.N., and D'Amico, N., ``Further results from the timing of the millisecond
  pulsars in 47 Tucanae'', {\em Mon. Not. R. Astron. Soc.}, {\bf 340},
  1359--1374, (2003). \epubtkKeywords{Radio astronomy, Pulsars, Binary systems}

\bibitem{fgr04}
Freire, P.C., Gupta, Y., Ransom, S.M., and Ishwara-Chandra, C.H., ``Giant
  Metrewave Radio Telescope Discovery of a Millisecond Pulsar in a Very
  Eccentric Binary System'', {\em Astrophys. J. Lett.}, {\bf 606}, L53--L56,
  (2004). \epubtkKeywords{Radio astronomy, Pulsars, Globular clusters}

\bibitem{fhn05}
Freire, P.C.C., Hessels, J.W.T., Nice, D.J., Ransom, S.M., Lorimer, D.R., and
  Stairs, I.H., ``The Millisecond Pulsars in NGC 6760'', {\em Astrophys. J.},
  {\bf 621}, 959--965, (2005). \epubtkKeywords{Radio astronomy, Pulsars, Binary
  systems}

\bibitem{fkl01}
Freire, P.C.C., Kramer, M., and Lyne, A.G., ``Determination of the orbital
  parameters of binary pulsars'', {\em Mon. Not. R. Astron. Soc.}, {\bf 322},
  885, (2001). Related online version (cited on 19 March 2001):
  \newline\url{http://arXiv.org/abs/astro-ph/0010463}. \epubtkKeywords{Pulsars,
  Neutron stars, Binary systems, Monte Carlo methods}

\bibitem{fst88}
Fruchter, A.S., Stinebring, D.R., and Taylor, J.H., ``A millisecond pulsar in
  an eclipsing binary'', {\em Nature}, {\bf 333}, 237--239, (1988).
  \epubtkKeywords{Pulsars, White dwarfs, Binary systems}

\bibitem{gm04}
Gil, J., and Melikidze, G.I., ``On the Giant Nano-subpulses in the Crab
  Pulsar'', in Camilo, F., and Gaensler, B.M., eds., {\em Young neutron stars
  and their environments}, Proceedings of an IAU Symposium on 14--17 July 2003
  in Sydney, Australia, vol. 218 of ASP Conference Series,  321, (Astronomical
  Society of the Pacific, San Francisco, U.S.A., 2004).
  \epubtkKeywords{Pulsars, Radio emission}

\bibitem{gol68}
Gold, T., ``Rotating neutron stars as the origin of the pulsating radio
  sources'', {\em Nature}, {\bf 218}, 731--732, (1968). \epubtkKeywords{Neutron
  stars, Pulsars}

\bibitem{gol69}
Gold, T., ``Rotating neutron stars and the nature of pulsars'', {\em Nature},
  {\bf 221}, 25--27, (1969). \epubtkKeywords{Radio astronomy, Pulsars}

\bibitem{gl98}
Gould, D.M., and Lyne, A.G., ``Multifrequency polarimetry of 300 radio
  pulsars'', {\em Mon. Not. R. Astron. Soc.}, {\bf 301}, 235--260, (1998).
  \epubtkKeywords{Radio astronomy, Pulsars}

\bibitem{go70}
Gunn, J.E., and Ostriker, J.P., ``On the nature of pulsars. III. Analysis of
  observations'', {\em Astrophys. J.}, {\bf 160}, 979--1002, (1970).
  \epubtkKeywords{Pulsar statistics, Pulsars, Galactic astronomy}

\bibitem{hm01}
Han, J.L., and Manchester, R.N., ``The shape of radio pulsar beams'', {\em Mon.
  Not. R. Astron. Soc.}, {\bf 320}, L35--L40, (2001). Related online version
  (cited on 19 March 2001):
  \newline\url{http://arXiv.org/abs/astro-ph/0010538}. \epubtkKeywords{Neutron
  stars, Pulsars}

\bibitem{han71}
Hankins, T.H., ``Microsecond Intensity Variation in the Radio Emission from CP
  0950'', {\em Astrophys. J.}, {\bf 169}, 487--494, (1971).
  \epubtkKeywords{Radio astronomy, Signal processing, Pulsars}

\bibitem{hkwe03}
Hankins, T.H., Kern, J.S., Weatherall, J.C., and Eilek, J.A., ``Nanosecond
  radio bursts from strong plasma turbulence in the Crab pulsar'', {\em
  Nature}, {\bf 422}, 141--143, (2003). \epubtkKeywords{Radio astronomy,
  Pulsars, Radio emission}

\bibitem{geo600}
Hannover University, ``GEO 600 home page'', project homepage, (2005). URL
  (cited on 29 January 2005): \newline\url{http://www.geo600.uni-hannover.de}.
  \epubtkKeywords{Gravitational wave detectors}

\bibitem{hp97}
Hansen, B., and Phinney, E.S., ``The pulsar kick velocity distribution'', {\em
  Mon. Not. R. Astron. Soc.}, {\bf 291}, 569--577, (1997).
  \epubtkKeywords{Neutron stars, Pulsars, Monte Carlo methods, Galactic
  astronomy}

\bibitem{hp98}
Hansen, B., and Phinney, E.S., ``Stellar forensics -- II. Millisecond pulsar
  binaries'', {\em Mon. Not. R. Astron. Soc.}, {\bf 294}, 569--581, (1998).
  \epubtkKeywords{Neutron stars, White dwarfs, Cooling}

\bibitem{hla93}
Harrison, P.A., Lyne, A.G., and Anderson, B., ``New Determinations of the
  Proper Motions of 44 Pulsars'', {\em Mon. Not. R. Astron. Soc.}, {\bf 261},
  113--124, (1993). \epubtkKeywords{Pulsars, Galactic astronomy}

\bibitem{nasaads}
Harvard-Smithsonian Center for Astrophysics, ``The NASA Astrophysics Data
  System'', web interface to database, (2001). URL (cited on 19 March 2001):
  \newline\url{http://adsabs.harvard.edu/}. \epubtkKeywords{Astronomy,
  Astronomical observations}

\bibitem{hel90}
Hellings, R.W., ``Thoughts on Detecting a Gravitational Wave Background with
  Pulsar Data'', in Backer, D.C., ed., {\em Impact of Pulsar Timing on
  Relativity and Cosmology}, Proceedings of the workshop, held at Berkeley,
  June 7--9, 1990, ~k1, (Center for Particle Astrophysics, Berkeley, U.S.A.,
  1990). \epubtkKeywords{Radio astronomy, Pulsars, Pulsar timing, Cosmic
  gravitational wave background}

\bibitem{hd83}
Hellings, R.W., and Downs, G.S., ``Upper limits on the isotropic gravitational
  radiation background from pulsar timing analysis'', {\em Astrophys. J.
  Lett.}, {\bf 265}, L39--L42, (1983). \epubtkKeywords{Radio astronomy,
  Pulsars, Pulsar timing}

\bibitem{hbp68}
Hewish, A., Bell, S.J., Pilkington, J.D.H., Scott, P.F., and Collins, R.A.,
  ``Observation of a rapidly pulsating radio source'', {\em Nature}, {\bf 217},
  709--713, (1968). \epubtkKeywords{Radio astronomy, Pulsars}

\bibitem{hil83}
Hills, J.G., ``The effects of Sudden Mass loss and a random kick velocity
  produced in a supernova explosion on the Dynamics of a binary star of an
  Arbitrary orbital Eccentricity'', {\em Astrophys. J.}, {\bf 267}, 322--333,
  (1983). \epubtkKeywords{Binary systems, Neutron stars, Pulsars}

\bibitem{hob05}
Hobbs, G., ``Pulsars and Gravitational Wave Detection'', (2005). URL (cited on
  22 January 2005): \newline\url{http://arXiv.org/abs/astro-ph/0412153}.
  \epubtkKeywords{Pulsars, Neutron stars}

\bibitem{hllk05}
Hobbs, G., Lorimer, D.R., Lyne, A.G., and Kramer, M., ``A statistical study of
  233 pulsar proper motions'', {\em Mon. Not. R. Astron. Soc.}, {\bf 360},
  974--992, (2005). \epubtkKeywords{Pulsars, Pulsar velocities}

\bibitem{hfs04}
Hobbs, G.B., Faulkner, A., Stairs, I.H., Camilo, F., Manchester, R.N., Lyne,
  A.G., Kramer, M., D'Amico, N., Kaspi, V.M., Possenti, A., McLaughlin, M.A.,
  Lorimer, D.R., Burgay, M., Joshi, B.C., and Crawford, F., ``The Parkes
  multibeam pulsar survey: IV. Discovery of 176 pulsars and parameters for 248
  previously known pulsars'', {\em Mon. Not. R. Astron. Soc.}, {\bf 352},
  1439--1472, (2004). \epubtkKeywords{Radio astronomy, Pulsars, Neutron stars,
  Binary systems}

\bibitem{hlk04}
Hobbs, G.B., Lyne, A.G., Kramer, M., Martin, C.E., and Jordan, C., ``Long-Term
  Timing Observations of 374 Pulsars'', {\em Mon. Not. R. Astron. Soc.}, {\bf
  353}, 1311--1344, (2004). \epubtkKeywords{Radio astronomy, Pulsars, Pulsar
  timing}

\bibitem{hr84}
Hogan, C.J., and Rees, M.J., ``Gravitational interactions of cosmic strings'',
  {\em Nature}, {\bf 311}, 109--114, (1984). \epubtkKeywords{Cosmic
  gravitational wave background}

\bibitem{hot04}
Hotan, A.W., ``Latest timing results for PSR J0437-4715'', personal
  communication. \epubtkKeywords{Pulsars, Pulsar timing}

\bibitem{hbo05}
Hotan, A.W., Bailes, M., and Ord, S.M., ``Geodetic Precession in PSR
  J1141-6545'', {\em Astrophys. J.}, {\bf 624}, 906--913, (2005). Related
  online version (cited on 22 January 2005):
  \newline\url{http://arXiv.org/abs/astro-ph/0412152}. \epubtkKeywords{Pulsars,
  Binary systems, Geodetic precession}

\bibitem{hul94}
Hulse, R.A., ``The discovery of the binary pulsar'', {\em Rev. Mod. Phys.},
  {\bf 66}, 699--710, (1994). \epubtkKeywords{Pulsars, Radio astronomy, Data
  analysis}

\bibitem{ht75a}
Hulse, R.A., and Taylor, J.H., ``Discovery of a pulsar in a binary system'',
  {\em Astrophys. J. Lett.}, {\bf 195}, L51--L53, (1975). \epubtkKeywords{Radio
  astronomy, Pulsars}

\bibitem{hun71}
Hunt, G.C., ``The rate of change of period of the pulsars'', {\em Mon. Not. R.
  Astron. Soc.}, {\bf 153}, 119--131, (1971). \epubtkKeywords{Pulsar timing,
  Radio astronomy, Pulsars}

\bibitem{virgo}
INFN, ``The Virgo Project'', project homepage, (2005). URL (cited on 29 January
  2005): \newline\url{http://www.virgo.infn.it}. \epubtkKeywords{Gravitational
  wave detectors}

\bibitem{ska}
International SKA Project Office, ``The Square Kilometre Array'', project
  homepage, (2005). URL (cited on 23 January 2005):
  \newline\url{http://www.skatelescope.org}. \epubtkKeywords{Radio astronomy}

\bibitem{jbv03}
Jacoby, B.A., Bailes, M., van Kerkwijk, M.H., Ord, S., Hotan, A., Kulkarni,
  S.R., and Anderson, S.B., ``PSR J1909-3744: A Binary Millisecond Pulsar with
  a Very Small Duty Cycle'', {\em Astrophys. J. Lett.}, {\bf 599}, L99--L102,
  (2003). \epubtkKeywords{Radio astronomy, Pulsars, Pulsar timing}

\bibitem{jb03}
Jaffe, A.H., and Backer, D.C., ``Gravitational Waves Probe the Coalescence Rate
  of Massive Black Hole Binaries'', {\em Astrophys. J.}, {\bf 583}, 616--631,
  (2003). \epubtkKeywords{Black holes, Pulsars, Gravitational waves}

\bibitem{jllw04}
Jenet, F.A., Lommen, A.N., Larson, S.L., and Wen, L., ``Constraining the
  properties fo supermassive black hole systems using pulsar timing:
  Application to 3C 66B'', {\em Astrophys. J.}, {\bf 606}, 799--803, (2004).
  \epubtkKeywords{Black holes, Pulsars, Gravitational waves}

\bibitem{jk91}
Johnston, H.M., and Kulkarni, S.R., ``On the detectability of pulsars in close
  binary systems'', {\em Astrophys. J.}, {\bf 368}, 504--514, (1991).
  \epubtkKeywords{Radio astronomy, Pulsars, Binary systems}

\bibitem{joh94}
Johnston, S., ``Evidence for a deficit of pulsars in the inner Galaxy'', {\em
  Mon. Not. R. Astron. Soc.}, {\bf 268}, 595--601, (1994).
  \epubtkKeywords{Pulsar statistics, Pulsars, Neutron stars, Galactic
  astronomy}

\bibitem{jlh93}
Johnston, S., Lorimer, D.R., Harrison, P.A., Bailes, M., Lyne, A.G., Bell,
  J.F., Kaspi, V.M., Manchester, R.N., D'Amico, N., Nicastro, L., and
  Shengzhen, J., ``Discovery of a very bright, nearby binary millisecond
  pulsar'', {\em Nature}, {\bf 361}, 613--615, (1993). \epubtkKeywords{Pulsars,
  Neutron stars, White dwarfs, Binary systems}

\bibitem{jlm92}
Johnston, S., Lyne, A.G., Manchester, R.N., Kniffen, D.A., D'Amico, N., Lim,
  J., and Ashworth, M., ``A High Frequency Survey of the Southern Galactic
  Plane for Pulsars'', {\em Mon. Not. R. Astron. Soc.}, {\bf 255}, 401--411,
  (1992). \epubtkKeywords{Radio astronomy, Pulsars}

\bibitem{jml92}
Johnston, S., Manchester, R.N., Lyne, A.G., Bailes, M., Kaspi, V.M., Qiao, G.,
  and D'Amico, N., ``PSR 1259--63: A binary radio pulsar with a Be star
  companion'', {\em Astrophys. J. Lett.}, {\bf 387}, L37--L41, (1992).
  \epubtkKeywords{Radio astronomy, Pulsars, Binary systems}

\bibitem{jl88}
Jones, A.W., and Lyne, A.G., ``Timing Observations of the Binary Pulsar PSR
  0655+64'', {\em Mon. Not. R. Astron. Soc.}, {\bf 232}, 473, (1988).
  \epubtkKeywords{Pulsars, Neutron stars, Binary systems, White dwarfs}

\bibitem{jlk03}
Joshi, B.C., Lyne, A.G., Kramer, M., Lorimer, D.R., Jordan, C.A., Holloway,
  A.J., Ikin, T.S., and Stairs, I.H., ``Coherent On-line Baseband Receiver for
  Astronomy'', in Bailes, M., Nice, D.J., and Thorsett, S.E., eds., {\em Radio
  Pulsars}, Proceedings of meeting held at Mediterranean Agonomic Institute of
  Chania, Crete, Greece, 26--29 August 2002, vol. 302 of ASP Conference Series,
   321--322, (Astronomical Society of the Pacific, San Fransisco, U.S.A.,
  2003). \epubtkKeywords{Radio astronomy, Pulsars, Coherent dedispersion}

\bibitem{jrs02}
Jouteux, S., Ramachandran, R., Stappers, B.W., Jonker, P.G., and van~der Klis,
  M., ``Searching for pulsars in close circular binary systems'', {\em Astron.
  Astrophys.}, {\bf 384}, 532--544, (2002). \epubtkKeywords{Binary pulsars,
  Pulsar searches, Signal processing}

\bibitem{mkv05b}
Kaaret, P., Morgan, E., and Vanderspek, R., ``Orbital parameters of HETE
  J1900.1-2455'', {\em Astron. Telegram}, {\bf 2005}(538), (June, 2005). URL
  (cited on 19 July 2005):
  \newline\url{http://www.astronomerstelegram.org/?read=538}.
  \epubtkKeywords{X-ray astronomy, Millisecond pulsars, Binary systems}

\bibitem{kal00}
Kalogera, V., ``Close Binaries with Two Compact Objects'', in Kramer, M., Wex,
  N., and Wielebinski, R., eds., {\em Pulsar Astronomy: 2000 and Beyond (IAU
  Colloquium 177)}, Proceedings of the 177th Colloquium of the IAU, held at the
  Max-Planck-Institut f\"ur Radioastronomie, Bonn, Germany, 30 August - 3
  September 1999, ASP Conference Series,  579--584, (Astronomical Society of
  the Pacific, San Francisco, U.S.A., 2000). \epubtkKeywords{Neutron stars,
  Pulsars, Binary systems, Monte Carlo methods, Gravitational wave sources}

\bibitem{kkl05b}
Kalogera, V., ``The Strongly Relativistic Double Pulsar and LISA'', lecture
  notes, Northwestern University, (2005). URL (cited on 19 July 2005):
  \newline\url{http://www.astro.northwestern.edu/Vicky/TALKS/LISA_0737.ppt}.
  Invited talk at the 5th International LISA Symposium, ESTEC, Noordwijk, The
  Netherlands, July 12--15, 2004. \epubtkKeywords{Gravitational wave detectors}

\bibitem{kkl04a}
Kalogera, V., Kim, C., Lorimer, D.R., Burgay, M., D'Amico, N., Possenti, A.,
  Manchester, R.N., Lyne, A.G., Joshi, B.C., McLaughlin, M.A., Kramer, M.,
  Sarkissian, J.M., and Camilo, F., ``The Cosmic Coalescence Rates for Double
  Neutron Star Binaries'', {\em Astrophys. J. Lett.}, {\bf 601}, L179--L182,
  (2004). \epubtkKeywords{Neutron stars, Binary systems, Gravitational wave
  sources}

\bibitem{kkl04b}
Kalogera, V., Kim, C., Lorimer, D.R., Burgay, M., D'Amico, N., Possenti, A.,
  Manchester, R.N., Lyne, A.G., Joshi, B.C., McLaughlin, M.A., Kramer, M.,
  Sarkissian, J.M., and Camilo, F., ``Erratum: ``The Cosmic Coalescence Rates
  for Double Neutron Star Binaries'''', {\em Astrophys. J. Lett.}, {\bf 614},
  L137--L138, (2004). \epubtkKeywords{Neutron stars, Binary systems,
  Gravitational wave sources}

\bibitem{kkl05}
Kalogera, V., Kim, C., Lorimer, D.R., Ihm, M., and Belczynski, C., ``The
  Galactic formation rate of eccentric neutron star white dwarf binaries'', in
  Rasio, F.A., and Stairs, I.H., eds., {\em Binary Radio Pulsars}, Proceedings
  of meeting held at the Aspen Center for Physics, USA, 12--16 January 2004,
  vol. 328 of ASP Conference Series,  261--267, (Astronomical Society of the
  Pacific, San Francisco, U.S.A., 2005). \epubtkKeywords{Radio astronomy,
  Pulsars}

\bibitem{kl00}
Kalogera, V., and Lorimer, D.~R., ``An Upper Limit on the Coalescence Rate of
  Double Neutron-Star Binaries in the Galaxy'', {\em Astrophys. J.}, {\bf 530},
  890--895, (2000). \epubtkKeywords{Pulsars, Pulsar statistics, Neutron stars}

\bibitem{knst01}
Kalogera, V., Narayan, R., Spergel, D.N., and Taylor, J.H., ``The Coalescence
  Rate of Double Neutron Star Systems'', {\em Astrophys. J.}, {\bf 556},
  340--356, (2001). Related online version (cited on 19 March 2001):
  \newline\url{http://arXiv.org/abs/astro-ph/0012038}. \epubtkKeywords{Neutron
  stars, Pulsars, Binary systems, Monte Carlo methods, Gravitational wave
  sources}

\bibitem{kas94}
Kaspi, V.M., {\em Applications of Pulsar Timing}, Ph.D. Thesis, (Princeton
  University, Princeton, U.S.A., 1994). \epubtkKeywords{Radio astronomy,
  Pulsars, Pulsar timing, General relativity, Cosmology, Fundamental physics}

\bibitem{kjb94}
Kaspi, V.M., Johnston, S., Bell, J.F., Manchester, R.N., Bailes, M., Bessell,
  M., Lyne, A.G., and D'Amico, N., ``A massive radio pulsar binary in the Small
  Magellanic Cloud'', {\em Astrophys. J. Lett.}, {\bf 423}, L43--L45, (1994).
  \epubtkKeywords{Radio astronomy, Pulsars, Binary systems}

\bibitem{klm00}
Kaspi, V.M., Lyne, A.G., Manchester, R.N., Crawford, F., Camilo, F., Bell,
  J.F., D'Amico, N., Stairs, I.H., McKay, N.P.F., Morris, D.J., and Possenti,
  A., ``Discovery of a young radio pulsar in a relativistic binary orbit'',
  {\em Astrophys. J.}, {\bf 534}, 321--327, (2000). \epubtkKeywords{Pulsars,
  Neutron stars, White dwarfs, Binary systems, Gravitational wave sources}

\bibitem{ktr94}
Kaspi, V.M., Taylor, J.H., and Ryba, M.F., ``High-Precision Timing of
  Millisecond Pulsars. III. Long-Term Monitoring of PSRs B1855+09 and
  B1937+21'', {\em Astrophys. J.}, {\bf 428}, 713--728, (1994).
  \epubtkKeywords{Pulsar timing, Radio astronomy, Pulsars, Fundamental physics,
  Cosmology}

\bibitem{kkl03}
Kim, C., Kalogera, V., and Lorimer, D.R., ``The Probability Distribution of
  Binary Pulsar Coalescence Rates. I. Double Neutron Star Systems in the
  Galactic Field'', {\em Astrophys. J.}, {\bf 584}, 985--995, (2003).
  \epubtkKeywords{Neutron stars, Binary systems, Gravitational wave sources}

\bibitem{kklw04}
Kim, C., Kalogera, V., Lorimer, D.R., and White, T., ``The Probability
  Distribution of Binary Pulsar Coalescence Rates. II. Neutron Star--White
  Dwarf Binaries'', {\em Astrophys. J.}, {\bf 616}, 1109--1117, (2004).
  \epubtkKeywords{Neutron stars, Binary systems, Gravitational wave sources}

\bibitem{krst88}
Kluzniak, W., Ruderman, M., Shaham, J., and Tavani, M., ``Nature and evolution
  of the eclipsing millisecond binary pulsar PSR 1957+20'', {\em Nature}, {\bf
  334}, 225--227, (1988). \epubtkKeywords{Binary pulsars, Ablation}

\bibitem{kgm04}
Kolonko, M., Gil, J., and Maciesiak, K., ``On the pulse-width statistics in
  radio pulsars'', {\em Astron. Astrophys.}, {\bf 428}, 943--951, (2005).
  \epubtkKeywords{Radio astronomy, Pulsars, Pulsar statistics}

\bibitem{kop97}
Kopeikin, S.M., ``Binary pulsars as detectors of ultralow-frequency
  gravitational waves'', {\em Phys. Rev. D}, {\bf 56}, 4455--4469, (1997).
  \epubtkKeywords{Cosmic gravitational wave background, Pulsars, Binary
  systems, Pulsar timing}

\bibitem{kra05}
Kramer, M., ``Latest timing results for PSR J0737-3039'', personal
  communication. \epubtkKeywords{Pulsars, Tests of relativistic gravity}

\bibitem{kbc04}
Kramer, M., Backer, D.C., Cordes, J.M., Lazio, T.J.W., Stappers, B.W., and
  Johnston, S., ``Strong-field tests of gravity using pulsars and black
  holes'', {\em New Astron. Rev.}, {\bf 48}, 993--1002, (2004).
  \epubtkKeywords{Black holes, Pulsars, Gravitational waves}

\bibitem{kbm03}
Kramer, M., Bell, J.F., Manchester, R.N., Lyne, A.G., Camilo, F., Stairs, I.H.,
  D'Amico, N., Kaspi, V.M., Hobbs, G., Morris, D.J., Crawford, F., Possenti,
  A., Joshi, B.C., McLaughlin, M.A., Lorimer, D.R., and Faulkner, A.J., ``The
  Parkes Multibeam Pulsar Survey - III. Young pulsars and the discovery and
  timing of 200 pulsars'', {\em Mon. Not. R. Astron. Soc.}, {\bf 342},
  1299--1324, (2003). \epubtkKeywords{Radio astronomy, Pulsars, Neutron stars,
  Binary systems}

\bibitem{kxc99}
Kramer, M., Xilouris, K.M., Camilo, F., Nice, D.J., Backer, D.C., Lorimer,
  D.R., Doroshenko, O.V., Lange, C., and Sallmen, S., ``Profile Instabilities
  of the Millisecond Pulsar J1022+1001'', {\em Astrophys. J.}, {\bf 520},
  324--334, (1999). \epubtkKeywords{Pulsars, Neutron stars}

\bibitem{kxl98}
Kramer, M., Xilouris, K.M., Lorimer, D.R., Doroshenko, O.V., Jessner, A.,
  Wielebinski, R., Wolszczan, A., and Camilo, F., ``The characteristics of
  millisecond pulsar emission: I. Spectra, pulse shapes and the beaming
  fraction'', {\em Astrophys. J.}, {\bf 501}, 270--285, (1998).
  \epubtkKeywords{Radio astronomy, Pulsars}

\bibitem{kapw91}
Kulkarni, S.R., Anderson, S.B., Prince, T.A., and Wolszczan, A., ``Old pulsars
  in the low-density globular clusters M13 and M53'', {\em Nature}, {\bf 349},
  47--49, (1991). \epubtkKeywords{Pulsars, Neutron stars}

\bibitem{lcw01}
Lange, C., Camilo, F., Wex, N., Kramer, M., Backer, D.C., Lyne, A.G., and
  Doroshenko, O.V., ``Precision timing measurements of PSR J1012+5307'', {\em
  Mon. Not. R. Astron. Soc.}, {\bf 326}, 274--282, (2001).
  \epubtkKeywords{Radio astronomy, Pulsar timing, Fundamental physics}

\bibitem{lp04}
Lattimer, J.M., and Prakash, M., ``The Physics of Neutron Stars'', {\em
  Science}, {\bf 304}, 536--542, (2004). \epubtkKeywords{Radio astronomy,
  Pulsars}

\bibitem{lmop87}
Lawson, K.D., Mayer, C.J., Osborne, J.L., and Parkinson, M.L., ``Variations in
  the Spectral Index of the Galactic Radio Continuum Emission in the Northern
  Hemisphere'', {\em Mon. Not. R. Astron. Soc.}, {\bf 225}, 307--327, (1987).
  \epubtkKeywords{Galactic astronomy}

\bibitem{le91}
Levinson, A., and Eichler, D., ``Can neutron stars ablate their companions?'',
  {\em Astrophys. J.}, {\bf 379}, 359--365, (1991). \epubtkKeywords{Binary
  pulsars, Ablation}

\bibitem{lwf04}
Lewandowski, W., Wolszczan, A., Feiler, G., Konacki, M., and
  So{\l}tysi{\'n}ski, T., ``Arecibo timing and single-pulse observations of
  eighteen pulsars'', {\em Astrophys. J.}, {\bf 600}, 905--913, (2004).
  \epubtkKeywords{Radio astronomy, Pulsars, Pulsar timing}

\bibitem{li02}
Li, X., ``Formation of Mildly Recycled Low-Mass Binary Pulsars from
  Intermediate-Mass X-Ray Binaries'', {\em Astrophys. J.}, {\bf 564}, 930--934,
  (2002). \epubtkKeywords{X-ray binaries, Accretion, Millisecond pulsars}

\bibitem{lpp96}
Lipunov, V.M., Postnov, K.A., and Prokhorov, M.E., ``The Scenario Machine:
  Binary Star Population Synthesis'', {\em Astrophys. Space Phys. Rev.}, {\bf
  9}, 1--178, (1996). Related online version (cited on 24 October 2005):
  \newline\url{http://xray.sai.msu.ru/~mystery/articles/review/}.
  \epubtkKeywords{Binary pulsars, Pulsar evolution}

\bibitem{lkd04}
L{\"o}hmer, O., Kramer, M., Driebe, T., Jessner, A., Mitra, D., and Lyne, A.G.,
  ``The parallax, mass and age of the PSR J2145-0750 binary system'', {\em
  Astron. Astrophys.}, {\bf 426}, 631--640, (2004). \epubtkKeywords{Radio
  astronomy, Pulsars, Pulsar timing, Astrometry}

\bibitem{lom02}
Lommen, A.N., ``New Limits on Gravitational Radiation using Pulsars'', in
  Becker, W., Lesch, H., and Truemper, J., eds., {\em Neutron Stars, Pulsars
  and Supernova Remnants}, Proceedings of 270. WE-Heraeus Seminar 21--25
  January 2002, Physikzentrum Bad Honnef, Germany, vol. 278 of MPE-Report,
  114--125, (Max Planck Institute for Extraterrestrial Physics, Garching,
  Germany, 2002). Related online version (cited on 24 January 2005):
  \newline\url{ftp://ftp.xray.mpe.mpg.de/people/web/MPE_Report_278.pdf}.
  \epubtkKeywords{Radio astronomy, Pulsars, Gravitational waves}

\bibitem{lb01}
Lommen, A.N., and Backer, D.C., ``Using pulsars to detect massive black hole
  binaries via gravitational radiation: Sagittarius A* and nearby galaxies'',
  {\em Astrophys. J.}, {\bf 562}, 297--302, (2001). \epubtkKeywords{Black
  holes, Pulsars, Gravitational waves}

\bibitem{lzb01}
Lommen, A.N., Zepka, A.F., Backer, D.C., McLaughlin, M.A., Cordes, J.M.,
  Arzoumanian, Z., and Xilouris, K.M., ``New Pulsars from an Arecibo Drift Scan
  Search'', {\em Astrophys. J.}, {\bf 545}, 1007--1014, (2001).
  \epubtkKeywords{Pulsars, Neutron stars}

\bibitem{lor95}
Lorimer, D.R., ``Pulsar statistics -- II. The local low-mass binary pulsar
  population'', {\em Mon. Not. R. Astron. Soc.}, {\bf 274}, 300--304, (1995).
  \epubtkKeywords{Neutron stars, Pulsars, Binary systems, Monte Carlo methods}

\bibitem{lor98e}
Lorimer, D.R., ``Binary and Millisecond Pulsars'', {\em Living Rev.
  Relativity}, {\bf 1}, lrr-1998-10, (1998). URL (cited on 19 March 2001):
  \newline\url{http://www.livingreviews.org/lrr-1998-10}. \epubtkKeywords{Radio
  astronomy, Pulsars}

\bibitem{lor01}
Lorimer, D.R., ``Binary and Millisecond Pulsars'', {\em Living Rev.
  Relativity}, {\bf 4}, lrr-2001-5, (2001). URL (cited on 22 January 2005):
  \newline\url{http://www.livingreviews.org/lrr-2001-5}. \epubtkKeywords{Radio
  astronomy, Pulsars}

\bibitem{lor04}
Lorimer, D.R., ``The Galactic Population and Birth Rate of Radio Pulsars'', in
  Camilo, F., and Gaensler, B.M., eds., {\em Young neutron stars and their
  environments}, Proceedings of an IAU Symposium on 14--17 July 2003 in Sydney,
  Australia, vol. 218 of ASP Conference Series,  105, (Astronomical Society of
  the Pacific, San Francisco, U.S.A., 2004). \epubtkKeywords{Pulsars, Pulsar
  statistics, Radial distribution}

\bibitem{lbdh93}
Lorimer, D.R., Bailes, M., Dewey, R.J., and Harrison, P.A., ``Pulsar
  statistics: the birthrate and initial spin periods of radio pulsars'', {\em
  Mon. Not. R. Astron. Soc.}, {\bf 263}, 403--415, (1993).
  \epubtkKeywords{Neutron stars, Pulsars, Monte Carlo methods, Galactic
  astronomy}

\bibitem{lcf03}
Lorimer, D.R., Camilo, F., Freire, P.C.C., Kramer, M., Lyne, A.G., Manchester,
  R.N., and D'Amico, N., ``Millisecond radio pulsars in 47 Tucanae'', in
  Bailes, M., Nice, D.J., and Thorsett, S.E., eds., {\em Radio Pulsars},
  Proceedings of meeting held at Mediterranean Agonomic Institute of Chania,
  Crete, Greece, 26--29 August 2002, vol. 302 of ASP Conference Series,
  363--366, (Astronomical Society of the Pacific, San Fransisco, U.S.A., 2003).
  \epubtkKeywords{Radio astronomy, Pulsars, Globular clusters}

\bibitem{lf05}
Lorimer, D.R., and Freire, P.C.C., ``Long-period binaries and the strong
  equivalence principle'', in Rasio, F.A., and Stairs, I.H., eds., {\em Binary
  Radio Pulsars}, Proceedings of meeting held at the Aspen Center for Physics,
  USA, 12--16 January 2004, vol. 328 of ASP Conference Series,  19--24,
  (Astronomical Society of the Pacific, San Francisco, U.S.A., 2005).
  \epubtkKeywords{Radio astronomy, Pulsars}

\bibitem{lk05}
Lorimer, D.R., and Kramer, M., {\em Handbook of Pulsar Astronomy}, vol.~4 of
  Cambridge Observing Handbooks for Research Astronomers, (Cambridge University
  Press, Cambridge, U.K.; New York, U.S.A, 2005), 1st edition.
  \epubtkKeywords{Radio astronomy, Neutron stars, Pulsars}

\bibitem{handbook}
Lorimer, D.R., and Kramer, M., {\em Handbook of Pulsar Astronomy}, vol.~4 of
  Cambridge Observing Handbooks for Research Astronomers, (Cambridge University
  Press, Cambridge, U.K.; New York, U.S.A., 2005). Related online version
  (cited on 18 January 2005):
  \newline\url{http://www.jb.man.ac.uk/pulsarhandbook}. \epubtkKeywords{Radio
  astronomy, Pulsars}

\bibitem{llb96}
Lorimer, D.R., Lyne, A.G., Bailes, M., Manchester, R.N., D'Amico, N., Stappers,
  B.W., Johnston, S., and Camilo, F., ``Discovery of four binary millisecond
  pulsars'', {\em Mon. Not. R. Astron. Soc.}, {\bf 283}, 1383--1387, (1996).
  \epubtkKeywords{Pulsars, Neutron stars, White dwarfs}

\bibitem{lma04}
Lorimer, D.R., McLaughlin, M.A., Arzoumanian, Z., Xilouris, K.M., Cordes, J.M.,
  Lommen, A.N., Fruchter, A.S., Chandler, A.M., and Backer, D.C., ``PSR
  J0609+2130: a disrupted binary pulsar?'', {\em Mon. Not. R. Astron. Soc.},
  {\bf 347}, L21--L25, (2004). \epubtkKeywords{Radio astronomy, Pulsars, Binary
  pulsars}

\bibitem{lnl95}
Lorimer, D.R., Nicastro, L., Lyne, A.G., Bailes, M., Manchester, R.N.,
  Johnston, S., Bell, J.F., D'Amico, N., and Harrison, P.A., ``Four new
  millisecond pulsars in the Galactic disk'', {\em Astrophys. J.}, {\bf 439},
  933--938, (1995). \epubtkKeywords{Pulsars, Neutron stars, White dwarfs,
  Binary systems}

\bibitem{lxf05}
Lorimer, D.R., Xilouris, K.M., Fruchter, A.S., Stairs, I.H., Camilo, F.,
  McLaughlin, M.A., Vazquez, A.M., Eder, J.E., Roberts, M.S.E., Hessels,
  J.W.T., and Ransom, S.M., ``Discovery of 10 pulsars in an Arecibo drift-scan
  survey'', {\em Mon. Not. R. Astron. Soc.}, {\bf 359}, 1524--1530, (2005).
  Related online version (cited on 19 July 2005):
  \newline\url{http://arXiv.org/abs/astro-ph/0504019}. \epubtkKeywords{Radio
  astronomy, Pulsars}

\bibitem{lylg95}
Lorimer, D.R., Yates, J.A., Lyne, A.G., and Gould, D.M., ``Multifrequency Flux
  density measurements of 280 pulsars'', {\em Mon. Not. R. Astron. Soc.}, {\bf
  273}, 411--421, (1995). \epubtkKeywords{Radio astronomy, Pulsars}

\bibitem{lzc95}
Lundgren, S.C., Zepka, A.F., and Cordes, J.M., ``A millisecond pulsar in a six
  hour orbit: PSR J0751+1807'', {\em Astrophys. J.}, {\bf 453}, 419--423,
  (1995). \epubtkKeywords{Pulsars, Neutron stars, Binary systems, White dwarfs}

\bibitem{lyn05}
Lyne, A.G., ``Binary Pulsar Discoveries in the Parkes Multibeam Survey'', in
  Rasio, F.A., and Stairs, I.H., eds., {\em Binary Radio Pulsars}, Proceedings
  of meeting held at the Aspen Center for Physics, USA, 12 Janaury - 16 January
  2004, vol. 328 of ASP Conference Series,  37--41, (Astronomical Society of
  the Pacific, San Francisco, U.S.A., 2005). \epubtkKeywords{Radio astronomy,
  Pulsars}

\bibitem{las82}
Lyne, A.G., Anderson, B., and Salter, M.J., ``The Proper Motions of 26
  Pulsars'', {\em Mon. Not. R. Astron. Soc.}, {\bf 201}, 503--520, (1982).
  \epubtkKeywords{Pulsars, Galactic astronomy}

\bibitem{lb90}
Lyne, A.G., and Bailes, M., ``The mass of the PSR 2303+46 system'', {\em Mon.
  Not. R. Astron. Soc.}, {\bf 246}, 15P--17P, (1990). \epubtkKeywords{Neutron
  stars, White dwarfs, Pulsars, Binary systems}

\bibitem{lbhb93}
Lyne, A.G., Biggs, J.D., Harrison, P.A., and Bailes, M., ``A long-period
  globular-cluster pulsar in an eclipsing binary system'', {\em Nature}, {\bf
  361}, 47--49, (1993). \epubtkKeywords{Pulsars, Neutron stars, Binary systems}

\bibitem{lbm87}
Lyne, A.G., Brinklow, A., Middleditch, J., Kulkarni, S.R., Backer, D.C., and
  Clifton, T.R., ``The discovery of a millisecond pulsar in the globular
  cluster M28'', {\em Nature}, {\bf 328}, 399--401, (1987).
  \epubtkKeywords{Pulsars, Neutron stars}

\bibitem{lbk04}
Lyne, A.G., Burgay, M., Kramer, M., Possenti, A., Manchester, R.N., Camilo, F.,
  McLaughlin, M., Lorimer, D.R., Joshi, B.C., Reynolds, J.E., and Freire,
  P.C.C., ``A Double-Pulsar System: A Rare Laboratory for Relativistic Gravity
  and Plasma Physics'', {\em Science}, {\bf 303}, 1153--1157, (2004).
  \epubtkKeywords{Radio astronomy, Pulsars, Tests of relativistic gravity}

\bibitem{lcm00}
Lyne, A.G., Camilo, F., Manchester, R.N., Bell, J.F., Kaspi, V.M., D'Amico, N.,
  McKay, N.P.F., Crawford, F., Morris, D.J., Sheppard, D.C., and Stairs, I.H.,
  ``The Parkes multibeam survey: PSR J1811--1736 -- a pulsar in a highly
  eccentric binary system'', {\em Mon. Not. R. Astron. Soc.}, {\bf 312},
  698--702, (2000). \epubtkKeywords{Radio astronomy, Pulsars, Neutron stars,
  Binary systems}

\bibitem{ls05}
Lyne, A.G., and Graham-Smith, F., {\em Pulsar Astronomy}, Cambridge
  Astrophysics Series, (Cambridge University Press, Cambridge, U.K.; New York,
  U.S.A, 2006), 3rd edition. in press. \epubtkKeywords{Radio astronomy, Neutron
  stars, Pulsars}

\bibitem{ll94}
Lyne, A.G., and Lorimer, D.R., ``High birth velocities of radio pulsars'', {\em
  Nature}, {\bf 369}, 127--129, (1994). \epubtkKeywords{Pulsars, Neutron stars,
  Fundamental physics, Gamma-ray bursts}

\bibitem{lm88}
Lyne, A.G., and Manchester, R.N., ``The shape of pulsar radio beams'', {\em
  Mon. Not. R. Astron. Soc.}, {\bf 234}, 477--508, (1988).
  \epubtkKeywords{Neutron stars, Pulsars}

\bibitem{lmd96}
Lyne, A.G., Manchester, R.N., and D'Amico, N., ``PSR B1745-20 and young pulsars
  in globular clusters'', {\em Astrophys. J. Lett.}, {\bf 460}, L41--L44,
  (1996). \epubtkKeywords{Pulsars, Neutron stars}

\bibitem{lmd90}
Lyne, A.G., Manchester, R.N., D'Amico, N., Staveley-Smith, L., Johnston, S.,
  Lim, J., Fruchter, A.S., Goss, W.M., and Frail, D.A., ``An eclipsing
  millisecond pulsar in the globular cluster Terzan 5'', {\em Nature}, {\bf
  347}, 650--652, (1990). \epubtkKeywords{Pulsars, Neutron stars, Binary
  systems}

\bibitem{lml98}
Lyne, A.G., Manchester, R.N., Lorimer, D.R., Bailes, M., D'Amico, N., Tauris,
  T.M., Johnston, S., Bell, J.F., and Nicastro, L., ``The Parkes Southern
  Pulsar Survey -- II. Final Results and Population Analysis'', {\em Mon. Not.
  R. Astron. Soc.}, {\bf 295}, 743--755, (1998). \epubtkKeywords{Radio
  astronomy, Pulsars, Monte Carlo methods, Galactic astronomy}

\bibitem{lmbm00}
Lyne, A.G., Mankelow, S.H., Bell, J.F., and Manchester, R.N., ``Radio pulsars
  in Terzan 5'', {\em Mon. Not. R. Astron. Soc.}, {\bf 316}, 491--493, (2000).
  \epubtkKeywords{Pulsars, Neutron stars}

\bibitem{lm89}
Lyne, A.G., and McKenna, J., ``PSR 1820 -- 11: a binary pulsar in a wide and
  highly eccentric orbit'', {\em Nature}, {\bf 340}, 367--369, (1989).
  \epubtkKeywords{Radio astronomy, Pulsars}

\bibitem{mhth05}
Manchester, R.N., Hobbs, G.B., Teoh, A., and Hobbs, M., ``The ATNF Pulsar
  Catalogue'', (2004). URL (cited on 22 January 2005):
  \newline\url{http://arXiv.org/abs/astro-ph/0412641}. \epubtkKeywords{Radio
  astronomy, Pulsars, Pulsar statistics}

\bibitem{ml77}
Manchester, R.N., and Lyne, A.G., ``Pulsar Interpulses -- two poles or one?'',
  {\em Mon. Not. R. Astron. Soc.}, {\bf 181}, 761--767, (1977).
  \epubtkKeywords{Neutron stars, Pulsars}

\bibitem{mlc01}
Manchester, R.N., Lyne, A.G., Camilo, F., Bell, J.F., Kaspi, V.M., D'Amico, N.,
  McKay, N.P.F., Crawford, F., Stairs, I.H., Possenti, A., Morris, D.J., and
  Sheppard, D.C., ``The Parkes multi-beam pulsar survey -- I. Observing and
  data analysis systems, discovery and timing of 100 pulsars'', {\em Mon. Not.
  R. Astron. Soc.}, {\bf 328}, 17--35, (2001). \epubtkKeywords{Radio astronomy,
  Pulsars, Neutron stars, Binary systems}

\bibitem{mld96}
Manchester, R.N., Lyne, A.G., D'Amico, N., Bailes, M., Johnston, S., Lorimer,
  D.R., Harrison, P.A., Nicastro, L., and Bell, J.F., ``The Parkes Southern
  Pulsar Survey I. Observing and data analysis systems and initial results'',
  {\em Mon. Not. R. Astron. Soc.}, {\bf 279}, 1235--1250, (1996).
  \epubtkKeywords{Radio astronomy, Data analysis, Pulsars, Galactic astronomy}

\bibitem{mncl80}
Manchester, R.N., Newton, L.M., Cooke, D.J., and Lyne, A.G., ``Detection of a
  pulsar in a long-period binary system'', {\em Astrophys. J. Lett.}, {\bf
  236}, L25--L27, (1980). \epubtkKeywords{Pulsars, Neutron stars, Binary
  systems, White dwarfs}

\bibitem{mt74}
Manchester, R.N., and Taylor, J.H., ``Period irregularities in pulsars'', {\em
  Astrophys. J. Lett.}, {\bf 191}, L63--L65, (1974). \epubtkKeywords{Radio
  astronomy, Pulsars, Pulsar timing}

\bibitem{mtem97}
Matsakis, D.N., Taylor, J.H., Eubanks, T.M., and Marshall, T., ``A statistic
  for describing pulsar and clock stabilities'', {\em Astron. Astrophys.}, {\bf
  326}, 924--928, (1997). \epubtkKeywords{Pulsar timing}

\bibitem{mpipsr}
Max Planck Institute for Radio Astronomy, ``MPIfR Pulsar Group'', project
  homepage, (1999). URL (cited on 20 March 2001):
  \newline\url{http://www.mpifr-bonn.mpg.de/div/pulsar/}. \epubtkKeywords{Radio
  astronomy, Pulsars}

\bibitem{mkl04}
McLaughlin, M.A., Kramer, M., Lyne, A.G., Lorimer, D.R., Stairs, I.H.,
  Possenti, A., Manchester, R.N., Freire, P.C.C., Joshi, B.C., Burgay, M.,
  Camilo, F., and D'Amico, N., ``The double pulsar system J0737-3039:
  Modulation of the radio emission from B by the radiation from A'', {\em
  Astrophys. J. Lett.}, {\bf 613}, L57--L60, (2004). \epubtkKeywords{Double
  pulsars}

\bibitem{mla05}
McLaughlin, M.A., Lorimer, D.R., Champion, D.J., Arzoumanian, Z., Backer, D.C.,
  Cordes, J.M., Fruchter, A.S., Lommen, A.N., and Xilouris, K.M., ``New Binary
  and Millisecond Pulsars from Arecibo Drift-Scan Searches'', in Rasio, F.A.,
  and Stairs, I.H., eds., {\em Binary Radio Pulsars}, Proceedings of meeting
  held at the Aspen Center for Physics, USA, 12 Janaury - 16 January 2004, vol.
  328 of ASP Conference Series,  43--49, (Astronomical Society of the Pacific,
  San Francisco, U.S.A., 2005). \epubtkKeywords{Radio astronomy, Pulsars}

\bibitem{mic91}
Michel, F.C., {\em Theory of Neutron Star Magnetospheres}, Theoretical
  Astrophysics, (University of Chicago Press, Chicago, U.S.A., 1991).
  \epubtkKeywords{Neutron stars, Fundamental physics}

\bibitem{mk84}
Middleditch, J., and Kristian, J., ``A search for young, luminous optical
  pulsars in extragalactic supernova remnants'', {\em Astrophys. J.}, {\bf
  279}, 157--161, (1984). \epubtkKeywords{Pulsars, Binary systems}

\bibitem{mkv05a}
Morgan, E., Kaaret, P., and Vanderspek, R., ``HETE J1900.1-2455 is a
  millisecond pulsar'', {\em Astron. Telegram}, {\bf 2005}(523), (June, 2005).
  URL (cited on 19 July 2005):
  \newline\url{http://www.astronomerstelegram.org/?read=523}.
  \epubtkKeywords{X-ray astronomy, Millisecond pulsars, Binary systems}

\bibitem{mhl02}
Morris, D.J., Hobbs, G., Lyne, A.G., Stairs, I.H., Camilo, F., Manchester,
  R.N., Possenti, A., Bell, J.F., Kaspi, V.M., D'Amico, N., McKay, N.P.F.,
  Crawford, F., and Kramer, M., ``The Parkes Multibeam Pulsar Survey - II.
  Discovery and timing of 120 pulsars'', {\em Mon. Not. R. Astron. Soc.}, {\bf
  335}, 275--290, (2002). \epubtkKeywords{Radio astronomy, Pulsars, Neutron
  stars, Binary systems}

\bibitem{nar87}
Narayan, R., ``The birthrate and initial spin period of single radio pulsars'',
  {\em Astrophys. J.}, {\bf 319}, 162--179, (1987). \epubtkKeywords{Neutron
  stars, Pulsars, Monte Carlo methods, Galactic astronomy}

\bibitem{nv83}
Narayan, R., and Vivekanand, M., ``Evidence for evolving elongated Pulsar
  Beams'', {\em Astron. Astrophys.}, {\bf 122}, 45--53, (1983).
  \epubtkKeywords{Neutron stars, Pulsars}

\bibitem{lisa}
NASA/ESA, ``Laser Interferometer Space Antenna'', project homepage, (2005). URL
  (cited on 24 January 2005): \newline\url{http://lisa.jpl.nasa.gov}.
  \epubtkKeywords{Gravitational wave detectors}

\bibitem{tama}
National Astronomical Observatory, ``TAMA project'', project homepage, (2005).
  URL (cited on 29 January 2005): \newline\url{http://tamago.mtk.nao.ac.jp}.
  \epubtkKeywords{Gravitational wave detectors}

\bibitem{aopsr}
National Astronomy and Ionosphere Center, ``Pulsars at the Arecibo
  Observatory'', project homepage, (1999). URL (cited on 7 December 2000):
  \newline\url{http://www.naic.edu/~pulsar}. \epubtkKeywords{Radio astronomy,
  Pulsars}

\bibitem{alfa}
National Astronomy and Ionosphere Center, ``Arecibo L-band feed array'',
  project homepage, (2005). URL (cited on 23 January 2005):
  \newline\url{http://alfa.naic.edu}. \epubtkKeywords{Radio astronomy}

\bibitem{arecibo}
National Astronomy and Ionosphere Center, ``The Arecibo Observatory'', project
  homepage, (2005). URL (cited on 29 January 2005):
  \newline\url{http://www.naic.edu}. \epubtkKeywords{Radio astronomy, Pulsars,
  Galactic astronomy, Extragalactic astronomy}

\bibitem{palfa}
National Astronomy and Ionosphere Center, ``Preliminary PALFA survey'', project
  homepage, (2005). URL (cited on 23 January 2005):
  \newline\url{http://www.naic.edu/~palfa}. \epubtkKeywords{Radio astronomy}

\bibitem{gcpsrs}
National Astronomy and Ionospheric Center, ``Pulsars in globular clusters'',
  web interface to database, (2005). URL (cited on 18 January 2005):
  \newline\url{http://www.naic.edu/~pfreire/GCpsr.html}. \epubtkKeywords{Radio
  astronomy, Pulsars}

\bibitem{gmrt}
National Centre for Radio Astonomy, ``The Giant Metre Wave Radio Telescope'',
  project homepage, (2005). URL (cited on 23 January 2005):
  \newline\url{http://www.gmrt.ncra.tifr.res.in}. \epubtkKeywords{Radio
  astronomy}

\bibitem{gbt}
National Radio Astronomy Observatory, ``The Green Bank Telescope'', project
  homepage, (2005). URL (cited on 23 January 2005):
  \newline\url{http://www.gb.nrao.edu}. \epubtkKeywords{Radio astronomy}

\bibitem{ran05}
National Radio Astronomy Observatory, ``Scott Ransom's home page'', personal
  homepage, (2005). URL (cited on 7 November 2005):
  \newline\url{http://www.cv.nrao.edu/~sransom}. \epubtkKeywords{Pulsars}

\bibitem{naf03}
Navarro, J., Anderson, S.B., and Freire, P.C.C., ``The Arecibo 430 MHz
  Intermediate Galactic Latitude Survey: Discovery of Nine Radio Pulsars'',
  {\em Astrophys. J.}, {\bf 594}, 943--951, (2003). \epubtkKeywords{Radio
  astronomy, Pulsars}

\bibitem{nbf95}
Navarro, J., de~Bruyn, A.G., Frail, D.A., Kulkarni, S.R., and Lyne, A.G., ``A
  very luminous binary millisecond pulsar'', {\em Astrophys. J. Lett.}, {\bf
  455}, L55--L58, (1995). \epubtkKeywords{Pulsars, Neutron stars, Binary
  systems, White dwarfs}

\bibitem{nll95}
Nicastro, L., Lyne, A.G., Lorimer, D.R., Harrison, P.A., Bailes, M., and
  Skidmore, B.D., ``PSR J1012+5307: a 5.26-ms pulsar in a 14.5-h binary
  system'', {\em Mon. Not. R. Astron. Soc.}, {\bf 273}, L68--L70, (1995).
  \epubtkKeywords{Neutron stars, White dwarfs, Binary systems}

\bibitem{nst96}
Nice, D.J., Sayer, R.W., and Taylor, J.H., ``PSR J 1518+4904: A Mildly
  Relativistic Binary Pulsar System'', {\em Astrophys. J. Lett.}, {\bf 466},
  L87--L90, (1996). \epubtkKeywords{Pulsars, Binary systems, Neutron stars,
  General relativity}

\bibitem{nst99}
Nice, D.J., Sayer, R.W., and Taylor, J.H., ``Timing Observations of the
  J1518+4904 Double Neutron Star System'', in Arzoumanian, Z., van~der Hooft,
  F., and van~den Heuvel, E.P.J., eds., {\em Pulsar Timing, General Relativity,
  and the Internal Structure of Neutron Stars}, Proceedings of the colloquium,
  Amsterdam, 24--27 September 1996,  79--83, (Royal Netherlands Academy of Arts
  and Sciences, Amsterdam, Netherlands, 1999). \epubtkKeywords{Pulsars, Neutron
  stars, Binary systems}

\bibitem{nss01}
Nice, D.J., Splaver, E.M., and Stairs, I.H., ``On the Mass and Inclination of
  the PSR J2019+2425 Binary System'', {\em Astrophys. J.}, {\bf 549}, 516--521,
  (2001). \epubtkKeywords{Pulsars, Neutron stars, Binary systems}

\bibitem{nt95}
Nice, D.J., and Taylor, J.H., ``PSRs J2019+2425 and J2322+2057 and the Proper
  Motions of Millisecond Pulsars'', {\em Astrophys. J.}, {\bf 441}, 429--435,
  (1995). \epubtkKeywords{Pulsars, Neutron stars, Galactic astronomy}

\bibitem{obv02}
Ord, S.M., Bailes, M., and van Straten, W., ``The Scintillation Velocity of the
  Relativistic Binary Pulsar PSR J1141-6545'', {\em Astrophys. J. Lett.}, {\bf
  574}, L75--L78, (2002). \epubtkKeywords{Pulsars, Binary systems, General
  relativity}

\bibitem{pac68}
Pacini, F., ``Rotating neutron stars, pulsars, and supernova remnants'', {\em
  Nature}, {\bf 219}, 145--146, (1968). \epubtkKeywords{Neutron stars, Pulsars}

\bibitem{pee93}
Peebles, P.J.E., {\em Principles of Physical Cosmology}, Princeton Series in
  Physics, (Princeton University Press, Princeton, U.S.A., 1993).
  \epubtkKeywords{Cosmology, Astrometry}

\bibitem{psrplanets}
Penn State University, ``Pulsar Planets'', personal homepage, (2001). URL
  (cited on 19 March 2001):
  \newline\url{http://www.astro.psu.edu/users/alex/pulsar_planets.htm}.
  \epubtkKeywords{Radio astronomy, Pulsars}

\bibitem{pet64}
Peters, P.C., ``Gravitational Radiation and the Motion of Two Point Masses'',
  {\em Phys. Rev.}, {\bf 136}, B1224--B1232, (1964). \epubtkKeywords{General
  relativity, Gravitational radiation, Binary systems}

\bibitem{pm63}
Peters, P.C., and Mathews, J., ``Gravitational Radiation from Point Masses in a
  Keplerian Orbit'', {\em Phys. Rev.}, {\bf 131}, 435--440, (1963).
  \epubtkKeywords{General relativity, Gravitational radiation, Binary systems}

\bibitem{pl04}
Pfahl, E., and Loeb, A., ``Probing the Spacetime Around Sgr A* with Radio
  Pulsars'', {\em Astrophys. J.}, {\bf 615}, 253--258, (2004).
  \epubtkKeywords{Pulsar populations}

\bibitem{phi91}
Phinney, E.S., ``The rate of neutron star binary mergers in the universe:
  Minimal predictions for gravity wave detectors'', {\em Astrophys. J. Lett.},
  {\bf 380}, L17--L21, (1991). \epubtkKeywords{Neutron stars, Pulsars, Binary
  systems, Gravitational wave sources}

\bibitem{phi92}
Phinney, E.S., ``Pulsars as Probes of Newtonian Dynamical Systems'', {\em
  Philos. Trans. R. Soc. London}, {\bf 341}, 39--75, (1992).
  \epubtkKeywords{Binary systems, Pulsars, Pulsar evolution}

\bibitem{pb81}
Phinney, E.S., and Blandford, R.D., ``Analysis of the Pulsar $P$--$\dot{P}$
  distribution'', {\em Mon. Not. R. Astron. Soc.}, {\bf 194}, 137--148, (1981).
  \epubtkKeywords{Pulsar statistics, Pulsars, Galactic astronomy}

\bibitem{pk94}
Phinney, E.S., and Kulkarni, S.R., ``Binary and millisecond pulsars'', {\em
  Annu. Rev. Astron. Astrophys.}, {\bf 32}, 591--639, (1994).
  \epubtkKeywords{Pulsars, Binary systems, Pulsar evolution}

\bibitem{ps91}
Phinney, E.S., and Sigurdsson, S., ``Ejection of pulsars and binaries to the
  outskirts of globular clusters'', {\em Nature}, {\bf 349}, 220--223, (1991).
  \epubtkKeywords{Pulsars, Neutron stars, Binary systems}

\bibitem{pv91}
Phinney, E.S., and Verbunt, F., ``Binary pulsars before spin-up and PSR 1820 --
  11'', {\em Mon. Not. R. Astron. Soc.}, {\bf 248}, 21P--23P, (1991).
  \epubtkKeywords{Binary systems, Neutron stars, Pulsars}

\bibitem{pv96}
Portegies~Zwart, S.F., and Verbunt, F., ``Population synthesis of high-mass
  binaries'', {\em Astron. Astrophys.}, {\bf 309}, 179, (1996).
  \epubtkKeywords{Binary pulsars, Pulsar evolution}

\bibitem{pdc05}
Possenti, A., D'Amico, N., Conongiu, A., Manchester, R.N., Sarkissian, J.M.,
  Camilo, F., and Lyne, A.G., ``A dozen new pulsars from the Parkes globular
  cluster search'', in Rasio, F.A., and Stairs, I.H., eds., {\em Binary Radio
  Pulsars}, Proceedings of meeting held at the Aspen Center for Physics, USA,
  12--16 January 2004, vol. 328 of ASP Conference Series,  189--194,
  (Astronomical Society of the Pacific, San Francisco, U.S.A., 2005).
  \epubtkKeywords{Radio astronomy, Pulsars}

\bibitem{pdm03}
Possenti, A., D'Amico, N., Manchester, R.N., Camilo, F., Lyne, A.G.,
  Sarkissian, J., and Corongiu, A., ``Three Binary Millisecond Pulsars in NGC
  6266'', {\em Astrophys. J.}, {\bf 599}, 475--484, (2003).
  \epubtkKeywords{Radio astronomy, Pulsars, Binary systems}

\bibitem{pakw91}
Prince, T.A., Anderson, S.B., Kulkarni, S.R., and Wolszczan, A., ``Timing
  observations of the 8 hour binary pulsar 2127+11C in the globular cluster
  M15'', {\em Astrophys. J. Lett.}, {\bf 374}, L41--L44, (1991).
  \epubtkKeywords{Pulsars, Neutron stars, Binary systems, Gravitational wave
  sources}

\bibitem{pripsr}
Princeton University, ``Princeton Pulsar Group'', project homepage, (1995). URL
  (cited on 20 March 2001): \newline\url{http://pulsar.princeton.edu/}.
  \epubtkKeywords{Radio astronomy, Pulsars, Astronomical observations}

\bibitem{mbeampsr}
Pulsar Group of the Australia Telescope National Facility (ATNF), ``Parkes
  Multibeam Pulsar Survey'', project homepage, (2000). URL (cited on 2 December
  2000): \newline\url{http://www.atnf.csiro.au/research/pulsar/pmsurv/}.
  \epubtkKeywords{Radio astronomy, Pulsars}

\bibitem{rr95}
Rajagopal, M., and Romani, R.W., ``Ultra-low-frequency gravitatational
  radiation from massive black hole binaries'', {\em Astrophys. J.}, {\bf 446},
  543--549, (1995). \epubtkKeywords{Black holes, Pulsars, Gravitational waves}

\bibitem{ran83}
Rankin, J.M., ``Toward an empirical theory of pulsar emission. I. Morphological
  taxonomy'', {\em Astrophys. J.}, {\bf 274}, 333--358, (1983).
  \epubtkKeywords{Neutron stars, Pulsars}

\bibitem{ran01}
Ransom, S.M., {\em New search techniques for binary pulsars}, Ph.D. Thesis,
  (Harvard University, Harvard, U.S.A., 2001). \epubtkKeywords{Radio astronomy,
  Pulsars, Data analysis}

\bibitem{rce03}
Ransom, S.M., Cordes, J.M., and Eikenberry, S.S., ``A New Search Technique for
  Short Orbital Period Binary Pulsars'', {\em Astrophys. J. Lett.}, {\bf 589},
  911--920, (2003). \epubtkKeywords{Binary pulsars, Pulsar searches, Signal
  processing}

\bibitem{rgh01}
Ransom, S.M., Greenhill, L., Herrnstein, J.R., Manchester, R.N., Camilo, F.,
  Eikenberry, S.S., and Lyne, A.G., ``A Binary Millisecond Pulsar in Globular
  Cluster NGC 6544'', {\em Astrophys. J. Lett.}, {\bf 546}, L25--L28, (2001).
  \epubtkKeywords{Radio astronomy, Pulsars, Binary systems}

\bibitem{rhs05b}
Ransom, S.M., Hessels, J.H., Stairs, I.H., Kaspi, V.M., and Backer, D.C., ``A
  20-cm survey for pulsars in globular clusters using the GBT and Arecibo'', in
  Rasio, F.A., and Stairs, I.H., eds., {\em Binary Radio Pulsars}, Proceedings
  of meeting held at the Aspen Center for Physics, USA, 12--16 January 2004,
  vol. 328 of ASP Conference Series,  199--205, (Astronomical Society of the
  Pacific, San Francisco, U.S.A., 2005). \epubtkKeywords{Radio astronomy,
  Pulsars}

\bibitem{rhs05}
Ransom, S.M., Hessels, J.W.T., Stairs, I.H., Freire, P.C.C., Kaspi, V.M.,
  Camilo, F., and Kaplan, D.L., ``Twenty-One Millisecond Pulsars in Terzan 5
  Using the Green Bank Telescope'', {\em Science}, {\bf 307}(5711), 892--896,
  (2005). \epubtkKeywords{Radio astronomy, Pulsars, Globular clusters}

\bibitem{rsb04}
Ransom, S.M., Stairs, I.H., Backer, D.C., Greenhill, L.J., Bassa, C.G.,
  Hessels, J.W.T., and Kaspi, V.M., ``GBT Discovery of Two Binary Millisecond
  Pulsars in the Globular Cluster M30'', {\em Astrophys. J.}, {\bf 604}, 328,
  (2004). \epubtkKeywords{Radio astronomy, Pulsars, Binary systems}

\bibitem{rv89}
Ratnatunga, K.U., and van~den Bergh, S., ``The rate of stellar collapses in the
  galaxy'', {\em Astrophys. J.}, {\bf 343}, 713--717, (1989).
  \epubtkKeywords{Supernovae, Galactic astronomy, Gravitational collapse}

\bibitem{rtj96}
Ray, P.S., Thorsett, S.E., Jenet, F.A., van Kerkwijk, M.H., Kulkarni, S.R.,
  Prince, T.A., Sandhu, J.S., and Nice, D.J., ``A Survey for Millisecond
  Pulsars'', {\em Astrophys. J.}, {\bf 470}, 1103--1110, (1996).
  \epubtkKeywords{Pulsars, Radio astronomy}

\bibitem{rc69b}
Richards, D.W., and Comella, J.M., ``The period of pulsar NP 0532'', {\em
  Nature}, {\bf 222}, 551--552, (1969). \epubtkKeywords{Radio astronomy,
  Pulsars}

\bibitem{rom89}
Romani, R.W., ``Timing a millisecond pulsar array'', in \"Ogelman, H., and
  van~den Heuvel, E.P.J., eds., {\em Timing Neutron Stars}, Proceedings of the
  NATO Advanced Study Institute on Timing Neutron Stars, \c{C}esme, Izmir,
  Turkey, 4--15 April 1988, vol. 262 of NATO ASI Series, Series C,  113--117,
  (Kluwer, Dordrecht, Netherlands; Boston, U.S.A., 1989). \epubtkKeywords{Radio
  astronomy, Pulsars, Pulsar timing, Cosmic gravitational wave background}

\bibitem{rom92}
Romani, R.W., ``Populations of low-mass black hole binaries'', {\em Astrophys.
  J.}, {\bf 399}, 621--626, (1992). \epubtkKeywords{Black holes, White dwarfs,
  Binary systems}

\bibitem{rt83}
Romani, R.W., and Taylor, J.H., ``An Upper Limit on the Stochastic Background
  of Ultralow-Frequency Gravitational Waves'', {\em Astrophys. J. Lett.}, {\bf
  265}, L35--L37, (1983). \epubtkKeywords{Cosmic gravitational wave background,
  Pulsars, Pulsar timing}

\bibitem{rt91a}
Ryba, M.F., and Taylor, J.H., ``High Precision Timing of Millisecond Pulsars.
  I. Astrometry and Masses of the PSR 1855+09 System'', {\em Astrophys. J.},
  {\bf 371}, 739--748, (1991). \epubtkKeywords{Radio astronomy, Pulsars, Pulsar
  timing, Fundamental physics}

\bibitem{rt91b}
Ryba, M.F., and Taylor, J.H., ``High Precision Timing of Millisecond Pulsars.
  II. Astrometry, Orbital Evolution, and Eclipses of PSR 1957+20'', {\em
  Astrophys. J.}, {\bf 380}, 557--563, (1991). \epubtkKeywords{Radio astronomy,
  Pulsars, Pulsar timing, Fundamental physics}

\bibitem{sbm97}
Sandhu, J.S., Bailes, M., Manchester, R.N., Navarro, J., Kulkarni, S.R., and
  Anderson, S.B., ``The Proper Motion and Parallax of PSR J0437--4715'', {\em
  Astrophys. J. Lett.}, {\bf 478}, L95--L98, (1997). \epubtkKeywords{Pulsars,
  Neutron stars, White dwarfs, Binary systems}

\bibitem{saz78}
Sazhin, M.V., ``Opportunities for detecting ultralong gravitational waves'',
  {\em Sov. Astron.}, {\bf 22}, 36--38, (1978). \epubtkKeywords{Cosmic
  gravitational wave background, Pulsars, Pulsar timing}

\bibitem{sch68}
Scheuer, P.A.G., ``Amplitude variations of pulsed radio sources'', {\em
  Nature}, {\bf 218}, 920--922, (1968). \epubtkKeywords{Pulsars, Galactic
  astronomy}

\bibitem{srs86}
Segelstein, D.J., Rawley, L.A., Stinebring, D.R., Fruchter, A.S., and Taylor,
  J.H., ``New millisecond pulsar in a binary system'', {\em Nature}, {\bf 322},
  714--717, (1986). \epubtkKeywords{Pulsars, Neutron stars, Binary systems}

\bibitem{sw04}
Seiradakis, J.H., and Wielebinski, R., ``Morphology and characteristics of
  radio pulsars'', {\em Astron. Astrophys. Rev.}, {\bf 12}, 239--271, (2004).
  \epubtkKeywords{Pulsars, Radio emission}

\bibitem{sl96}
Shemar, S.L., and Lyne, A.G., ``Observations of pulsar glitches'', {\em Mon.
  Not. R. Astron. Soc.}, {\bf 282}, 677--690, (1996). \epubtkKeywords{Pulsar
  timing, Radio astronomy, Pulsars}

\bibitem{shk70}
Shklovskii, I.S., ``Possible causes of the secular increase in pulsar
  periods'', {\em Sov. Astron.}, {\bf 13}, 562--565, (1970).
  \epubtkKeywords{Neutron stars, Fundamental physics}

\bibitem{sp95}
Sigurdsson, S., and Phinney, E.S., ``Dynamics and interactions of binaries and
  neutron stars in globular clusters'', {\em Astrophys. J. Suppl. Ser.}, {\bf
  99}, 609--635, (1995). \epubtkKeywords{Binary dynamics, Globular clusters}

\bibitem{srh03}
Sigurdsson, S., Richer, H.B., Hansen, B.M., Stairs, I.H., and Thorsett, S.E.,
  ``A Young White Dwarf Companion to Pulsar B1620-26: Evidence for Early Planet
  Formation'', {\em Science}, {\bf 301}, 193--196, (2003).
  \epubtkKeywords{Pulsars}

\bibitem{sb76}
Smarr, L.L., and Blandford, R.D., ``The binary pulsar: Physical processes,
  possible companions and evolutionary histories'', {\em Astrophys. J.}, {\bf
  207}, 574--588, (1976). \epubtkKeywords{Accretion, Binary systems, Neutron
  stars, Pulsars}

\bibitem{smi03}
Smith, F.G., ``The radio emission from pulsars'', {\em Rep. Prog. Phys.}, {\bf
  66}, 173--238, (2003). \epubtkKeywords{Pulsars, Radio emission}

\bibitem{sns05}
Splaver, E.M., Nice, D.J., Stairs, I.H., Lommen, A.N., and Backer, D.C.,
  ``Masses, Parallax, and Relativistic Timing of the PSR J1713+0747 Binary
  System'', {\em Astrophys. J.}, {\bf 620}, 405--415, (2005).
  \epubtkKeywords{Pulsars, Binary systems}

\bibitem{sta05b}
Stairs, I.~H., ``Overview of Pulsar Tests of General Relativity'', in Rasio,
  F.A., and Stairs, I.H., eds., {\em Binary Radio Pulsars}, Proceedings of
  meeting held at the Aspen Center for Physics, USA, 12--16 January 2004, vol.
  328 of ASP Conference Series,  3--18, (Astronomical Society of the Pacific,
  San Francisco, U.S.A., 2005). \epubtkKeywords{Radio astronomy, Pulsars}

\bibitem{sta05}
Stairs, I.H., ``Prospects of measuring trigonometric parallax for PSR
  B1534+12'', personal communication. \epubtkKeywords{Pulsars, Tests of
  relativistic gravity}

\bibitem{sta98}
Stairs, I.H., {\em Observations of Binary and Millisecond Pulsars with a
  Baseband Recording System}, Ph.D. Thesis, (Princeton University, Princeton,
  U.S.A., 1998). \epubtkKeywords{Radio astronomy, Signal processing, Pulsars}

\bibitem{sta03}
Stairs, I.H., ``Testing General Relativity with Pulsar Timing'', {\em Living
  Rev. Relativity}, {\bf 6}, lrr-2003-5, (2003). URL (cited on 22 January
  2005): \newline\url{http://www.livingreviews.org/lrr-2003-5}.
  \epubtkKeywords{Radio astronomy, Pulsars}

\bibitem{sta04}
Stairs, I.H., ``Pulsars in Binary Systems: Probing Binary Stellar Evolution and
  General Relativity'', {\em Science}, {\bf 304}, 547--552, (2004).
  \epubtkKeywords{Radio astronomy, Pulsars}

\bibitem{sac98}
Stairs, I.H., Arzoumanian, Z., Camilo, F., Lyne, A.G., Nice, D.J., Taylor,
  J.H., Thorsett, S.E., and Wolszczan, A., ``Measurement of Relativistic
  Orbital Decay in the PSR B1534+12 Binary System'', {\em Astrophys. J.}, {\bf
  505}, 352--357, (1998). \epubtkKeywords{Pulsars, Neutron stars, Binary
  systems, Observational tests of relativity theory}

\bibitem{sfl05}
Stairs, I.H., Faulkner, A.J., Lyne, A.G., Kramer, M., Lorimer, D.R.,
  McLaughlin, M.A., Manchester, R.N., Hobbs, G.B., Camilo, F., Possenti, A.,
  Burgay, M., D'Amico, N., Freire, P.C.C., and Gregory, P.C., ``Discovery of
  three wide-orbit binary pulsars'', {\em Astrophys. J.}, submitted, (2005).
  \epubtkKeywords{Radio astronomy, Pulsars, Binary systems, Tests of
  relativity}

\bibitem{sml01}
Stairs, I.H., Manchester, R.N., Lyne, A.G., Kaspi, V.M., Camilo, F., Bell,
  J.F., D'Amico, N., Kramer, M., Crawford, F., Morris, D.J., McKay, N.P.F.,
  Lumsden, S.L., Tacconi-Garman, L.E., Cannon, R.D., Hambly, N.C., and Wood,
  P.R., ``PSR J1740-3052: a pulsar with a massive companion'', {\em Mon. Not.
  R. Astron. Soc.}, {\bf 325}, 979--988, (2001). \epubtkKeywords{Radio
  astronomy, Pulsars}

\bibitem{sml03}
Stairs, I.H., Manchester, R.N., Lyne, A.G., Kramer, M., Kaspi, V.M., Camilo,
  F., and D'Amico, N., ``The massive binary pulsar J1740-3052'', in Bailes, M.,
  Nice, D.J., and Thorsett, S.E., eds., {\em Radio Pulsars}, Proceedings of
  meeting held at Mediterranean Agonomic Institute of Chania, Crete, Greece,
  26--29 August 2002, vol. 302 of ASP Conference Series,  85--88, (Astronomical
  Society of the Pacific, San Fransisco, U.S.A., 2003). \epubtkKeywords{Radio
  astronomy, Pulsars}

\bibitem{sst00}
Stairs, I.H., Splaver, E.M., Thorsett, S.E., Nice, D.J., and Taylor, J.H., ``A
  Baseband Recorder for Radio Pulsar Observations'', {\em Mon. Not. R. Astron.
  Soc.}, {\bf 314}, 459--467, (2000). Related online version (cited on 19 March
  2001): \newline\url{http://arXiv.org/abs/astro-ph/9912272}.
  \epubtkKeywords{Radio astronomy, Signal processing, Pulsars}

\bibitem{sttw02}
Stairs, I.H., Thorsett, S.E., Taylor, J.H., and Wolszczan, A., ``Studies of the
  Relativistic Binary Pulsar PSR B 1534+12: I. Timing Analysis'', {\em
  Astrophys. J.}, {\bf 581}, 501--508, (2002). \epubtkKeywords{Pulsars, Tests
  of relativistic gravity}

\bibitem{sn96}
Standish, E.M., and Newhall, X.X., ``New accuracy levels for solar system
  ephemerides. (Lecture)'', in Ferraz-Mello, S., Morando, B., and Arlot, J.-E.,
  eds., {\em Dynamics, Ephemerides, and Astrometry of the Solar System},
  Proceedings of the 172nd Symposium of the International Astronomical Union,
  held in Paris, France, 3--8 July, 1995, ~29, (Kluwer, Dordrecht, Netherlands;
  Boston, U.S.A., 1996). \epubtkKeywords{Ephemerides, Astrometry}

\bibitem{sbl96}
Stappers, B.W., Bailes, M., Lyne, A.G., Manchester, R.N., D'Amico, N., Tauris,
  T.M., Lorimer, D.R., Johnston, S., and Sandhu, J.S., ``Probing the Eclipse
  Region of a Binary Millisecond Pulsar'', {\em Astrophys. J. Lett.}, {\bf
  465}, L119--L122, (1996). \epubtkKeywords{Pulsars, Neutron stars, Binary
  systems}

\bibitem{spw87}
Stella, L., Priedhorsky, W., and White, N.E., ``The discovery of a 685 second
  orbital period from the X-ray source 4U 1820--30 in the globular cluster NGC
  6624'', {\em Astrophys. J. Lett.}, {\bf 312}, L17--L21, (1987).
  \epubtkKeywords{Binary systems, Neutron stars}

\bibitem{scenario}
Sternberg Astronomical Institute, Moscow State University, ``Scenario Machine
  Engine'', project homepage, (2001). URL (cited on 19 March 2001):
  \newline\url{http://xray.sai.msu.ru/sciwork/scenario.html}.
  \epubtkKeywords{Binary systems, Neutron stars, White dwarfs, Black holes,
  Monte Carlo methods}

\bibitem{srtr90}
Stinebring, D.R., Ryba, M.F., Taylor, J.H., and Romani, R.W., ``Cosmic
  Gravitational-Wave Background: Limits from Millisecond Pulsar Timing'', {\em
  Phys. Rev. Lett.}, {\bf 65}, 285--288, (1990). \epubtkKeywords{Cosmic
  gravitational wave background, Pulsars, Pulsar timing}

\bibitem{simy03}
Sudou, H., Iguchi, S., Murata, Y., and Yoshiaki, T., ``Orbital motion in the
  radio galaxy 3C 66B: evidence for a supermassive black hole binary'', {\em
  Science}, {\bf 300}, 1263--1265, (2003). \epubtkKeywords{Radio astronomy,
  Black holes}

\bibitem{swinpsr}
Swinburne University of Technology, ``Centre for Astrophysics and
  Supercomputing'', institutional homepage, (2000). URL (cited on 8 December
  2000): \newline\url{http://astronomy.swin.edu.au}. \epubtkKeywords{Pulsars}

\bibitem{tm98}
Tauris, T.M., and Manchester, R.N., ``On the evolution of pulsar beams'', {\em
  Mon. Not. R. Astron. Soc.}, {\bf 298}, 625--636, (1998).
  \epubtkKeywords{Neutron stars, Pulsars}

\bibitem{ts99}
Tauris, T.M., and Savonije, G.J., ``Formation of millisecond pulsars. I.
  Evolution of low-mass X-ray binaries with {$P_{\rm orb}>2$} days'', {\em
  Astron. Astrophys.}, {\bf 350}, 928--944, (1999). \epubtkKeywords{Binary
  pulsars, Pulsar evolution}

\bibitem{tay91}
Taylor, J.H., ``Millisecond Pulsars: Nature's Most Stable Clocks'', {\em Proc.
  IEEE}, {\bf 79}, 1054--1062, (1991). \epubtkKeywords{Radio astronomy,
  Pulsars, Pulsar timing, Fundamental physics}

\bibitem{tay94}
Taylor, J.H., ``Binary Pulsars and Relativistic Gravity'', {\em Rev. Mod.
  Phys.}, {\bf 66}, 711--719, (1994). \epubtkKeywords{Pulsars, Neutron stars,
  Observational tests of relativity theory}

\bibitem{tc93}
Taylor, J.H., and Cordes, J.M., ``Pulsar Distances and the Galactic
  Distribution of Free Electrons'', {\em Astrophys. J.}, {\bf 411}, 674--684,
  (1993). \epubtkKeywords{Radio astronomy, Pulsars, Galactic astronomy}

\bibitem{tm77}
Taylor, J.H., and Manchester, R.N., ``Galactic distribution and evolution of
  pulsars'', {\em Astrophys. J.}, {\bf 215}, 885--896, (1977).
  \epubtkKeywords{Pulsars, Neutron stars, Pulsar statistics, Galactic
  astronomy}

\bibitem{tw82}
Taylor, J.H., and Weisberg, J.M., ``A New Test of General Relativity:
  Gravitational Radiation and the Binary Pulsar PSR 1913+16'', {\em Astrophys.
  J.}, {\bf 253}, 908--920, (1982). \epubtkKeywords{Pulsars, Tests of
  relativistic gravity}

\bibitem{tw89}
Taylor, J.H., and Weisberg, J.M., ``Further experimental tests of relativistic
  gravity using the binary pulsar PSR 1913+16'', {\em Astrophys. J.}, {\bf
  345}, 434--450, (1989). \epubtkKeywords{Pulsars, Tests of relativistic
  gravity}

\bibitem{nobpr1993}
The Nobel Foundation, ``The Nobel Prize in Physics 1993'', institutional
  homepage, (June, 2000). URL (cited on 19 March 2001):
  \newline\url{http://www.nobel.se/physics/laureates/1993}.
  \epubtkKeywords{Radio astronomy, Pulsars, Tests of relativistic gravity}

\bibitem{tacl99}
Thorsett, S.E., Arzoumanian, Z., Camilo, F., and Lyne, A.G., ``The Triple
  Pulsar System PSR B1620--26 in M4'', {\em Astrophys. J.}, {\bf 523},
  763--770, (1999). \epubtkKeywords{Radio astronomy, Pulsars, Fundamental
  physics}

\bibitem{tamt93}
Thorsett, S.E., Arzoumanian, Z., McKinnon, M.M., and Taylor, J.H., ``The Masses
  of Two Binary Neutron Star Systems'', {\em Astrophys. J. Lett.}, {\bf 405},
  L29--L32, (1993). \epubtkKeywords{Pulsars, Neutron stars, Binary systems}

\bibitem{tat93}
Thorsett, S.E., Arzoumanian, Z., and Taylor, J.H., ``PSR B1620-26: A binary
  radio pulsar with a planetary companion?'', {\em Astrophys. J. Lett.}, {\bf
  412}, L33--L36, (1993). \epubtkKeywords{Radio astronomy, Pulsars, Fundamental
  physics}

\bibitem{tc99}
Thorsett, S.E., and Chakrabarty, D., ``Neutron Star Mass Measurements. I. Radio
  Pulsars'', {\em Astrophys. J.}, {\bf 512}, 288--299, (1999).
  \epubtkKeywords{Pulsars, Binary systems, Neutron stars, White dwarfs}

\bibitem{td96}
Thorsett, S.E., and Dewey, R.J., ``Pulsar Timing Limits on Very Low Frequency
  Stochastic Gravitational Radiation'', {\em Phys. Rev. D}, {\bf 53}, 3468,
  (1996). \epubtkKeywords{Cosmic gravitational wave background, Pulsars, Pulsar
  timing}

\bibitem{tsb99}
Toscano, M., Sandhu, J.S., Bailes, M., Manchester, R.N., Britton, M.C.,
  Kulkarni, S.R., Anderson, S.B., and Stappers, B.W., ``Millisecond pulsar
  velocities'', {\em Mon. Not. R. Astron. Soc.}, {\bf 307}, 925--933, (1999).
  \epubtkKeywords{Pulsars, Neutron stars, Galactic astronomy}

\bibitem{tri68}
Trimble, V., ``Motions and Structure of the Filamentary Envelope of the Crab
  Nebula'', {\em Astrophys. J.}, {\bf 73}, 535--547, (1968).
  \epubtkKeywords{Neutron stars, Galactic astronomy}

\bibitem{ty93}
Tutukov, A.V., and Yungelson, L.R., ``The merger rate of neutron star and black
  hole binaries'', {\em Mon. Not. R. Astron. Soc.}, {\bf 260}, 675--678,
  (1993). \epubtkKeywords{Neutron stars, Black holes, Binary systems}

\bibitem{ubcpsr}
University of British Columbia, ``Ingrid Stairs / UBC Pulsar Group'', personal
  homepage, (2005). URL (cited on 22 January 2005):
  \newline\url{http://www.astro.ubc.ca/people/stairs}. \epubtkKeywords{Radio
  astronomy, Pulsars}

\bibitem{cagpsr}
University of Cagliari, ``The Italian Pulsar Group'', project homepage, (2005).
  URL (cited on 18 January 2005):
  \newline\url{http://pulsar.ca.astro.it/~pulsar}. \epubtkKeywords{Radio
  astronomy, Pulsars}

\bibitem{bkypsr}
University of California Berkeley Astronomy Department, ``Berkeley Pulsar
  Group'', project homepage, (2000). URL (cited on 20 March 2001):
  \newline\url{http://astro.berkeley.edu/~mpulsar/}. \epubtkKeywords{Radio
  astronomy, Pulsars}

\bibitem{jodpsr}
University of Manchester, ``Jodrell Bank Observatory Pulsar Group'', project
  homepage, (1997). URL (cited on 20 March 2001):
  \newline\url{http://www.jb.man.ac.uk/~pulsar/}. \epubtkKeywords{Radio
  astronomy, Pulsars}

\bibitem{epndb}
University of Manchester, ``The European Pulsar Network Data Archive'', web
  interface to database, (2005). URL (cited on 17 January 2005):
  \newline\url{http://www.jb.man.ac.uk/~pulsar/Resources/epn}.
  \epubtkKeywords{Radio astronomy, Pulsars, Astronomical observations}

\bibitem{vt91}
van~den Bergh, S., and Tammann, G.A., ``Galactic and Extragalactic Supernova
  Rates'', {\em Annu. Rev. Astron. Astrophys.}, {\bf 29}, 363--407, (1991).
  \epubtkKeywords{Supernovae, Galactic astronomy, Extragalactic astronomy,
  Gravitational collapse}

\bibitem{vdh94}
van~den Heuvel, E.P.J., ``The binary pulsar PSR J2145--0750: a system
  originating from a ow or intermediate mass X-ray binary with a donor star on
  the asymptotic giant branch?'', {\em Astron. Astrophys.}, {\bf 291},
  L39--L42, (1994). \epubtkKeywords{Neutron stars, Pulsars, White dwarfs,
  Binary systems}

\bibitem{vbjj05}
van Kerkwijk, M.H., Bassa, C.G., Jacoby, B.A., and Jonker, P.G., ``Optical
  studies of companions to millisecond pulsars'', in Rasio, F.A., and Stairs,
  I.H., eds., {\em Binary Radio Pulsars}, Proceedings of meeting held at the
  Aspen Center for Physics, USA, 12 Janaury - 16 January 2004, vol. 328 of ASP
  Conference Series,  357--369, (Astronomical Society of the Pacific, San
  Francisco, U.S.A., 2005). \epubtkKeywords{Optical astronomy, Pulsars, White
  dwarfs}

\bibitem{vk95}
van Kerkwijk, M.H., and Kulkarni, S.R., ``Spectroscopy of the white-dwarf
  companions of PSR B 0655+64 and 0820+02'', {\em Astrophys. J. Lett.}, {\bf
  454}, L141--L144, (1995). \epubtkKeywords{Pulsars, Neutron stars, Binary
  systems, White dwarfs}

\bibitem{vk99}
van Kerkwijk, M.H., and Kulkarni, S.R., ``A Massive White Dwarf Companion to
  the Eccentric Binary Pulsar System PSR B2303+46'', {\em Astrophys. J. Lett.},
  {\bf 516}, L25--L28, (1999). \epubtkKeywords{Neutron stars, White dwarfs,
  Pulsars, Binary systems}

\bibitem{van03}
van Straten, W., {\em High-Precision Timing and Polarimetry of PSR
  J0437--4715}, Ph.D. Thesis, (Swinburne University of Technology, Swinburne,
  Australia, 2003). \epubtkKeywords{Radio astronomy, Pulsars, Data analysis}

\bibitem{vn81}
Vivekanand, M., and Narayan, R., ``A New Look at Pulsar Statistics -- Birthrate
  and Evidence for Injection'', {\em J. Astrophys. Astron.}, {\bf 2}, 315--337,
  (1981). \epubtkKeywords{Pulsar statistics, Pulsars, Galactic astronomy}

\bibitem{wei96}
Weisberg, J.M., ``The Galactic Electron Density Distribution'', in Johnston,
  S., Walker, M.A., and Bailes, M., eds., {\em Pulsars: Problems and Progress
  (IAU Colloquium 160)}, Proceedings of the 160th Colloquium of the IAU, held
  at the Research Centre for Theoretical Astrophysics, University of Sydney,
  Australia, 8--12 January 1996, vol. 105 of ASP Conference Series,  447--454,
  (Astronomical Society of the Pacific, San Francisco, U.S.A., 1996).
  \epubtkKeywords{Pulsars, Galactic astronomy}

\bibitem{wt05}
Weisberg, J.M., and Taylor, J.H., ``The Relativistic Binary Pulsar B1913+16:
  Thirty Years of Observations and Analysis'', in Rasio, F.A., and Stairs,
  I.H., eds., {\em Binary Radio Pulsars}, Proceedings of meeting held at the
  Aspen Center for Physics, USA, 12 Janaury - 16 January 2004, vol. 328 of ASP
  Conference Series,  25--31, (Astronomical Society of the Pacific, San
  Francisco, U.S.A., 2005). \epubtkKeywords{Radio astronomy, Pulsars}

\bibitem{wex95}
Wex, N., ``The second post-Newtonian motion of compact binary-star systems with
  spin'', {\em Class. Quantum Grav.}, {\bf 12}, 983--1005, (1995).
  \epubtkKeywords{Relativistic gravity}

\bibitem{wz97}
White, N.E., and Zhang, W., ``Millisecond X-Ray Pulsars in Low-mass X-ray
  Binaries'', {\em Astrophys. J. Lett.}, {\bf 490}, L87--L90, (1997).
  \epubtkKeywords{Accretion, Neutron stars, Pulsars, Binary systems}

\bibitem{wij05}
Wijnands, R., ``Accretion-Driven Millisecond X-ray Pulsars'', in Lowry, J.A.,
  ed., {\em Trends in Pulsar Research}, (Nova Science, Hauppauge, U.S.A.,
  2005). Related online version (cited on 23 January 2005):
  \newline\url{http://arXiv.org/abs/astro-ph/0501264}. in press.
  \epubtkKeywords{X-ray astronomy, Millisecond pulsars, Binary systems}

\bibitem{wv98}
Wijnands, R., and van~der Klis, M., ``A millisecond pulsar in an X-ray binary
  system'', {\em Nature}, {\bf 394}, 344--346, (1998).
  \epubtkKeywords{Accretion, Neutron stars, Pulsars, Binary systems}

\bibitem{willtalk}
Will, C.M., ``Einstein's Relativity Put to Nature's Test: A Centennial
  Perspective'', lecture notes, American Physical Society, (1999). URL (cited
  on 8 December 2000):
  \newline\url{http://www.apscenttalks.org/pres_masterpage.cfm?nameID=99}.
  \epubtkKeywords{Tests of relativistic gravity}

\bibitem{wil01}
Will, C.M., ``The Confrontation between General Relativity and Experiment'',
  {\em Living Rev. Relativity}, {\bf 4}, lrr-2001-4, (2001). URL (cited on 24
  January 2005): \newline\url{http://www.livingreviews.org/lrr-2003-5}.
  \epubtkKeywords{General relativity}

\bibitem{wol91a}
Wolszczan, A., ``A nearby 37.9 ms radio pulsar in a relativistic binary
  system'', {\em Nature}, {\bf 350}, 688--690, (1991). \epubtkKeywords{Radio
  astronomy, Pulsars, Neutron stars, Binary systems}

\bibitem{wol94}
Wolszczan, A., ``Confirmation of Earth-mass planets orbiting the millisecond
  pulsar PSR 1257+12'', {\em Science}, {\bf 264}, 538--542, (1994).
  \epubtkKeywords{Pulsars, Pulsar timing, Fundamental physics}

\bibitem{wdk00}
Wolszczan, A., Doroshenko, O.V., Konacki, M., Kramer, M., Jessner, A.,
  Wielebinski, R., Camilo, F., Nice, D.J., and Taylor, J.H., ``Timing
  Observations of Four Millisecond Pulsars with the Arecibo and Effelsberg
  Radio Telescopes'', {\em Astrophys. J.}, {\bf 528}, 907--912, (2000).
  \epubtkKeywords{Radio astronomy, Pulsars}

\bibitem{wf92}
Wolszczan, A., and Frail, D.A., ``A planetary system around the millisecond
  pulsar PSR 1257+12'', {\em Nature}, {\bf 355}, 145--147, (1992).
  \epubtkKeywords{Radio astronomy, Pulsars}

\bibitem{yjb05}
York, T., Jackson, N., Browne, I.W.A., Wucknitz, O., and Skelton, J.E., ``The
  Hubble constant from the gravitational lens CLASS B0218+357 using the
  Advanced Camera for Surveys'', {\em Mon. Not. R. Astron. Soc.}, {\bf 357},
  124--134, (2005). \epubtkKeywords{Pulsars}

\end{thebibliography}

\end{document}